\documentclass[reprint,aps,prc,superscriptaddress,nofootinbib]{revtex4-1}
\usepackage{tikz}\usetikzlibrary{decorations.pathmorphing}
\usepackage{newtxtext}\usepackage[varvw,bigdelims]{newtxmath}
\usepackage[colorlinks=true,allcolors=blue]{hyperref}
\usepackage{amsmath,amssymb}
\usepackage{bm}
\usepackage{graphicx}
\date{September 24, 2019}

\begin{document}

\title{
Comparison of heavy-ion transport simulations: Collision integral with pions and $\Delta$~resonances~in~a~box
}

\author{Akira Ono}
\email{ono@nucl.phys.tohoku.ac.jp}
\affiliation{Department of Physics, Tohoku University, Sendai 980-8578, Japan}

\author{Jun Xu}
\email{xujun@zjlab.org.cn}
\affiliation{Shanghai Advanced Research Institute, Chinese Academy of Sciences, Shanghai 201210, China}
\affiliation{Shanghai Institute of Applied Physics, Chinese Academy of Sciences, Shanghai 201800, China}

\author{Maria Colonna}
\affiliation{INFN-LNS, Laboratori Nazionali del Sud, 95123 Catania, Italy}

\author{Pawel Danielewicz}
\affiliation{National Superconducting Cyclotron Laboratory and Department of Physics and Astronomy, Michigan State University, East Lansing, Michigan 48824, USA}

\author{Che Ming Ko}
\affiliation{Cyclotron Institute and Department of Physics and
Astronomy, Texas A\&M University, College Station, Texas 77843, USA}

\author{Manyee Betty Tsang}
\affiliation{National Superconducting Cyclotron Laboratory and
Department of Physics and Astronomy, Michigan State
University, East Lansing, Michigan 48824, USA}

\author{Yong-Jia Wang}
\affiliation{School of Science, Huzhou University, Huzhou 313000,
China}

\author{Hermann Wolter}
\affiliation{Physics Department, University of Munich, D-85748 Garching, Germany}

\author{Ying-Xun Zhang}
\affiliation{China Institute of Atomic Energy, Beijing 102413, China}
\affiliation{Guangxi Key Laboratory Breeding Base of Nuclear Physics and Technology, Guilin 541004, China}

\author{Lie-Wen Chen}
\affiliation{School of Physics and Astronomy and Shanghai Key
Laboratory for Particle Physics and Cosmology, Shanghai Jiao Tong
University, Shanghai 200240, China}

\author{Dan Cozma}
\affiliation{IFIN-HH, Reactorului 30, 077125 M\v{a}gurele-Bucharest,
Romania}

\author{Hannah Elfner}
\affiliation{GSI Helmholtzzentrum f\"{u}r Schwerionenforschung, Planckstr.\ 1, 64291 Darmstadt, Germany}
\affiliation{Institute for Theoretical Physics, Goethe University, Max-von-Laue-Strasse 1, 60438 Frankfurt am Main, Germany}
\affiliation{Frankfurt Institute for Advanced Studies, Johann Wolfgang Goethe University, Ruth-Moufang-Strasse 1, 60438 Frankfurt am Main, Germany}

\author{Zhao-Qing Feng}
\affiliation{School of Physics and Optoelectronics, South China University of Technology, Guangzhou 510640, China}

\author{Natsumi Ikeno}
\affiliation{Department of Life and Environmental Agricultural Sciences, Tottori University, Tottori 680-8551, Japan}
\affiliation{RIKEN Nishina Center, 2-1 Hirosawa, Wako, Saitama 351-0198, Japan}

\author{Bao-An Li}
\affiliation{Department of Physics and Astronomy, Texas A\&M University-Commerce, Commerce, Texas 75429-3011, USA}

\author{Swagata Mallik}
\affiliation{Physics Group, Variable Energy Cyclotron Centre, 1/AF Bidhan Nagar, Kolkata 700064, India}

\author{Yasushi Nara}
\affiliation{Akita International University, Akita 010-1292, Japan}

\author{Tatsuhiko Ogawa}
\affiliation{Research Group for Radiation Transport Analysis, Japan Atomic Energy Agency, Shirakata, Tokai, Ibaraki 319-1195, Japan}

\author{Akira Ohnishi}
\affiliation{Yukawa Institute for Theoretical Physics, Kyoto University, Kyoto 606-8502, Japan}

\author{Dmytro Oliinychenko}
\affiliation{Lawrence Berkeley National Laboratory, 1 Cyclotron Rd, Berkeley, CA 94720, USA}

\author{Jun Su}
\affiliation{Sino-French Institute of Nuclear Engineering \& Technology, Sun Yat-sen University, Zhuhai 519082, China}

\author{Taesoo Song}
\affiliation{GSI Helmholtzzentrum f\"{u}r Schwerionenforschung, Planckstr.\ 1, 64291 Darmstadt, Germany}

\author{Feng-Shou Zhang}
\affiliation{Key Laboratory of Beam Technology and Material Modification of Ministry of Education, College of Nuclear Science and Technology, Beijing Normal University, Beijing 100875, China}
\affiliation{Beijing Radiation Center, 100875 Beijing, China}

\author{Zhen Zhang}
\affiliation{Sino-French Institute of Nuclear Engineering \& Technology, Sun Yat-sen University, Zhuhai 519082, China}

\begin{abstract}
  \begin{description}
  \item[Background] Simulations by transport codes are indispensable for extracting valuable physical information from heavy-ion collisions.  Pion observables such as the $\pi^-/\pi^+$ yield ratio are expected to be sensitive to the symmetry energy at high densities.
  \item[Purpose] To evaluate, understand and reduce the uncertainties in transport-code results originating from different approximations in handling the production of $\Delta$ resonances and pions.
  \item[Methods] We compare ten transport codes under controlled conditions for a system confined in a box, with periodic boundary conditions, and initialized with nucleons at saturation density and at 60 MeV temperature.  The reactions $NN\leftrightarrow N\Delta$ and $\Delta\leftrightarrow N\pi$ are implemented, but the Pauli blocking and the mean-field potential are deactivated in the present comparison.  Thus these are cascade calculations including pions and $\Delta$ resonances.  Results are compared to those from the two reference cases of a chemically equilibrated ideal gas mixture and of the rate equation.
  \item[Results] For the numbers of $\Delta$ and $\pi$, deviations from the reference values are observed in many codes, and they depend significantly on the size of the time step.  These deviations are tied to different ways in ordering the sequence of reactions, such as collisions and decays, that take place in the same time step.  Better agreements with the reference values are seen in the reaction rates and the number ratios among the isospin species of $\Delta$ and $\pi$.  Both the reaction rates and the number ratios are, however, affected by the correlations between particle positions, which are absent in the Boltzmann equation, but are induced by the way particle scatterings are treated in many of the transport calculations.  The uncertainty in the transport-code predictions of the $\pi^-/\pi^+$ ratio, after letting the existing $\Delta$ resonances decay, is found to be within a few percent for the system initialized at $n/p = 1.5$.
  \item[Conclusions] The uncertainty in the final $\pi^-/\pi^+$ ratio in this simplified case of particles in a box is sufficiently small so that it does not strongly impact constraining the high-density symmetry energy from heavy-ion collisions.  Most of the sources of uncertainties have been understood, and individual codes may be further improved in future applications.  This investigation will be extended in the future to heavy-ion collisions to ensure the problems identified here remain under control.
  \end{description}
\end{abstract}

\maketitle

\section{Introduction\label{sec:intro}}

Heavy-ion collisions provide a unique opportunity to study in the laboratory the nuclear equation of state for a wide range of densities, temperatures and neutron--proton asymmetries.  However, in the evolution, transient partially out-of-equilibrium states are produced in the collisions, and this requires theoretical models to extract the nuclear equation of state from measured observables.  For collisions at incident energies between the Fermi-energy regime and several GeV/nucleon, transport equations are usually used to model the full quantum many-body dynamics under different levels of approximations, such as truncations of many-body correlations and semiclassical approximations.

Ideally, the determination of physical quantities from heavy-ion collisions should be independent of the numerical implementation of the transport equations.  Because of the complexity of transport equations and the numerical algorithms employed in individual transport codes, particularly the invoked statistical sampling and finite phase-space resolutions, careful checks of their accuracies are essential.  A first comparison of transport calculations at energies around 1 GeV/nucleon focusing on meson production was published in Ref.~\cite{kolomeitsev2005}.  Aiming at improved descriptions of heavy-ion collisions at energies between the Fermi-energy regime and several hundred MeV/nucleon, efforts have continued over the past years to compare and evaluate many different transport codes.  The result of the comparison of 19 transport codes in $\mathrm{Au}+\mathrm{Au}$ collisions at 100 and 400   MeV/nucleon was published in Ref.~\cite{xu2016}.  In this case, the differences among the results of transport codes seem to be originating in a complicated way from various sources, such as the differences in the initialization of the system, in the treatment of Pauli blocking of the two-nucleon ($NN$) collision term and to a lesser extent in the numerical integration in solving the propagation of nucleons in the mean-field potential.  In order to disentangle these different sources of uncertainties, it has been decided to perform comparisons under controlled conditions for systems confined in a box.  The first result of the box comparison was published in Ref.~\cite{yxzhang2018} where 15 transport codes were compared concentrating on the $NN$ elastic collision term without mean-field potentials, in a system with an initial Fermi-Dirac distribution at the temperature of either $T=0$ or 5 MeV.  One of the important findings there was that the differences among the codes are mainly due to inaccuracy in the evaluated Pauli-blocking factor, which is tightly linked to the fluctuations in the representation of the phase space in transport codes by a finite number of elements, e.g., Monte Carlo particles or so-called test particles.

It was proposed first by Li \cite{bali2002,bali2008} that the $\pi^-/\pi^+$ ratio of the yields of charged pions could be a sensitive probe of the nuclear symmetry energy at high densities, which has since stimulated many theoretical and experimental efforts.  However, divergent constraints on the nuclear symmetry energy were obtained so far by using different transport codes \cite{xiao2009,zqfeng2010,xie2013,hong2014} based on the same $\mathrm{Au}+\mathrm{Au}$ experimental data from the FOPI collaboration \cite{reisdorf2007}.  Recently, experiments were carried out with exotic beams of Sn isotopes at RIKEN/RIBF to measure charged pions from collisions of nuclei with various neutron-to-proton ratios.  To obtain meaningful physical information from measured pion data, it is an urgent and extremely important task to provide reliable predictions on the production of pions based on transport theories.  It should be noted that the $\pi^-/\pi^+$ ratio is expected to depend not only on the nuclear equation of state, but also on other physical ingredients such as the potentials for pions and $\Delta$ resonances \cite{ferini2005,xu2010,xu2013,song2015,zhenzhang2017,zhenzhang2018,bali2015,cozma2016,cozma2017,zqfeng2015,zqfeng2017}, the in-medium $NN$ cross sections \cite{guo2014ibuu04,cozma2016}, and the cluster correlations \cite{ikeno2016,ikeno2016erratum}.  It may also depend on the treatment of Pauli blocking \cite{ikeno2016erratum} and the momentum dependence of the nuclear mean field.  For reliable discussions on these physical problems by comparing the calculated results to the experimental data, we should first evaluate and hopefully eliminate uncertainties in the calculated results originating from unphysical sources.  Ideally, all transport codes should give the same result when the same physical ingredients are specified, or the differences should be understood as resulting from the different strategies used in implementing them.

In the present work, we carry out the comparison of transport codes for the simplified case of pions and $\Delta$ resonances in a box without mean-field potentials.  After a brief introduction of the participating codes in Sec.~\ref{sec:codes}, the conditions imposed on the calculations are described in Sec.~\ref{sec:homework}.  We allow the $NN\leftrightarrow N\Delta$ and $\Delta\leftrightarrow N\pi$ processes as well as elastic scatterings of two baryons.  The system is initialized with nucleons using the relativistic Boltzmann distribution at the temperature $T=60$ MeV.  In the early stage of a real heavy-ion collision, the relative momentum of the colliding nuclei determines the amount of inelastic collisions.  In the present system in a box, we simulate this effect by this rather high temperature.  Expecting that the Pauli blocking is not particularly important because of the high temperature, unlike in the situation of Ref.~\cite{yxzhang2018}, we turn off the Pauli blocking\footnote{We also ignore the Bose-Einstein enhancement factor in the collision term.} in all transport codes used in this comparison, so that the differences tied to other issues may be revealed clearly.  A comparison in more realistic situations of heavy-ion collisions is currently in progress.  The benefit of the present comparison in a box is that we know exactly all the physical quantities of this thermally and chemically equilibrated system to which the solution of transport equations should converge after a sufficiently long time.  In fact, we will see that some of the reaction rates and the specific ratios of the chemical composition of particles are reproduced rather well by all transport codes.  However, for some other important quantities, we also find unexpectedly large differences among the code results and relative to the equilibrium values.  Without going into great detail, in Sec.~\ref{sec:digest}, we give an overview of the most important aspects of the results.  Although uncertainties in the transport-code results may be judged superficially from these results, a real understanding of its implications requires a deeper understanding of the transport equations and the methods used in solving them.  After reviewing and preparing some theoretical backgrounds in Sec.~\ref{sec:tptheory}, we dedicate the later sections (Sec.~\ref{sec:ndelta}, Sec.~\ref{sec:ndeltapi} and Sec.~\ref{sec:noneq}) to thorough analyses.  We finally conclude that most of the remaining differences among the results of the transport codes are well understood as originating from their different methods of modeling, such as different implementations of common numerical methods and as intentions to represent different physics details.  In particular, these are mainly related to the processes for $\Delta$ resonances and pions, which were not studied in the former comparisons presented in Refs.~\cite{xu2016,yxzhang2018}.

\section{Participating codes\label{sec:codes}}

\begin{table}
\caption{\label{tab:codes} Code acronyms, code type (BUU or QMD), correspondents and representative references for the ten codes participating in the present comparison.  BUU(p) and BUU(f) denote BUU codes that employ the parallel-ensemble and full-ensemble methods, respectively.}
\begin{tabular}{llll}
  \hline\hline
  Acronym & Type & Code correspondents & Ref. \\
  \hline
  BUU-VM \footnote{BUU code developed jointly at VECC and McGill.}
          & BUU(p)& Mallik &  \cite{mallik2014,mallik2015,mallik2015hybrid}\\
  IBUU
          & BUU(p)& Xu, Chen, Li & \cite{bali2008,bali1997,bali1998,lwchen2014} \\
  IQMD-BNU
          & QMD & Su, F.S.~Zhang & \cite{su2011,su2013,su2014}\\
  IQMD-IMP \footnote{Also known as LQMD in literature.}
          & QMD & Feng  &\cite{zqfeng2011,zqfeng2012}\\
  JAM
          & QMD & Ikeno, Ono, Nara, Ohnishi & \cite{nara1999,ikeno2016}\\
  JQMD
          & QMD & Ogawa & \cite{niita1995,ogawa2015}\\
  pBUU
          & BUU(f)&  Danielewicz & \cite{danielewicz2000,danielewicz1991}\\
  RVUU
          & BUU(p)& Song, Z.~Zhang, Ko & \cite{song2015,ko1988,ko1996}\\
  SMASH
          & BUU(f)& Oliinychenko, Elfner & \cite{weil2016}\\
  TuQMD
          & QMD & Cozma & \cite{khoa1992,maheswari1998,shekhter2003,cozma2013}\\
    \hline
    \hline
    \end{tabular}
\end{table}

Table \ref{tab:codes} lists the 10 transport codes that participated in the present comparison.  There are two types of transport theories that are widely used for heavy-ion collisions in the energy region considered in the present work.  One type aims to solve the Boltzmann--Uehling--Uhlenbeck (BUU) equation for the time evolution of the one-body phase-space distribution.  One set of BUU codes employed in practice represent the phase-space distribution by using the test particle method.  The solution to the BUU equation is then obtained by following the motions of these test particles in the mean field and the collisions between them.  These codes are called the full-ensemble BUU codes if all pairs of test particles are considered for the possibility of collisions.  There is another set of BUU codes, called the parallel-ensemble BUU codes, in which test particles are grouped into sub-ensembles with each containing the same number of test particles as that of the physical particles, and collisions are considered only within each sub-ensemble.\footnote{In general, the mean-field potentials and the Pauli-blocking factors are calculated by using the test particles in all sub-ensembles.}  The other type of transport theory employed is the quantum molecular dynamics (QMD) model that puts more emphasis on many-body correlations.  In this approach, wave packets, with each of them corresponding to a nucleon, move classically under the forces between them, which approximately corresponds to the propagation in the mean field.  Wave packets can also collide and are scattered to random directions, which is similar to how the collision term is handled in the BUU codes.  In the present comparison, since we turn off the mean-field interaction and the Pauli blocking in the collision term, the parallel-ensemble BUU codes are expected to work equivalently to the QMD codes.

Most of the participating codes are developed for studying heavy-ion collisions in the energy regime where the mean-field effects are indispensable.  Since the propagation of particles in these codes is described by solving their equations of motion using certain time step $\Delta t$, one would naturally ask what is the number of particle collisions and their ordering during this time step.  For sufficiently small $\Delta t$, the ordering of particle collisions should not matter much.  In the previous comparison study presented in Ref.~\cite{yxzhang2018}, where only $NN$ elastic collisions were considered and the $\Delta t$ was taken to be 0.5 or 1 fm/$c$, no significant differences were found among the results from different codes.  The role of the time step in the integration of the transport equation was already studied in the early development of transport codes, e.g.~in Ref.~\cite{aichelin1985}.  In the present case, however, we find unexpectedly strong $\Delta t$ dependence in the results from many of these codes.  One of the main outcomes of the present work is that we understand how this issue is caused by the adopted prescriptions for handling the sequence of particle collisions and decays of resonances.

Another key concept to understand the transport-code results is the correlation induced inevitably by the geometrical prescription used for treating particle collisions.  In many transport codes, a pair of particles is assumed to collide at their closest approach if the distance is within the range of the cross section.  Although this seems to be a physically reasonable prescription, it is not quite identical to the collision term in the BUU equation that does not include particle correlations. For example, when two particles have collided, transport codes forbid them to repeat collisions with each other, but they can still collide again after one of them is scattered by some other particle around them.  Such higher-order correlations exist in the calculations of most transport codes.  We have seen in the previous comparison study in Ref.~\cite{yxzhang2018} that particle correlations enhance the $NN$ elastic collision rate in many codes, although impacts of this enhancement on observables are still not clear.  In the present study with the inclusion of pions and $\Delta$ resonances, we find that particle correlations can affect observables such as the $\pi^-/\pi^+$ ratio.  The correlation can, in principle, be a true physical effect, but we find that it sometimes affects the results as if the isospin symmetry were broken in transport codes.  We will clarify how this can happen in these codes and how strongly it may affect some important observables.

\section{Homework description\label{sec:homework}}

The participating codes were to carry out box calculations for the present comparison under the conditions specified below.

\subsection{Common setup}

The system should be confined in a box with periodic boundary conditions in the same way as in Ref.~\cite{yxzhang2018}.  The dimensions of the cubic box are $L_k=20$ fm with $k=x,y,z$.  A particle that leaves the box on one side should be regarded as entering it from the opposite side with the same momentum.  The only necessary change in the code is to redefine the separation $\Delta r_{ij,k}=r_{i,k}-r_{j,k}$ between two points $\bm{r}_i$ and $\bm{r}_j$ to $\mathop{\mathrm{modulo}}(\Delta r_{ij,k}+\frac12L_k,\ L_k)-\frac12L_k$, where the modulo function is the remainder after division by $L_k$, defined to take a value between 0 and $L_k$.  This method is completely sufficient and can cope with all aspects of calculations, as long as the characteristic lengths, such as the collision distance $\sqrt{\sigma/\pi}$, are shorter than $\frac12L_k$.  When a particle $i$ has moved out of the box, the code may optionally shift the coordinate into the box as $\mathop{\mathrm{modulo}}(r_{i,k},L_k)$.

The system is initialized with 1280 nucleons and without any other particles, which corresponds to the baryon number density $\rho=0.16\ \text{fm}^{-3}$ with the box size $L_k=20$ fm.  We study two cases of an isospin symmetric system, initialized with 640 neutrons and 640 protons [$\delta=(N-Z)/(N+Z)=0$], and an isospin asymmetric system, initialized with 768 neutrons and 512 protons ($\delta=0.2$ or $N/Z=1.5$, which is of the order of asymmetries reachable in real heavy-ion collisions).  The positions of nucleons should be uniformly distributed in the box at initialization.  The momenta of nucleons should be initialized by following the relativistic Boltzmann distribution
\begin{equation}
f(\bm{p})\propto e^{-(1/T)\sqrt{m_N^2+\bm{p}^2}},
\end{equation}
with the temperature parameter $T=60$ MeV and the nucleon mass $m_N$. 

For the box calculations considered in this paper, we deactivate nuclear mean field and electromagnetic interactions on any particle.  We also turn off Pauli blocking of the final states of a collision.  We further assume isotropic elastic scatterings with a constant cross section $\sigma_{\text{el}}=40$ mb for any pair of two baryons, i.e.~for $NN$, $N\Delta$ and $\Delta\Delta$, which help to thermalize these baryons.  Inelastic cross sections are described later in detail.  Any artificial threshold or cut on the c.m.~energy or distance should not be implemented.  Unphysical scatterings must be removed, i.e., after a collision happened for a pair of two particles, the same pair should not collide again until one of them collides with some other particle.  For the nucleon and pion masses, they are taken to be $m_N=0.938$ GeV and $m_\pi=0.139$ GeV, respectively.\footnote{
  The pBUU code used slightly different masses.
}

In all calculations, the system should be evolved from $t=0$ to 150 fm/$c$.  However, for the first 10 fm/$c$, we require to let the system evolve only with elastic scatterings for relaxation.  For the time step, a value of $\Delta t=0.5$ or 1 fm/$c$ was recommended.\footnote{Unless otherwise stated, we show the results with the time step that the code authors chose following this recommendation.  We will also show the results with $\Delta t=0.2$ fm/$c$ later.}  For QMD codes and parallel-ensemble BUU codes, simulations from 1000 events are carried out in each case, but only 10 events with 100 test particles per physical particle are required for the full-ensembled BUU codes.  An exception in the latter case has been pBUU, operated for the comparison with 1 event at 1000 test particles per physical particle (see Sec.~\ref{sec:pBUU}).

\subsection{$NN\leftrightarrow N\Delta$ processes\label{sec:formula-De2P0}}

We choose the $NN\to N\Delta$ cross section to be isotropic so that it agrees with the energy-dependent parametrization given in Ref.~\cite{bertsch1988} for the isospin-averaged cross section.  Considering the isospin dependence, it is given by\footnote{
The notation $\sigma_{NN\to N\Delta(m)}$ stands for the differential cross section to produce a $\Delta$ particle with a specific mass $m$, and it may also be written as $2\pi d\sigma_{NN\to N\Delta}/dm$.
}
\begin{equation}
  \label{eq:sigma-NNND}
  \begin{split}
    \sigma_{NN\to N\Delta(m)}&=
    \tfrac43C_{NNN\Delta}
    \frac{20\ \text{mb}\times(\sqrt{s}-M_{\text{th}})^2}{(0.015\ \text{GeV}^2)+(\sqrt{s}-M_{\text{th}})^2}
    \\ &\quad\times
    P(m,s)
  \end{split}
\end{equation}
for $\sqrt{s}>M_{\text{th}} = 2m_N+m_\pi$, and it is zero for $\sqrt{s}\le M_{\text{th}}$.  The isospin Clebsh-Gordan factor is
\begin{equation}
  C_{NNN\Delta}
  =\begin{cases}
    \frac34 & \mbox{for}\ nnp\Delta^-,\ ppn\Delta^{++}\\
    \frac14 & \mbox{for}\ nnn\Delta^0,\ ppp\Delta^+,\ npn\Delta^+,\ npp\Delta^0\\
    0 & \mbox{otherwise}.
  \end{cases}
\end{equation}
The last factor $P(m,s)$ in Eq.~\eqref{eq:sigma-NNND} represents a normalized probability distribution for the mass of produced $\Delta$ and is taken to be
\cite{danielewicz1991}
\begin{equation}
  \label{eq:mass-sampling}
  P(m,s)=\frac{p^*(s,m_N,m)m A(m)}
  {\int_{m_N+m_\pi}^{\sqrt{s}-m_N}p^*(s,m_N,m')m'A(m')\frac{dm'}{2\pi}}
\end{equation}
for $m\in[m_N+m_\pi,\ \sqrt{s}-m_N]$, and $P(m,s)=0$ otherwise.  Here $A(m)$ is the spectral function of $\Delta$, to be defined below, and $p^*(s,m_N,m)$ is the momentum of a particle in the c.m.~frame of the two particles with the given masses [see Eq.~\eqref{eq:pincm} below].  Because of this $p^*$ factor, the distribution $P(m,s)$ vanishes at the upper bound $m=\sqrt{s}-m_N$.  The probability distribution is normalized as
\begin{equation}
  \int_{m_N+m_\pi}^{\sqrt{s}-m_N}
  P(m,s)\frac{dm}{2\pi}=1.
\end{equation}

For the spectral function $A(m)$ of $\Delta$, we take a Breit-Wigner form
\begin{equation}
  \label{eq:Aspect}
  A(m)=\frac{1}{\tilde{A}}\frac{4M_\Delta^2\Gamma(m)}
  {(m^2-M_\Delta^2)^2+M_\Delta^2\Gamma^2(m)}
\end{equation}
with $M_\Delta=1.232$ GeV and a mass-dependent width parameter
\begin{equation}
  \label{eq:Gamma}
  \Gamma(m)=\frac{0.47 q^3}{m_\pi^2+0.6 q^2},
\end{equation}
where $q=p^*(m^2,m_N,m_\pi)$ is the pion momentum in a $\Delta\to N\pi$ decay.  With the normalization factor $\tilde{A}=0.95$, the spectral function is approximately normalized,
\begin{equation}
\int_{m_N+m_\pi}^\infty A(m)\frac{dm}{2\pi}\approx 1.
\end{equation}
Since $\Gamma(m)$ vanishes at the threshold $m=m_N+m_\pi$, the distribution $P(m,s)$ of Eq.~\eqref{eq:mass-sampling}, as well as the spectral function $A(m)$, also vanishes at the threshold.

The $N\Delta\to NN$ cross section is related to the $NN\to N\Delta$ cross section by the detailed balance condition,
\begin{equation}
  \label{eq:detbal}
  \sigma_{N\Delta(m)\to NN}
  =\frac{1}{1+\delta_{NN}}
  \frac{g_N}{g_\Delta A(m)}
  \frac{p^{*2}_{NN}}{p^{*2}_{N\Delta(m)}}
  \sigma_{NN\to N\Delta(m)},
\end{equation}
with $p^*_{NN}=p^*(s,m_N,m_N)$ and $p^*_{N\Delta(m)}=p^*(s,m_N,m)$.
The spin degeneracy factors are $g_N=2$ and $g_\Delta=4$.  It is also possible to define a transition matrix element in this context by
\begin{equation}
\begin{split}
  \frac{|\mathcal{M}_{NN\to N\Delta(m)}|^2}{16\pi s}
  &=
  \frac{\frac43\times
    20\ \text{mb}\times(\sqrt{s}-M_{\text{th}})^2}{(0.015\ \text{GeV}^2)+(\sqrt{s}-M_{\text{th}})^2}
  \\
  & \times
  \frac{p^*(s,m_N,m_N) m}
  {\int_{m_N+m_\pi}^{\sqrt{s}-m_N}p^*(s,m_N,m')m'A(m')\frac{dm'}{2\pi}},
\end{split}
\end{equation}
and to express the cross sections in a symmetric form as
\begin{align}
  \sigma_{NN\to N\Delta(m)}
  &=C_{NNN\Delta}
  \frac{|\mathcal{M}_{NN\to N\Delta(m)}|^2}{16\pi s}
  \frac{p^*_{N\Delta(m)}}{p^*_{NN}} A(m),
  \\
  \sigma_{N\Delta(m)\to NN}
  &=\frac{C_{NNN\Delta}}{1+\delta_{NN}}
  \frac{|\mathcal{M}_{N\Delta(m)\to NN}|^2}{16\pi s}
  \frac{p^*_{NN}}{p^*_{N\Delta(m)}},
\end{align}
with the relation for the matrix elements,
\begin{equation}
  g_N g_N|\mathcal{M}_{NN\to N\Delta(m)}|^2
  =g_N g_\Delta|\mathcal{M}_{N\Delta(m)\to NN}|^2.
\end{equation}

\subsection{$\Delta\leftrightarrow N\pi$ processes\label{sec:formula-De2Pe}}

In addition to the processes described above, the decay of $\Delta\rightarrow N\pi$ and its inverse process $N\pi\rightarrow\Delta$ are also taken into account in the present study.  Any other processes to produce pions, such as the s-wave pion production are, however, turned off in this homework study.  The pion absorption processes other than $N\pi\to\Delta$ are also turned off.

The rate for the decay $\Delta\rightarrow N\pi$ to a specific channel is
\begin{equation}
\label{eq:Gamma_c}
\Gamma_{\Delta(m)\rightarrow N\pi}=C_{\Delta N\pi}\Gamma(m),
\end{equation}
where the total decay width $\Gamma(m)$ is the same as that in the $\Delta$ spectral function $A(m)$ [Eqs.~\eqref{eq:Gamma} and \eqref{eq:Aspect}].  The isospin Clebsh-Gordan factor is
\begin{equation}
  C_{\Delta N\pi}
  =\begin{cases}
    1 & \mbox{for}\ \Delta^-\leftrightarrow n\pi^-,\ \Delta^{++}\leftrightarrow p\pi^+\\
    \frac23 & \mbox{for}\ \Delta^0\leftrightarrow n\pi^0,\ \Delta^+\leftrightarrow p\pi^0\\
    \frac13 & \mbox{for}\ \Delta^0\leftrightarrow p\pi^-,\ \Delta^+\leftrightarrow n\pi^+\\
    0 & \mbox{otherwise}.
  \end{cases}
\end{equation}
The $N\pi\to\Delta$ cross section, related to the $\Delta\to N\pi$ rate by detailed balance, is
\begin{equation}
\begin{split}
  \sigma_{N\pi\to\Delta(m)}
  &=\frac{g_\Delta}{g_N g_\pi}C_{\Delta N\pi}
  \frac{\pi}{[p^*(s,m_N,m_\pi)]^2}\Gamma(\sqrt{s})A(\sqrt{s})
  \\
  &\quad\times2\pi\delta(m-\sqrt{s})
\end{split}
\end{equation}
with $g_\pi=1$.  The mass of produced $\Delta$ is determined by the energy $\sqrt{s}$ in the c.m.~frame, as expressed by the delta function on the right-hand side of the equation above.

\section{Digest of results\label{sec:digest}}
\subsection{Numbers of $\Delta$ and $\pi$}

\begin{figure*}
  \includegraphics[width=\textwidth]{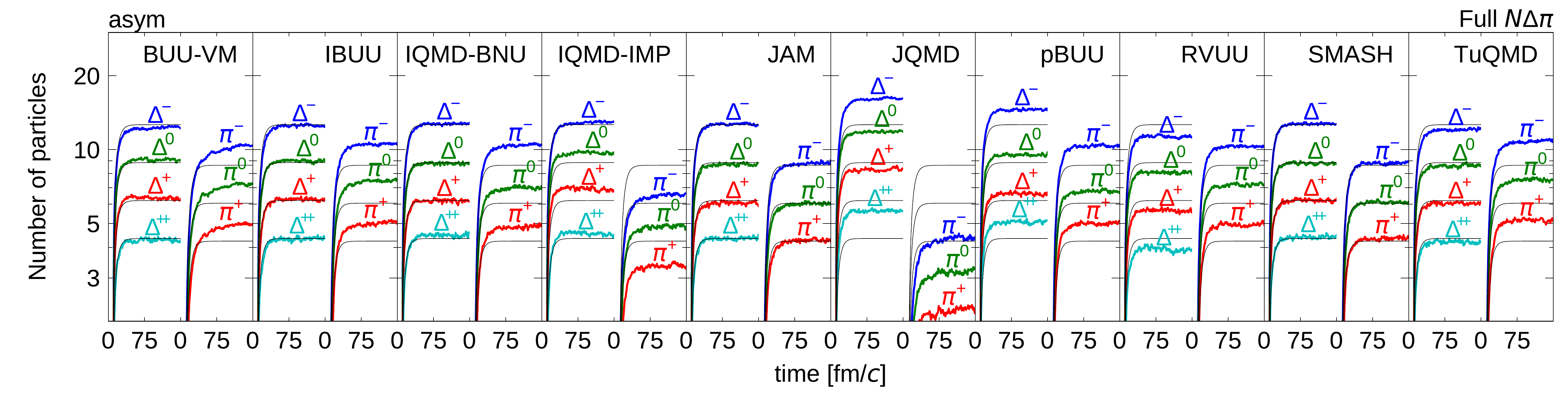}
  \caption{\label{fig:nat1a} Time evolution of the numbers of $\Delta$ and $\pi$ in an asymmetric ($\delta=0.2$) full-$N\Delta\pi$ system.  Results from the rate equation are represented by thin lines.}
\end{figure*}

Although our main goal in the experimental context may be the prediction of the $\pi^-/\pi^+$ ratio, we here start showing basic information on the absolute numbers of particles.  Figure \ref{fig:nat1a} shows the time evolution of the numbers of particles in the case of an asymmetric system ($\delta=0.2$).  The results from different codes are shown in different panels by colored thick lines.  For the time interval of $0<t<150$ fm/$c$ in the calculations, the evolutions of $\Delta$ and $\pi$ are shown side by side in each panel.  After the production of $\Delta$ resonances sets in at $t=10$ fm/$c$ in our homework condition, the numbers of $\Delta$ and $\pi$ increase and reach equilibrium rather quickly in the time scale shown here.  As a reference, results from the rate equation are shown by thin lines in each panel.  The rate equation, which is described in Appendix~\ref{sec:rateeq}, assumes thermally equilibrated momentum distributions at any instant but without the assumption of chemical equilibrium.  As a result, results from the rate equation do not have to agree quantitatively with those from the Boltzmann equation simulated by transport codes, in particular at early times.  However, both results should agree at late times with those of an ideal relativistic Boltzmann-gas mixture at chemical equilibrium, which can be easily calculated exactly as in Appendix~\ref{sec:equil}.  In the first part of this section, we focus on these equilibrated particle numbers at late times.

At a first glance of Fig.~\ref{fig:nat1a}, we find deviations in the numbers of $\Delta$ and $\pi$ ($N_\Delta$ and $N_\pi$) among different codes and with respect to the reference case of the rate equation.  Many codes (BUU-VM, IBUU, IQMD-BNU, pBUU, RVUU and TuQMD) overestimate $N_\pi$ by 20\% or more, while IQMD-IMP and JQMD underestimate it.  The deviations of $N_\Delta$ are not as serious as those of $N_\pi$ in most codes, but it seems difficult to find any systematic rule tying the deviations of $N_\pi$ and $N_\Delta$.  However, some codes (JAM and SMASH) agree with the reference case relatively well for both $N_\pi$ and $N_\Delta$.

The difference in $N_\Delta$ and $N_\pi$ among codes may be a serious issue because it may affect the predictions for heavy-ion collisions, where the number of finally emitted pions is related to $N_\Delta$ and $N_\pi$ at intermediate times.  Furthermore, the difference in $N_\Delta$ and $N_\pi$ can affect the dynamics of heavy-ion collisions, when these particles are propagated under the mean-field potential.  For example, since pions move rapidly because of their light masses, the codes with high $N_\pi$ are expected to predict rapid escape of many pions from the high-density region of heavy-ion collisions, while the codes with low $N_\pi$, at the cost of high $N_\Delta$, may predict that pions are emitted later and more equilibrated because of $\Delta$ particles moving slower and staying longer in the dense region of the reaction.

Such deviations in $N_\pi$ and $N_\Delta$ are surprising in view of the simple setup of the present homework with only the collision term without mean field and Pauli blocking.  In principle, we cannot draw any reasonable conclusion until we can understand the origin of these deviations and their impacts on other observables, by undertaking detailed analyses in the later sections, and delving into the characteristics of individual codes.  In this section, we thus only put forward statements that will be supported by the detailed analyses.

As mentioned in Sec.~\ref{sec:codes}, many codes rely on time steps to solve the transport equation.  If we consider the fact that the $N\pi\to\Delta$ cross section is large and the lifetime of $\Delta$ is not very long, the results may depend on the value of the time step $\Delta t$.  In the present comparison, many codes use $\Delta t=0.5$ fm/$c$ except the BUU-VM and JQMD codes that use a larger value of $\Delta t=1.0$ fm/$c$.  The large deviations of $N_\pi$ and $N_\Delta$ from JQMD in Fig.~\ref{fig:nat1a} are likely due to this choice of $\Delta t$.  On the other hand, two of the participating codes (JAM and SMASH) do not rely on time steps owing to their numerical method, in particular when the mean-field interaction is turned off.  These are called time-step--free codes in the present paper.  It is probably not accidental that these codes reproduce the true equilibrium values of $N_\pi$ and $N_\Delta$ very well as shown in Fig.~\ref{fig:nat1a}.  Thus the treatment of time steps is a key issue in interpreting transport-code results.  Detailed analyses are required to understand the different ways deviations emerge for different codes.  In the later sections, we will find that the deviations in $N_\pi$ and $N_\Delta$, which strongly depend on $\Delta t$ (as we will see in Fig.~\ref{fig:dt_nats}), are mainly due to the different ways collisions and decays are ordered within the same time step.

\subsection{Isotopic ratios}

In spite of the significant deviations in the absolute numbers $N_\pi$ and $N_\Delta$ from these transport codes, one can see in Fig.~\ref{fig:nat1a} that the ratios among isospin species of $\pi$ and $\Delta$ are more or less as expected, i.e., the lines for particle numbers tend to be equally spaced in these semi-logarithmic plots as they should be in the ideal Boltzmann-gas mixture under chemical equilibrium.  Thus one may still hope that transport codes can predict the isotopic ratios of these particles faithfully.  The charged pion ratio $\pi^-/\pi^+$ observed in heavy-ion collisions is expected to be sensitive to the high-density symmetry energy, since it depends on the neutron-to-proton ratio ($n/p$) in the compressed region.  To reliably constrain the high-density symmetry energy from measured $\pi^-/\pi^+$ ratio, transport codes are required to accurately describe the mechanism through which the information on $n/p$ is reflected in the observed $\pi^-/\pi^+$ ratio.  In the present comparison, this problem is studied under the simple condition of nuclear matter in a box without the ambiguities due to the treatments of mean-field potentials, in-medium effects and the Pauli-blocking factors.  Only after this is understood in a code, it can reliably predict the $\pi^-/\pi^+$ ratio for heavy-ion collisions.  To obtain a stringent constraint on the characteristics of nuclear symmetry energy at high density, beyond a rough discrimination between the soft and stiff density dependencies, an accuracy of at least 5\% is needed for the predicted $\pi^-/\pi^+$ ratio from a transport code.\footnote{The qualitative statements in the present paper on the agreement of results depend on this target accuracy that we choose here.  Quantitative results are always given, so the statements on the quality can be translated depending on the purpose.}

For the isotopic ratios to assess among $\pi^-$, $\pi^0$ and $\pi^+$, and among $\Delta^-$, $\Delta^0$, $\Delta^+$ and $\Delta^{++}$, we select the following three ratios of particle numbers:
\begin{subequations}
  \label{eq:piratios}
\begin{align}
  \mbox{$\pi$ ratio}
  &= \pi^-/\pi^+ \\
  \mbox{$\Delta$($\pi$-like) ratio}
  &= \frac{\Delta^-+\tfrac13\Delta^0}{\Delta^{++}+\tfrac13\Delta^+}\\
  \mbox{$\pi$-like ratio}
  &= \frac{\pi^-+\Delta^-+\tfrac13\Delta^0}{\pi^++\Delta^{++}+\tfrac13\Delta^+}
\end{align}
\end{subequations}
These ratios are expected to depend strongly on the $n/p$ ratio, e.g.~$\pi^-/\pi^+=(n/p)^2$ for the $\pi$ ratio in the chemically equilibrated ideal Boltzmann-gas mixture, which is expected to be realized in the transport models without the Pauli blocking and the Bose-Einstein enhancement.  The $\pi$-like ratio is intermediate between the $\Delta$($\pi$-like) ratio and the $\pi$ ratio.  It corresponds to the observed $\pi^-/\pi^+$ ratio if the equilibrated particles suddenly froze out and the decay of $\Delta$ resonances were included.

\begin{figure*}
  \includegraphics[width=\textwidth]{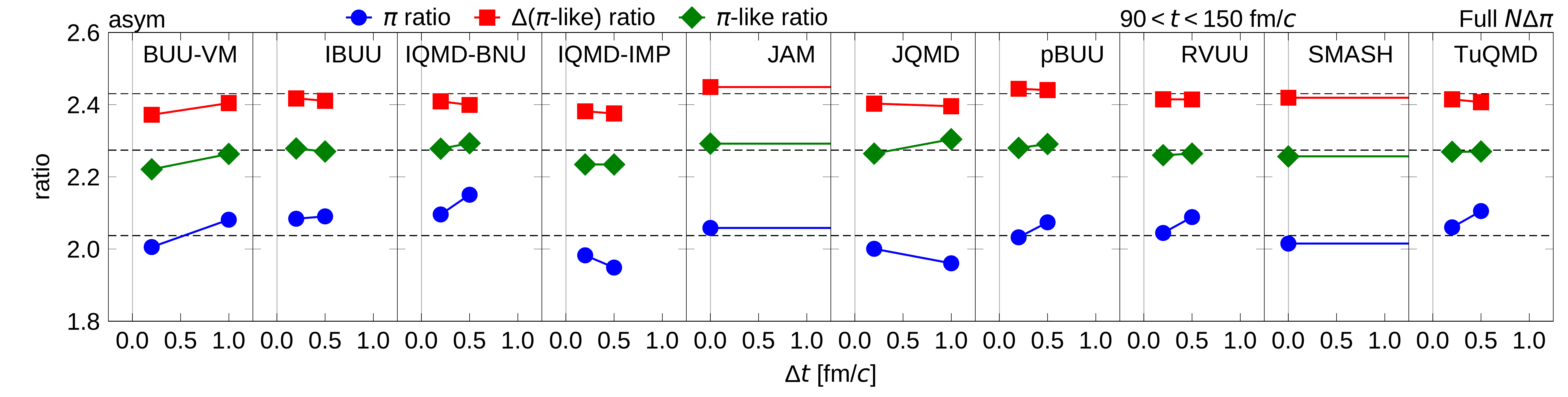}
  \caption{\label{fig:dt_piratioa_late}
    The time-step dependence of isotopic ratios, see Eq.~\eqref{eq:piratios}, in an asymmetric ($\delta=0.2$) full-$N\Delta\pi$ system, calculated from the particle numbers averaged over late times $90<t<150$ fm/$c$.  The result with the homework time-step size $\Delta t$ chosen by the code and that with $\Delta t=0.2$ fm/$c$ are connected by a line to guide the eye, for each of the three charged pion ratios [Blue circles: the $\pi$ ratio; green diamonds: the $\pi$-like ratio; and red squares: the $\Delta$($\pi$-like) ratio].  The horizontal dashed lines in each panel indicate the values for the ideal Boltzmann-gas mixture.
  }
\end{figure*}

In Fig.~\ref{fig:dt_piratioa_late}, these ratios obtained by averaging the particle numbers over the times $90<t<150$ fm/$c$, in which they are expected to have been equilibrated, are plotted with symbols.  For the codes relying on time steps, the result with $\Delta t$ chosen by the code and that with $\Delta t=0.2$ fm/$c$ are connected by a line to guide the eye.  The results from the time-step--free codes (JAM and SMASH) agree relatively well with the ratios for the ideal Boltzmann-gas mixture in chemical equilibrium (horizontal dashed lines).  For the $\pi$ ratio (blue circles), the results from different codes spread around the expected value in the range of $\pm7\%$.  The situation is better at smaller $\Delta t$.  In particular, many codes seem to converge almost to the correct value if the $\Delta t$ dependence of the ratios is linear.  For the $\Delta$($\pi$-like) ratios (red squares), the agreement with the expected value is better than that for the $\pi$ ratio in many cases even for large $\Delta t$, and the dependence on $\Delta t$ is not as strong as for the $\pi$ ratio, with more codes underestimating than overestimating this ratio.  For the $\pi$-like ratio (green diamonds), which is most directly related to observables measured in heavy-ion collisions, a relatively good agreement of $\pm2\%$ is found among all transport codes even with large $\Delta t$.  This is rather surprising in view of the larger deviations in the $\pi$ ratio (blue circles) and in the absolute numbers of $\Delta$ and $\pi$ in Fig.~\ref{fig:nat1a}.  The reason for this good agreement in the $\pi$-like ratio is given in the detailed analyses in later sections.  These results thus suggest that transport codes can reliably predict the equilibrated value of the $\pi$-like ratio in the box configuration.

\begin{figure*}
  \includegraphics[width=\textwidth]{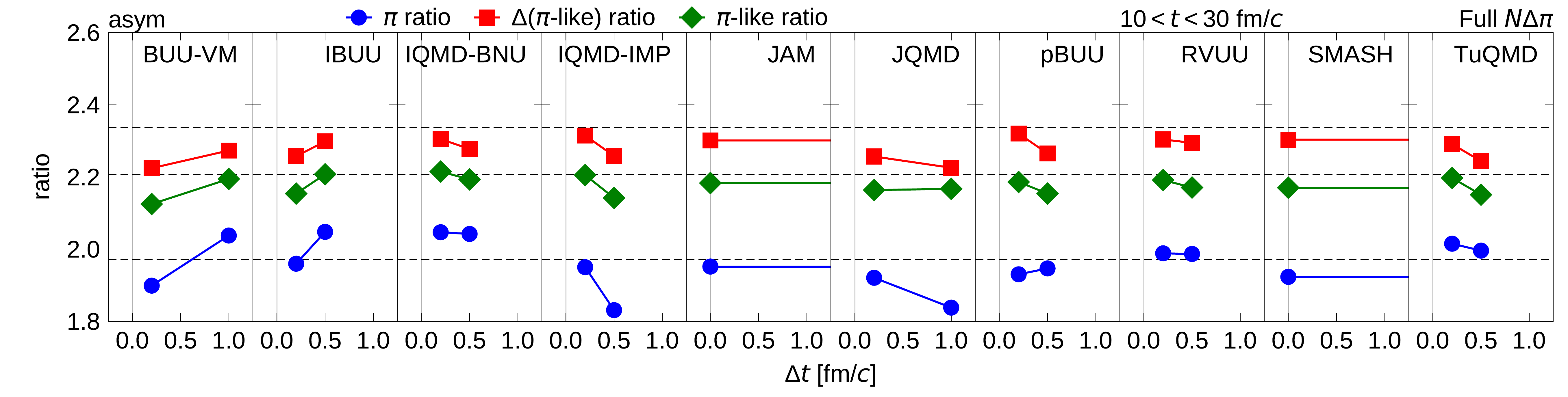}
  \caption{\label{fig:dt_piratioa_early}
    The time-step dependence of isotopic ratios, as in Fig.~\ref{fig:dt_piratioa_late}, calculated from the particle numbers averaged over early times $10<t<30$ fm/$c$.  Here the horizontal dashed lines in each panel indicate the values from the solution of the rate equation.
  }
\end{figure*}

In heavy-ion collisions where pions are produced, the violent phase of the reaction ends within a few tens of fm/$c$, and therefore the box comparison at early times is as important as that of later time when the system reaches equilibrium.  Figure~\ref{fig:dt_piratioa_early} shows a similar comparison of the isotopic ratios of the numbers of particles averaged over the early times $10<t<30$ fm/$c$.  Now the reference value from the rate equation is shown by the horizontal dashed line for each ratio.  As mentioned above, the rate equation does not assume chemical equilibrium but assumes thermal equilibration of momentum distributions, and therefore the transport-code results do not need to agree exactly with this reference value.  In fact, these ratios predicted by transport codes are often slightly lower than the reference values, which may indicate some real dynamical effects.  The behaviors of these ratios calculated at early times are similar to those at late times (Fig.~\ref{fig:dt_piratioa_late}) in some aspects, but there are also differences.  For the $\pi$ ratio (blue circles), deviations of more than $\pm7\%$ are found among the transport-code results for large $\Delta t$. Although many results tend to converge for smaller $\Delta t$, they do not compare as well as in the case of late times.  For the $\Delta$($\pi$-like) ratio (red squares) and the $\pi$-like ratio (green diamonds), we observe somewhat unorderly changes in predictions when $\Delta t$ is changed.  Compared to the case at late times, there seems to be an additional effect of $\Delta t$ dependence that affects the three ratios similarly in most codes.  For example, when $\Delta t$ is reduced, the $\Delta$($\pi$-like) and $\pi$-like ratios in BUU-VM and IBUU decrease more strongly, and those in IQMD-BNU, IQMD-IMP, JQMD, pBUU, RVUU and TuQMD increase more strongly than at late times.

The two full-ensemble BUU codes, namely pBUU (at $\Delta t\to 0$) and SMASH, agree well with each other for all three isotopic ratios at both early and late times.  The JAM results are close to those of the full-ensemble BUU codes.  The other QMD and parallel ensemble BUU codes show qualitatively different trends in the $\Delta t$ dependence, as mentioned above.  For those codes that predict similar values of the $\pi$-like ratio at late times, they do not necessarily agree very well with each other at early times.  When the results are linearly extrapolated to $\Delta t\to 0$,  the deviations of the three ratios from those in pBUU, SMASH and JAM become larger with few exceptions.  The differences among different codes are still within a few percent level for the $\pi$-like ratio, though the results are not as reliable as at late times because of the remaining $\Delta t$ dependence.

The high-density symmetry energy may be constrained to some degree even with the uncertainty of a few percent in the transport-code results for the $\pi$-like ratio.  However, a fundamental understanding is desirable, in particular if the uncertainty depends on whether the system is at equilibrium or not.  The detailed analyses in later sections suggest that the correlations induced by the geometrical method prescribed for collisions need to be better controlled.  Although correlations can in general exist physically, we will see later that those induced in transport codes sometimes can violate the isospin symmetry.  The correlations are expected to be the strongest in the limit of $\Delta t\to 0$.  A relation between the correlations and the non-equilibrium effects seems to cause these complicated behaviors of the isotopic ratios, in particular at early times.

\subsection{Guide to the following sections}

The agreement of the $\pi$-like ratio predicted by the 10 participating codes, within errors of a few percent level, is almost satisfactory, at least under the studied conditions and for our physical purpose.  However, it is still desirable to understand the origin of the remaining deviations, in order to justify the robustness of such an agreement against the change of conditions, and also in order to improve individual codes to further reduce errors.  This requires a detailed knowledge of the methods to handle the processes for $\Delta$ and $\pi$ in the transport codes, as reviewed and explained in Sec.~\ref{sec:tptheory}, and detailed analyses of the calculated results as performed in Secs.~\ref{sec:ndelta}, \ref{sec:ndeltapi} and \ref{sec:noneq}.  A summary of the performance of codes in the present box comparison is found in Fig.~\ref{fig:dtsummary} in Sec.~\ref{sec:ndeltapi-summary}.  Conclusions derived from such analyses are summarized in Sec.~\ref{sec:concl}.

\section{Transport approaches\label{sec:tptheory}}
\subsection{Boltzmann equation}
Without mean-field potentials in the present code comparison, the Boltzmann equation for the phase-space distribution function $f_\alpha(\bm{r},\bm{p},t)$ is
\begin{equation}
  \label{eq:boltz}
  \frac{\partial f_\alpha(\bm{r},\bm{p},t)}{\partial t}
  +\frac{\bm{p}}{\sqrt{m_\alpha^2+\bm{p}^2}}\cdot
  \frac{\partial f_\alpha(\bm{r},\bm{p},t)}{\partial\bm{r}}
  =I_\alpha(\bm{r},\bm{p},t),
\end{equation}
where the index $\alpha$ labels the different particle species, and $m_\alpha$ is the rest mass of species $\alpha$.  In the present study, we include $\Delta(1232)$ resonances besides nucleons and pions, so that $\alpha\in N\cup\pi\cup\Delta$ with
\begin{align}
  N&=\{n,\ p\},\\ \pi&=\{\pi^-,\ \pi^0,\ \pi^+\},\\
  \Delta &= \{\Delta^-,\ \Delta^0,\ \Delta^+,\ \Delta^{++}\}.
\end{align}
In our study only the $\Delta$ resonance is characterized by a spectral function. As it is usually done in transport simulations with particles of finite width, we treat the spectral function of such a particle as a mass distribution, such that the mass takes continuous values within the mass distribution.  In the following, as well as in Eq.~\eqref{eq:boltz}, we thus interpret $\Delta$ resonances with different masses $m$ as different particle species,
\begin{equation}
\begin{split}
\Delta^-&=\{\Delta^-(m);\ m>m_N+m_\pi\},\\
  \Delta^0&=\{\Delta^0(m);\ m>m_N+m_\pi\},\\
  \Delta^+&=\{\Delta^+(m);\ m>m_N+m_\pi\},\\
  \Delta^{++}&=\{\Delta^{++}(m);\ m>m_N+m_\pi\}.
\end{split}
\end{equation}
Each particle specified by an index $\alpha$ has a definite mass $m_\alpha$ and satisfies the relativistic dispersion relation $E=\sqrt{m_\alpha^2+\bm{p}^2}$.  A summation over the index $\alpha\in\Delta$ then includes an integration over the mass of $\Delta$.  

The collision term $I_\alpha(\bm{r},\bm{p},t)$ in Eq.(\ref{eq:boltz}) generally includes different types of two-particle collisions and decays,
\begin{equation}
  I_\alpha = \sum_\beta\sum_{\gamma\le\delta}I_{\alpha\beta\leftrightarrow\gamma\delta}
  +\sum_\beta\sum_\gamma I_{\alpha\beta\leftrightarrow\gamma}
  +\sum_{\beta\le\gamma}I_{\alpha\leftrightarrow\beta\gamma}
\end{equation}
with each term expressed in terms of cross sections ($\sigma$) and/or decay rates ($\Gamma$) as\footnote{
  The $\Omega$ integration is conventionally over the $4\pi$ solid angles.  The angle-integrated total cross section is related to the differential cross section by $\sigma_{\gamma\delta\rightarrow\alpha\beta}=(1+\delta_{\alpha\beta})^{-1}\int d\Omega(d\sigma_{\gamma\delta\rightarrow\alpha\beta}/d\Omega)$, where $\delta_{\alpha\beta}=1$, if $\alpha$ and $\beta$ are identical particles, and $\delta_{\alpha\beta}=0$, otherwise.  We do not consider here resonances decaying into two identical particles.
}
\begin{gather}
  \label{eq:icoll-22}
  \begin{split}
    I_{\alpha\beta\leftrightarrow\gamma\delta}
    &=\frac{g_\gamma g_\delta}{g_\alpha}\int\frac{d\bm{p}_2}{(2\pi)^3}
    \int d\Omega\, v'_{34}\frac{d\sigma_{\gamma\delta\to\alpha\beta}}{d\Omega}
    \frac{f_\gamma f_\delta}{1+\delta_{\gamma\delta}} \\
    &\quad -g_\beta\int\frac{d\bm{p}_2}{(2\pi)^3}
    v'_{12}\sigma_{\alpha\beta\to\gamma\delta}
    f_\alpha f_\beta,
  \end{split}\\
  \label{eq:icoll-21}
  \begin{split}
    I_{\alpha\beta\leftrightarrow\gamma}
    = 
    \frac{g_\gamma}{g_\alpha}
    \int \frac{d\Omega}{4\pi}\Gamma'_{\gamma\to\alpha\beta}f_\gamma
    -g_\beta\int\frac{d\Omega}{4\pi}
    v'_{12}\sigma_{\alpha\beta\rightarrow\gamma}f_\alpha f_\beta,
  \end{split}\\
  \label{eq:icoll-12}
  \begin{split}
    I_{\alpha\leftrightarrow\beta\gamma}
    = \frac{g_\beta g_\gamma}{g_\alpha}
    \int \frac{d\Omega}{4\pi}v'_{23}\sigma_{\beta\gamma\to\alpha}
    f_\beta f_\gamma
    -\Gamma'_{\alpha\to\beta\gamma}f_\alpha.
  \end{split}
\end{gather}
The degeneracy factors $g_\alpha$ are for spins, i.e., $g_\alpha=2$ for $\alpha\in N$, $g_\alpha=1$ for $\alpha\in\pi$, and $g_\alpha=4$ for $\alpha\in\Delta$.  The abbreviations $f_\alpha$, $f_\beta$, $f_\gamma$ and $f_\delta$ are for $f_\alpha(\bm{r},\bm{p},t)$, $f_\beta(\bm{r},\bm{p}_2,t)$, $f_\gamma(\bm{r},\bm{p}_3,t)$ and $f_\delta(\bm{r},\bm{p}_4,t)$, respectively.  The subscripts in $v'$, whose definition is given below, correspond to those in the momentum vectors $\bm{p}$ ($\equiv\bm{p}_1$), $\bm{p}_2$, $\bm{p}_3$ and $\bm{p}_4$.  Here, we do not consider the possible Pauli blocking of the final states.  In each integrand, the energy and momentum conservations have to be imposed on the momentum vectors, and the solid angle $\Omega$ represents the direction of a momentum vector in the c.m.~frame of the collision or decay.  The decay rate in the computational frame ($\Gamma'$), which in the present study is the rest frame of the box, is related to that in the rest frame of the decaying particle ($\Gamma$) by a Lorentz factor, e.g.,
\begin{equation}
  \label{eq:GammaInCompFrame}
  \Gamma'_{\gamma\to\alpha\beta}
  =(m_\gamma/E_3)\Gamma_{\gamma\to\alpha\beta}
\end{equation}
with $E_3=\sqrt{m_\gamma^2+\bm{p}_3^2}$.  The quantity $v'$, which agrees with the relative velocity of the colliding particles for colinear motion, is linked to the relative velocity $v^*$ observed in the c.m.~frame of the colliding particles according to the relation
\begin{equation}
  \label{eq:vprime}
  v'_{34}=p^*(s,m_\gamma,m_\delta)\frac{\sqrt{s}}{E_3E_4}
  =v^*_{34}\frac{E_3^*E_4^*}{E_3E_4},
\end{equation}
where $\sqrt{s}=E_3^*+E_4^*$ is the total energy in the c.m.~frame of the colliding particles, and the momentum $p^*$ of a particle in that frame is given by
\begin{equation}
  \label{eq:pincm}
  p^*(s,m,m')=\frac{\sqrt{s-(m+m')^2}\sqrt{s-(m-m')^2}}{2\sqrt{s}}.
\end{equation}

\subsection{Test particles}

To solve the Boltzmann equation numerically, the distribution functions are represented in terms of finite elements, so-called test particles \cite{cywong1982}, as
\begin{equation}\label{eq:testp}
  f_\alpha(\bm{r},\bm{p},t)=\frac{(2\pi)^3}{g_\alpha N_{\text{tp}}}
  \sum_i \delta_{\alpha\alpha_i}
  G\bigl(\bm{r}-\bm{r}_i(t)\bigr)\tilde{G}\bigl(\bm{p}-\bm{p}_i(t)\bigr),
\end{equation}
where $N_{\text{tp}}$ is the number of test particles per physical particle, and $g_\alpha$ is the spin degeneracy factor.  Each test particle $i$ of particle species $\alpha_i$ has its time-dependent coordinate $\bm{r}_i$ and momentum $\bm{p}_i$, and contributes to the distribution function with the shape functions $G$ and $\tilde{G}$, which can be $\delta$ functions or normalized Gaussian functions.  Since reactions and decays are considered here, test particles may change their identities $\alpha_i$ as well as may be created and annihilated. Note that we follow the convention of $\hbar=c=1$ in the present study.

The test particles can be regarded as samples randomly taken from the distribution functions, and therefore some fluctuations are induced as a result of the finite value of $N_{\text{tp}}$.  If there were no collision term in the Boltzmann equation [Eq.~\eqref{eq:boltz}], the solution would be obtained from the classical deterministic motions of test particles.  With the collision term, one may in principle consider an ensemble of final states for a collision, e.g.~populating different reaction channels and scattering angles, by splitting the test particles with suitably reduced weights assigned to them.  In practice, however, only one sample is randomly selected for the final state of a collision or a decay, so that the number $N_{\text{tp}}$ is kept constant.  Of course, the fluctuations induced by the finite number of test particles are expected to disappear in the limit of $N_{\text{tp}}\to\infty$.

The BUU codes aim to solve the Boltzmann equation [Eq.~\eqref{eq:boltz}] by choosing a relatively large but finite number for $N_{\text{tp}}$ such as $N_{\text{tp}}=100$.  The choice of $N_{\text{tp}}$ in BUU is an issue in the trade-off between the numerical accuracy and the computational time.  On the other hand, the QMD codes adopt $N_{\text{tp}}=1$, i.e., each test particle corresponds to a physical particle, so that large fluctuations are induced and the exact solution of the Boltzmann equation is not accurately reproduced.  This is an intention of the QMD model to go beyond the Boltzmann equation by incorporating physical fluctuations and correlations.  Physical fluctuations can also be introduced to the Boltzmann or the BUU equation by an additional fluctuation term, which leads to the Boltzmann-Langevin equation.  There exist some codes which implement such a term approximately \cite{colonna1998,napolitani2013}. In practice, the finite number of test particles also contributes to fluctuations.  We may naively expect that the difference between BUU and QMD is not so important in the present case, though it is important in the general cases when including the Pauli blocking and mean field, with the representation of the distribution function affecting the time evolution, e.g.~as seen in Ref.~\cite{yxzhang2018}.

\subsection{Numerical integration with time steps\label{sec:dtinteg}}

Most of the participating codes in the present study solve the Boltzmann equation approximately by introducing time steps of a finite size $\Delta t$.  If $\Delta t$ is sufficiently small, the details of the method described below would not affect the results.  However, in the results of the present work, we find that common choices of $\Delta t$, such as $\Delta t=0.5$ fm/$c$, may not be small enough, and the results may depend on the adopted numerical prescriptions.

The Boltzmann equation [Eq.~\eqref{eq:boltz}] may be formally integrated for a time interval $[t_k-\tfrac12\Delta t,\ t_k+\tfrac12\Delta t]$ during the $k$-th time step as
\begin{equation}
  \label{eq:dtinteg-decomp}
  \begin{split}
  f(t_k+\tfrac12\Delta t) &= f(t_k-\tfrac12\Delta t)
  + P[f,\ t_k-\tfrac12\Delta t,t_k]\\
  &+\int_{t_k-\frac12\Delta t}^{t_k+\frac12\Delta t}I[f(\tau)]d\tau
  + P[f,\ t_k,\ t_k+\tfrac12\Delta t],
\end{split}
\end{equation}
where the index of the particle species $\alpha$ and the phase-space coordinates $(\bm{r},\bm{p})$ are suppressed in $f_\alpha(\bm{r},\bm{p},t)$ and others for brevity, while the dependence on the distribution function is indicated explicitly for the collision term $I[f]$.  The integral for the propagation term during the time interval $[\tau_1,\tau_2]$,
\begin{equation}
  P[f,\tau_1,\tau_2] = -\int_{\tau_1}^{\tau_2}\frac{\bm{p}}{\sqrt{m_\alpha^2+\bm{p}^2}}
  \cdot\frac{\partial f(\tau)}{\partial \bm{r}}d\tau,
\end{equation}
represents the free motions of particles in the present study.  It can include the mean-field term in general.  The integral for the collision term $I[f(\tau)]$ is more complicated.  With the distribution function $f(t_k-\tfrac12\Delta t)$ known at the beginning of the $k$-th time step but $f(\tau)$ not known for $\tau$ in the interval $[t_k-\tfrac12\Delta t,\ t_k+\tfrac12\Delta t]$, some approximations are necessary to evaluate the integrals over $\tau$ for the propagation and the collisions.

By using the test particle representation [Eq.~\eqref{eq:testp}] for the phase-space distribution functions in Eqs.~\eqref{eq:icoll-22}, \eqref{eq:icoll-21} and \eqref{eq:icoll-12}, 
the collision integral $I[f(\tau)]$ can be written as a sum $\sum_{q=1}^Q I^{(q)}[f(\tau)]$ with each term $I^{(q)}$ corresponding to a specific pair of two colliding test particles, i.e.\ $q=(i,j)$, or a test particle that can decay $q=i$.  The loss and gain terms due to the same collision or decay should be included in the same term $I^{(q)}$.  The integral for the collision term in Eq.~\eqref{eq:dtinteg-decomp} can then be expressed as
\begin{equation}
  \label{eq:dtinteg-colldecomp}
    \int_{t_k-\frac12\Delta t}^{t_k+\frac12\Delta t} I[f(\tau)]d\tau
  =
  \sum_{q=1}^Q\int_{t_k-\frac12\Delta t}^{t_k+\frac12\Delta t} I^{(q)}[f(\tau)]d\tau.
\end{equation}

To calculate the time evolution of the system, the terms on the right-hand side of Eq.~\eqref{eq:dtinteg-decomp} are evaluated sequentially from left to right using Eq.~\eqref{eq:dtinteg-colldecomp} normally by staggering the integration of mean-field and collision terms.  First, the propagation term $P[f,\ t_k-\tfrac12\Delta t,\ t_k]$ is evaluated as if there are no collisions and decays, so that we can define
\begin{equation}
f_k^{(0)}=f(t_k-\tfrac12\Delta t)+P[f,\ t_k-\tfrac12\Delta t,\ t_k]
\end{equation}
and evaluate it by letting test particles move along the classical trajectories.  Next, the collision terms are evaluated following the sequence $q=1,2,\dots,Q$ as
\begin{equation}
  \label{eq:dtinteg-seq}
  f_k^{(q)}=f_k^{(q-1)}
  +\int_{t_k-\frac12\Delta t}^{t_k+\frac12\Delta t}I^{(q)}[f^{(q-1)}(\tau)]d\tau,
\end{equation}
where the function $f^{(q-1)}(\tau)$, defined for $\tau\in[t_k-\tfrac12\Delta t,\ t_k+\tfrac12\Delta t]$, is determined by the free propagation (even in the case with mean field) with the condition $f^{(q-1)}(\tau)=f_k^{(q-1)}$ at $\tau=t_k$.  In the numerical calculation, one of the possible outcomes, e.g.~the reaction channel and the scattering angle, in a collision is determined randomly in the implementation of the collision integral of Eq.~\eqref{eq:dtinteg-seq}, so that $f_k^{(q)}$ is always represented by test particles.  The momenta of test particles are usually changed by a collision $I^{(q)}$, while the spatial coordinates are not.  Particle identities may be changed by a collision or a decay, such as a baryon from $N$ to $\Delta$, and a meson, such as the pion, may be created or annihilated.  Finally, $f_k^{(Q)}$ is propagated by $P[f,\ t_k,\ t_k+\tfrac12\Delta t]$ to obtain $f(t_k+\tfrac12\Delta t)$.  In practical calculations, this final propagation and the first propagation in the next time step may occur at the same time because of the propagation $P_{k+1}=P[f,t_k,t_{k+1}]$ from $t_k$ to $t_{k+1}=t_k+\Delta t$.

The results of this widely adopted method of solving the Boltzmann equation may lead to errors most likely of the linear order in $\Delta t$.  However, in some special cases such as the $NN$ collision rates in the nucleon gas in a box studied in Ref.~\cite{yxzhang2018}, the inaccuracy due to the finite value of $\Delta t$ seems to have little impact.  On the other hand, using a finite number of test particles causes another kind of deviation of transport-code results from the solution of the Boltzmann equation due to the correlations induced by collisions, as we discussed in Ref.~\cite{yxzhang2018}.  In the present work, they are found to affect the results in a rather surprising way.  The essential difference from the case of Ref.~\cite{yxzhang2018} is that baryons can change their identities and pions can get created or absorbed in the inelastic processes $NN\leftrightarrow N\Delta$ and $N\pi\leftrightarrow\Delta$.  In the next two subsections, we give considerations on these issues, which are indispensable for understanding the results of transport codes in the present work.  In particular, the potential sources of violation of isospin symmetry are discussed.

\subsection{Sequence of collisions and decays}\label{sec:seq}

\begin{figure*}
  \centering
  \includegraphics[scale=0.35]{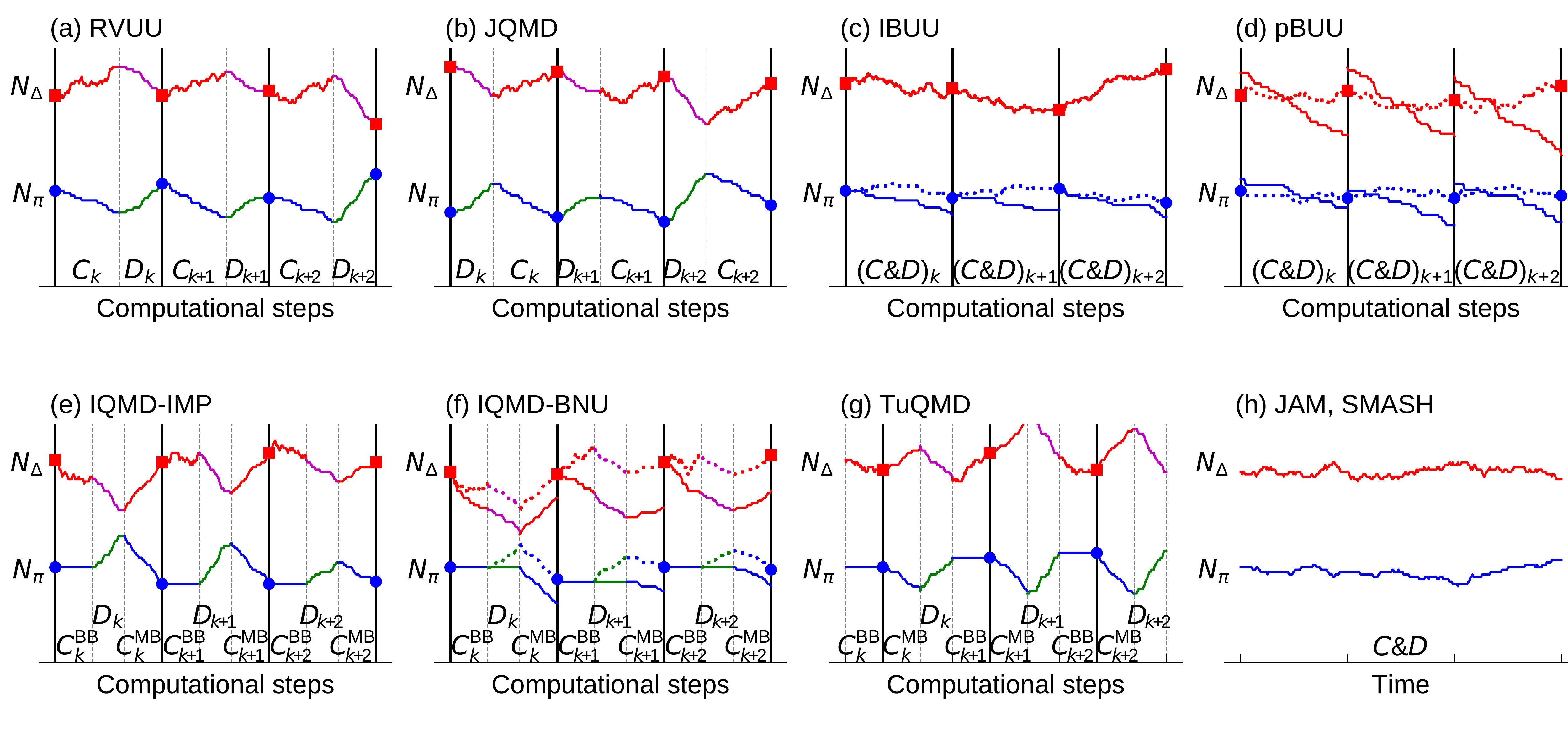}
  \caption{\label{fig:seqsim} Strategies for collision--decay sequence: Illustrative evolution of the numbers of $\Delta$ and $\pi$ particles due to collisions ($C$) and decays ($D$) during three time-step intervals ($k$, $k+1$ and $k+2$), for different computational strategies in the codes.  Each panel (a)--(h) represents a strategy for handling the collisions and decays in the indicated code.  BUU-VM combines the strategies represented in (a) and (b).  The vertical solid lines indicate the times when particles are propagated (usually at the time-step boundaries), and the symbols show the numbers $N_\Delta$ and $N_\pi$ of propagated $\Delta$ and $\pi$ particles, respectively.  Solid lines indicate the number of particles that actually take part in $C$ and $D$, while dotted lines include stealth particles which have been created but are not yet allowed to interact or decay. These data are obtained for a simplified system consisting of 125 particles that can change identities as $\text{`}N\text{'}\leftrightarrow\text{`}\Delta\text{'}\leftrightarrow\text{`}\pi\text{'}$.  The constant conversion probabilities per time step ($\Delta t \sim 1$ fm/$c$) for the conversions $\text{`}N\text{'}\leftrightarrow\text{`}\Delta\text{'}$ and $\text{`}\Delta\text{'}\leftrightarrow\text{`}\pi\text{'}$ are chosen to be similar to the reaction rates for $NN\leftrightarrow N\Delta$ and $\Delta\leftrightarrow N\pi$, respectively, in the system of ideal gas mixture studied in the homework.  The numbers $N_\Delta$ and $N_\pi$ are also similar to those in the latter system, but with $N_\Delta$ shifted upward by 25 relatively to $N_\pi$ to avoid the overlap of lines in the figure.  Given the qualitative insights to be gained here, numerical scales are suppressed.  }
\end{figure*}

The evaluation of the collision term using Eq.~\eqref{eq:dtinteg-seq} apparently depends on the order of the sequence in which collisions and decays are considered in a given time interval.  To study this effect, we denote by $C_k$ the list of collision pairs and by $D_k$ the list of unstable particles during the $k$-th time step.  Although the way the collision pairs within $C_k$ are ordered can be an important issue, we first discuss the issue on the ordering of $C_k$ and $D_k$.

There are various ways to decide the sequence of collisions and decays, and they are depicted for various methodologies in Fig.~\ref{fig:seqsim}.  In panel (a), the sequence is chosen to be $(C_k,D_k)$ during a time step, i.e., collisions occur first for all pairs in some order and are then followed by decays.  The horizontal axis in this figure shows the progress of the sequence $q$ in Eq.~\eqref{eq:dtinteg-seq} or the computational steps.  The two lines for $N_\Delta$ and $N_\pi$ indicate an illustrative example of the change of the numbers of $\Delta$ and $\pi$, respectively, during the progress of the sequence.  As particle numbers have achieved approximate equilibrium in this case, $N_\Delta$ increases on average during the collisions in $C_k$ sequence and decreases by the decays in $D_k$.  The number $N_\pi$ always monotonically decreases in $C_k$ processes and increases in $D_k$.  It should be noted that the particles are actually propagated by $P_{k+1}$ after all the processes in $C_k$ and $D_k$ have completed.  The symbols on the vertical solid lines indicate the numbers of these propagated particles.

In another method, corresponding to the results shown in panel (b) of Fig.~\ref{fig:seqsim}, the sequence $(D_k,C_k)$, which is opposite to the case of (a), is chosen.  By comparing (a) and (b), we can easily expect that the method (b) gives relatively large $N_\Delta$ and small $N_\pi$ compared to the method (a).  The difference between these two methods might seem nothing more than the issue of at which stage the numbers of particles are counted, because the sequence $(D_k,C_{k+1},D_{k+1},C_{k+2},\dots)$ in the method (a) could be equivalent to the sequence $(D_k,C_k,D_{k+1},C_{k+1},\dots)$ in the method (b).  However, since the particles at the time-step boundaries are propagated, these methods can result in different evolutions of the system.

The reasons for these potential inaccuracies in these methods can be argued in the following way.  In method (a), pions produced in $D_k$ cannot be absorbed during the same time step, and this thus leads to too large an $N_\pi$.  Since $\Delta$ particles produced in $C_k$ decay in $D_k$ during the full time-step interval $\Delta t$ as if they had existed since the beginning of the time step [see Eq.~\eqref{eq:dtinteg-seq}], $N_\Delta$ is thus reduced. The opposite arguments apply to the method (b), i.e., $\Delta$ particles created in $C_k$ have no chance to decay in the same time step and pions produced in $D_k$ can interact in $C_k$ as if it had existed since the beginning of the time step, resulting in too large an $N_\Delta$ and too small an $N_\pi$ in the method (b).

These shortcomings of methods (a) and (b) may be avoided by treating collisions and decays in a more democratic way.  In the method shown in panel (c) of Fig.~\ref{fig:seqsim}, collisions and decays are mixed by inserting the decays of particles at different places in the list $C_k$ of collision pairs.  This sequence is denoted by $(C\&D)_k$ in panel (c).  A technical difficulty in this method is how to handle the list of reactions in a sequence when a pion is created in the process $\Delta\to N\pi$.  In principle, it should be reasonable to update the list in some way to allow the collisions of the created pion.  However, in the specific code (IBUU) of panel (c), the collisions of a created pion are ignored until the next time step.  The appearance of such stealth pions will weaken the absorption of pions and thus the method may overpredict $N_\pi$.  In the figure, the solid lines do not include stealth particles, while the dotted lines show the numbers including stealth particles.

Other methods illustrated in panels (d)--(h) in Fig.~\ref{fig:seqsim} are discussed later after more context is built up.

The ordering of particle pairs in $C_k$ can also be an issue, particularly if it can cause conflicts with the symmetries of the system, such as the isospin symmetry and the forward--backward symmetry in collisions of two identical nuclei.  Many of the participating codes usually construct a list of particles at the initialization of an event and then make the list of collision pairs by taking particles from the particle list in a fixed order.  For example, in box simulations, some codes may make the particle list by first listing all the protons and then neutrons.  Then in $C_k$ for a time step, $pp$ collisions tend to take place in the early part of the sequence, while $nn$ collisions occur in the later part.  Although these details do not seem to affect the results in Ref.~\cite{yxzhang2018}, where only elastic $NN$ collisions are considered, they may cause problems when particles can change identities by collisions.  For example, when a $\Delta^{++}$ particle is produced from $pp\to n\Delta^{++}$ in an early part of $C_k$, it can be absorbed via $\Delta^{++}n\to pp$ by colliding with one of the neutrons later in $C_k$ in the same time step.  On the other hand, after the creation of a $\Delta^-$ particle from $nn\to p\Delta^-$ in a later part of $C_k$, it cannot be absorbed via $p\Delta^-\to nn$ in the same time step because collisions with protons have already been included in $C_k$.  This difference between $\Delta^{++}$ and $\Delta^-$ interactions induces an unphysical asymmetry between their numbers that are propagated after $C_k$.  In the present work, many code authors have noticed this problem and modified their codes before obtaining the final results.  For example, the problem can be avoided if the list of collision pairs is obtained by taking particles from the particle list in a random order, which has been done already in Ref.~\cite{yxzhang2018} within TuQMD.  Some codes have chosen instead to randomly or evenly order the protons and neutrons in the particle list at initialization.  More on this can be found in Sec.~\ref{sec:codespec} for code-specific details.

There is another type of codes, corresponding to panel (h) of Fig.~\ref{fig:seqsim}.  These codes do not assume any predefined order of the collision--decay sequence, and therefore are free from the problems discussed above, so the formulation in Sec.~\ref{sec:dtinteg} does not apply to them.  In these codes (JAM and SMASH), collisions and decays take place according to their event times, and each particle is propagated between the two events.  A collision happens when the distance between the two particles is minimum in their center-of-mass frame.  The time for the decay of each unstable particle is determined randomly according to the decay rate, when it is created in the final state of a collision.  After every event of collision or decay, the list of future events is then updated.  These codes are thus time-step free as far as the combination of free propagation and collision term is concerned.

For transport codes developed in early days of heavy-ion collisions, it is often difficult to find in the literature the precise description of the employed numerical methods.  However, the Vlasov--Uehling--Uhlenbeck code, which was available in the floppy disk attached to the book containing Ref.~\cite{hartnack1993}, already used a method to process collisions and decays in a proper order, as in cascade codes for high-energy heavy-ion collisions \cite{cugnon1980,kodama1984}.  The same approach was taken in the original code of IQMD \cite{hartnack1989,aichelin1991} and in UrQMD \cite{bass1998}.  These codes have influenced some later codes such as JAM \cite{nara1999} and SMASH \cite{weil2016}.  On the other hand, for heavy-ion collisions at lower energies where the mean-field interaction is essential, the methods to process collisions and decays in a predefined order within a time step have been widely used\footnote{In the IQMD-BNU and IQMD-IMP codes, the collision procedure in the original IQMD code \cite{hartnack1989,aichelin1991} was replaced by their own procedures with time steps, which treat collisions and decays in a predefined order.}, considering the numerical cost and the simplicity of the code structure.  Most likely the results do not depend much on the method in many cases, e.g.\ as seen in Ref.~\cite{yxzhang2018}.  However, it does not seem to have been addressed in the literature that the difference in the ordering of the collision--decay sequence affects the results strongly for pion production.  Another issue, that is to be discussed in the next subsection, i.e.\ the consequences of correlations induced by the geometrical method for collisions, also does not seem to have been discussed in the literature.

\subsection{Correlations induced by collisions\label{sec:hocorr}}

To evaluate each term in Eq.~\eqref{eq:dtinteg-colldecomp} for the pair $q=(i,j)$ of particles, many codes use the geometrical conditions to determine if collisions can occur, as introduced by Cugnon \cite{cugnon1980} and reviewed in Ref.~\cite{bertsch1988} by Bertsch and Das Gupta, possibly with some modifications.  In this method, each pair $(i,j)$ of particles would undergo a collision when they reach the closest approach in the two-particle c.m.~frame and if the distance $d^*_{ij}$ at this time in that frame is within the interaction range, $d^*_{ij}\le\sqrt{\sigma_{\text{tot}}/\pi}$.  In BUU codes employing the full-ensemble method, the pair should be considered for test particles and the distance condition should be $d^*_{ij}\le\sqrt{(\sigma_{\text{tot}}/N_{\text{tp}})/\pi}$.  The evaluation of the integral $I^{(q)}$ in Eq.~\eqref{eq:dtinteg-seq} in a transport code corresponds to letting the pair collide when the distance condition is satisfied during this time-step interval $[t_k-\tfrac12\Delta t,\ t_k+\tfrac12\Delta t]$ in the computational frame.  There are some variants of the distance condition as described in Ref.~\cite{yxzhang2018} for the case of including only $NN$ elastic collisions.  In the case there are several collision channels for the pair $(i,j)$, one should use the total cross section $\sigma_{\text{tot}}$ in the distance condition.  When a collision occurs, a channel is then selected based on the ratio of its partial cross section to the total cross section.  A scattered particle thereafter changes its momentum and possibly its identity, e.g., from a nucleon to a $\Delta$ particle with certain mass, and it may later also collide with other particles with its new properties in the sequence for the same time step, with the exception of the IQMD-BNU code (see Sec.~\ref{sec:IQMD-BNU}).

It should be noted that all collisions occur locally in the Boltzmann equation since the collision terms [Eqs.~\eqref{eq:icoll-22}, \eqref{eq:icoll-21} and \eqref{eq:icoll-12}] include distribution functions only at a single spatial coordinate $\bm{r}$.  With the geometrical condition for collisions, two particles are separated by some distance when they collide.  Although this difference from the Boltzmann equation is unavoidable in solving transport equations using the test particle method, it can be eliminated by taking the limit of $N_{\text{tp}}\to \infty$ in full-ensemble BUU codes.  On the other hand, nuclear interactions are of finite range in nature, whose effect is, however, not accounted for in the Boltzmann equation.  The above non-local collisions induced in the test particle realization of transport models are closely related to the problem of correlations induced by collisions to be discussed below in detail.

\begin{figure}
\includegraphics{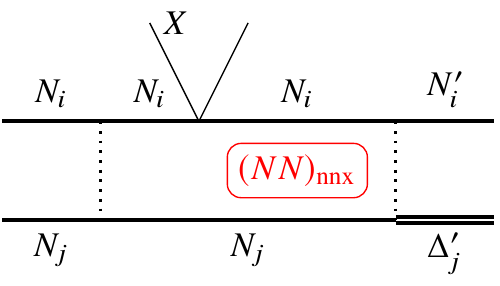}
\caption{\label{fig:nnx}  The higher-order correlation $(NN)_{\text{nnx}}$ induced between $N_i$ and $N_j$ after the $N_iN_j$ elastic collision and the scattering of $N_i$ (or $N_j$) by another particle $X$.  This correlation enhances the possibility of the second $N_iN_j$ collision leading to $N_iN_j\to N_i'\Delta_j'$.
}
\end{figure}

A typical case of what we have termed higher-order correlations in Ref.~\cite{yxzhang2018} is depicted in Fig.~\ref{fig:nnx}.  Right after the elastic collision of two nucleons $N_i$ and $N_j$,\footnote{The subscript attached to a particle denotes the index in the particle list.} they are spatially close to each other when collisions are treated using the geometrical method.  These two particles are not allowed to repeat collisions because all the interactions between them should have been taken into account by the first $N_iN_j$ collision.  Such simple correlations thus lead to spurious collisions, which can be technically avoided if one lets each particle carry the identifier of its most recent collision.  For example, after the first $N_iN_j\to N_iN_j$ collision, $N_i$ and $N_j$ have the same collision identifier that would forbid them to collide with each other again unless after one of them has been scattered by some other particle $X$, resulting in a new collision identifier.  If the scattering by $X$ happens soon after the first $N_iN_j$ collision, the two nucleons are still close in space and thus have a higher chance of undergoing the second $N_iN_j$ collision than what is expected for uncorrelated two nucleons.  This higher-order correlation is denoted by $(NN)_{\text{nnx}}$ in the present paper.\footnote{We use a notation to characterize the correlation by indicating the correlated particles in the parentheses and the processes causing the correlation in the subscript.  The particles in the latter processes are shown in the subscript after transforming the particle names into lowercase roman characters, such as $N\to\text{n}$, $\Delta\to\text{d}$, $X\to\text{x}$ and $\pi\to\text{p}$, to avoid confusing them with the correlated particles.}  The correlation will enhance the chance of the second $N_iN_j$ collision, which affects the $N_iN_j\to N_i'\Delta_j$ reaction rate in our present case as well as the elastic collision rate. 

Although correlations do not exist in the Boltzmann equation, they may exist in the true quantum many-body problem.  However, the correlations induced by the geometrical method, e.g.\ in QMD codes, is of classical nature because the uncertainty relation is ignored.  Some investigation of non-local effects in transport equations have been done by Morawetz et al.\ \cite{morawetz2001}.  Furthermore, the correct procedure in quantum mechanics e.g.\ for the process of Fig.~\ref{fig:nnx} is of course to first calculate the amplitude for the whole process $N_iN_jX\to N_i'\Delta_j'X$, integrating over the intermediate states, and then to square it to obtain the probability.  In transport codes, this is replaced by the three independent stochastic processes for $N_iN_j\to N_iN_j$, $N_iX\to N_iX$ and $N_iN_j\to N_i'\Delta_j'$.  The validity of this approximation is not known well in general.  However, the isospin violation we will discuss below is a direct unphysical consequence of this kind of approximation, under which the isospin coupling cannot be treated in the quantum mechanical way.

\subsubsection{$N\Delta$ correlation without the effects of pions\label{sec:nnxd}}

As a simple example of correlations specifically tied to the present investigation, we consider those for a $\Delta$ particle.  Because of the not very long lifetime of $\Delta$, some existing $\Delta$ particles should have been created not too long ago in $N_iN_j\to N_i'\Delta_j'$ reactions.  Since converting $N_j$ to $\Delta_j'$ reduces the kinetic energy, the relative velocity between $N_i'$ and $\Delta_j'$ becomes smaller, and therefore $N_i'$ is likely closer to $\Delta_j'$ in space.  Thus, for an existing $\Delta$, its chance to find a nucleon nearby is larger than what is expected from the one-body nucleon density.  Such a correlation is not present in the Boltzmann equation.

\begin{figure}
\includegraphics{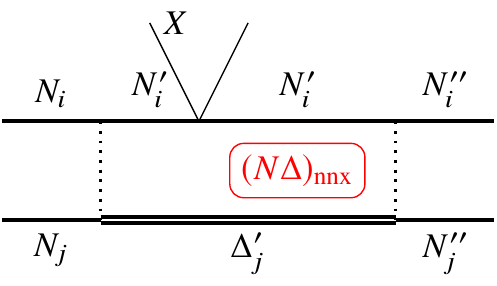}
\caption{\label{fig:nnxd}  The higher-order correlation $(N\Delta)_{\text{nnx}}$ induced between $N_i'$ and $\Delta_j'$ after $N_iN_j\to N_i'\Delta_j'$ and the scattering of $N_i'$ (or $\Delta_j'$) by another particle $X$.  This correlation enhances the possibility of the $N_i'\Delta_j'$ collision leading to $N_i'\Delta_j'\to N_i''N_j''$.
}
\end{figure}

However, after the $N_iN_j\to N_i'\Delta_j'$ reaction, $\Delta_j'$ should not be directly absorbed by $N_i'$ by the $N_i'\Delta_j'\to N_i''N_j''$ reaction because all the interactions between particles $i$ and $j$ should have been taken into account by the first $N_iN_j$ collision unless any other particles participated.  Although such spurious repetition of collisions are forbidden in transport codes as mentioned above, higher-order correlations may still be present.  As depicted in Fig.~\ref{fig:nnxd}, the pair of $N'_i$ and $\Delta'_j$ can undergo another collision if one of them has been scattered by another particle $X$.  If not enough time has passed after the first $N_iN_j\to N'_i\Delta'_j$ reaction, the two particles $N'_i$ and $\Delta'_j$ are still close in space and thus have a large chance to undergo the process $N_i'\Delta_j'\to N_i''N_j''$.  This higher-order correlation between $N_i'$ and $\Delta_j'$ is denoted by $(N\Delta)_{\text{nnx}}$.  In case the total cross section of the first $N_iN_j$ collision is small compared to the $N'_i\Delta'_j$ cross section, whether the two particles have a significant $(N\Delta)_{\text{nnx}}$ correlation depends on the mean free time $\tau_{\text{free}}$ for $N'_i$ and $\Delta'_j$ after the first reaction, their relative velocity $v_{\text{rel}}$, and the cross section $\sigma_{\text{tot}}[N'_i\Delta'_j]$.  In QMD codes and in BUU codes with the parallel-ensemble method, the condition for reinteraction is
\begin{equation}
\label{eq:hocorr-cond}
  v_{\text{rel}}\tau_{\text{free}} \lesssim\sqrt{\sigma_{\text{tot}}[N'_i\Delta'_j]/\pi}.
\end{equation}
For BUU codes using the full-ensemble method, the effect of higher-order correlations between test particles is expected to be weak because $\sigma_{\text{tot}}[N'_i\Delta'_j]$ is replaced by $\sigma_{\text{tot}}[N'_i\Delta'_j]/N_{\text{tp}}$.  We note that such higher-order correlations are not considered in the Boltzmann equation. Although such correlations can in principle physically exist in some way, they are not necessarily induced correctly by the prescription described here.

The existence of higher-order correlations is an issue essentially independent of the numerical integration with a finite time step (Sec.~\ref{sec:dtinteg}) and the issue of the collision--decay sequence (Sec.~\ref{sec:seq}).  Correlations should exist even in the limit of $\Delta t\to0$.  In fact, it is rather expected that the effect of correlations may be weak with a choice of a large $\Delta t$ since the same pair of particles is only allowed to collide once in each time step.

As seen in Ref.~\cite{yxzhang2018}, higher-order correlations can lead to higher collision rates.  Although it is not \textit{a priori} clear how correlations may affect the actual dynamics in heavy-ion collisions and of systems in a box, they need to be well understood to not cause significant unphysical effects such as the violation of important symmetries.  An example is the potential risk of violation of isospin symmetry due to the $(N\Delta)_{\text{nnx}}$ correlation.  Let us consider a series of processes such as $n_in_j\to p_i\Delta_j^-$, $p_iX\to p_iX$ (or $\Delta_j^-X\to\Delta_j^-X$) and then $p_i\Delta_j^-\to n_in_j$.  Because the cross section $\sigma_{\text{tot}}[p_j\Delta_j^-]$ is large, the correlation enhances the absorption of $\Delta^-$ and therefore suppresses the number of $\Delta^-$ (and $\Delta^{++}$ for the same reason).  On the other hand, for a $\Delta^0$ particle produced from either $n_in_j\to n_i\Delta^0_j$ or $n_ip_j\to p_i\Delta^0_j$, the total cross section $\sigma_{\text{tot}}[n_i\Delta^0_j]$ or $\sigma_{\text{tot}}[p_i\Delta^0_j]$ is not as large as $\sigma_{\text{tot}}[p\Delta^-]$ due to the isospin Clebsh-Gordan coefficients for these inelastic channels.  Therefore, the suppression of the numbers of $\Delta^0$ and $\Delta^+$ is expected to be weaker than that of $\Delta^-$ and $\Delta^{++}$.  This asymmetry among the different species of $\Delta$ arises even in isospin-symmetric systems, which contradicts the isospin symmetry.  In applications of transport codes, it is important to ensure that such a violation of isospin symmetry is not so large as to significantly affect the physical observables.

The environment around the colliding pair of particles can influence the strength of correlations.  Let us consider a neutron-rich environment and nucleon collisions that yield a $\Delta$, such as $p_ip_j\to n_i\Delta^{++}_j$.  The superfluous $n_i\Delta^{++}_j$ collision potentially leading to $n_i\Delta^{++}_j\to p_ip_j$ occurs in the presence of many potential $n\Delta^{++}_j$ collisions where $n$ are uncorrelated with $\Delta^{++}_j$.  However, if the collision producing a $\Delta$ is $n_in_j\to p_i\Delta^-_j$, then the superfluous $p_i\Delta^-_j$ collision occurs in the presence of fewer $p\Delta^-_j$ collisions.  The latter superfluous collision represents a greater relative error than a superfluous collision for the case of $n_i\Delta^{++}_j$, tilting isospin symmetry.  This possible asymmetry in the effect of correlations on $\Delta^{++}$ and $\Delta^-$ (or the asymmetry on $\Delta^+$ and $\Delta^0$ for a similar reason) does not necessarily imply a violation of isospin symmetry because it originates from the asymmetry of the environment around colliding particles.

\subsubsection{Correlations with the participation of pions\label{sec:npid}}

\begin{figure}
\includegraphics{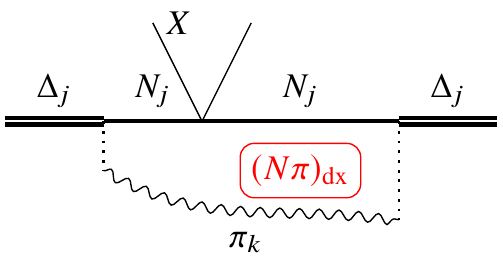}

\caption{\label{fig:npidx}  The higher-order correlation $(N\pi)_{\text{dx}}$ induced between $N_j$ and $\pi_k$ after $\Delta_j\to N_j\pi_k$ and the scattering of $N_j$ by another particle $X$.  This correlation may enhance the 
possibility for the $N_j\pi_k\to \Delta_j$ reaction.
}
\end{figure}

As for the $(N\Delta)_{\text{nnx}}$ correlation (Fig.~\ref{fig:nnxd}), higher-order correlations between a pion and a nucleon can be induced through the $\Delta\leftrightarrow N\pi$ reactions with the participation of an extra particle $X$ as in Fig.~\ref{fig:npidx}.  This $(N\pi)_{\text{dx}}$ correlation is expected to enhance the absorption of $\pi_k$ by $N_j$, and this effect is the strongest for $N_j\pi_k=n\pi^-$ or $p\pi^+$ because of their largest cross sections among the different isospin channels.  A result from this correlation is an enhanced production of $\Delta^-$ and $\Delta^{++}$, which constitutes a violation of isospin symmetry.  However, this correlation may become weaker if the pion $\pi_k$ is more likely absorbed by one of many other surrounding nucleons before the sequence $N_jX\to N_jX$ and $N_j\pi_k\to\Delta_j$ can take place as shown in Fig.~\ref{fig:npidx}.  The strength of this $(N\pi)_{\text{dx}}$ correlation may depend on the details of the prescriptions in the code, such as the way the position of the pion $\pi_k$, relative to that of $N_j$, is determined.

\begin{figure}
\includegraphics{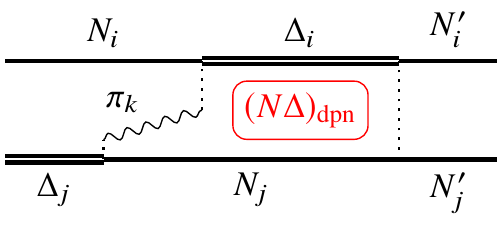}

\caption{\label{fig:npid1}  The correlation $(N\Delta)_{\text{dpn}}$ induced between $N_j$ and $\Delta_i$ after the pion transfer from $\Delta_j$ to $N_i$. This correlation enhances the possibility of the $N_j\Delta_i$ collision such as $N_j\Delta_i\to N_j'N_i'$.
}
\end{figure}

There are other types of correlations that do not require the participation of another particle $X$ to have an impact.  Besides the $(N\Delta)_{\text{nnx}}$ correlation (Fig.~\ref{fig:nnxd}), the correlation between $N$ and $\Delta$ can be induced when a pion is transferred from a $\Delta$ particle to a nucleon as in Fig.~\ref{fig:npid1}.  After the decay of $\Delta_j$, the pion $\pi_k$ is absorbed by $N_i$ from the surrounding nucleons.  Since the pion absorption $N_i\pi_k\to\Delta_i$ can happen immediately after the decay of $\Delta$ due to the strong pion absorption, $\Delta_i$ and $N_j$ can often be spatially close to each other, leading to a stronger correlation.  No participating code forbids this interaction of $\Delta_j$ and $N_j$, though it is technically possible to forbid it by marking the produced $\Delta$ with a suitable collision identifier (see also footnote \ref{footnote:collid} for an analogous case).  This $(N\Delta)_{\text{dpn}}$ correlation may affect the isospin symmetry through at least three possible effects.  First, $\Delta^{++}$ and $\Delta^-$ from the pion transfer reactions $\Delta_j\to N_j\pi_k$ and $N_i\pi_k\to\Delta_i$ cannot be absorbed by a proton and a neutron, respectively, because the corresponding reaction $\Delta N\to NN$ is not possible, thus reducing the chance of a correlation effect when $\Delta_i$ is $\Delta^-$ or $\Delta^{++}$.  Second, since the $N\pi$ cross section is large, the distance between $N_i$ and $\pi_k$ in Fig.~\ref{fig:npid1} is not generally very small when treating collisions by using the geometrical method as in most transport models.  This weakens the correlation between $\Delta_i$ and $N_j$ when the $N_i\pi_k$ in the intermediate state has the largest cross section.  Third, the impact of the $(N\Delta)_{\text{dpn}}$ correlation for the $N\Delta\to NN$ process in Fig.~\ref{fig:npid1} on the isospin violation can be similar to that of the $(N\Delta)_{\text{nnx}}$ correlation in Fig.~\ref{fig:nnxd}.  While the numbers of $\Delta^0$ and $\Delta^+$, relative to $\Delta^-$ and $\Delta^{++}$, are suppressed by the first two effects, they are enhanced by the last effect.

\begin{figure}
\includegraphics{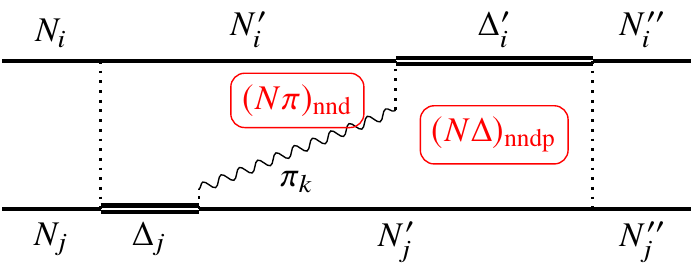}

\caption{\label{fig:npid2}  The correlation $(N\pi)_{\text{nnd}}$ between $\pi_k$ and $N_i'$ after a $N_iN_j\to N_i'\Delta_j$ collision and the $\Delta_j\to N_j'\pi_k$ decay.  It gives rise to the $(N\Delta)_{\text{nndp}}$ correlation between $N_j'$ and $\Delta_i'$, after the absorption of $\pi_k$ by $N_i'$.  These correlations may enhance the possibility of a $\Delta_i'N_j'$ collision yielding $\Delta_i'N_j'\to N_i''N_j''$.
}
\end{figure}

Depending on the way a transport code treats particle collisions during a time step, a pion may be absorbed not only by one of the uncorrelated surrounding nucleons, but also by the nucleon ($N_i'$) that triggered its production, as shown in Fig.~\ref{fig:npid2}.  After the production of $\pi_k$ from $\Delta_j\to N_j'\pi_k$, its absorption again by $N'_j$ is forbidden\footnote{When $\pi_k$ is created in some codes, its position is chosen in such as way that $\pi_k$ cannot be directly absorbed by $N'_j$ in the geometrical method.  Alternatively, the same collision/decay identifier may be given to $\pi_k$ and $N'_j$ to forbid the direct absorption.} since it is evidently a spurious process.  However, since the lifetime of $\Delta$ is short and the relative velocity between $N_i'$ and $\Delta_j$ is usually low, it is likely that $N_i'$ is still spatially close to $\pi_k$.  Most of the participating codes allow the direct absorption by $N_i'$ in $N_i'\pi_k\to\Delta_i'$ (with an exception of the JAM code).\footnote{\label{footnote:collid}Only a single code (JAM) requires a scattering of $N'_i$ or $\Delta_j$ by some other particle $X$, before $N'_i+\pi_k\to\Delta'_i$ is allowed.  This is implemented by using the same collision identifier of $\Delta_j$ for $N_j'$ and $\pi_k$.  In general, the appropriate treatment should depend on the cross sections used in individual codes to describe $NN\to NN\pi$ and $NN\to NN$ scatterings.  In the present work, however, since a common set of cross sections is specified in the homework, the correlations affect the results differently depending on the treatment.}  With such a correlation $(N\pi)_{\text{nnd}}$, the $N\pi\to\Delta$ rate is expected to be higher than in the case without correlations, resulting in a lower number of $\pi$ and a higher number of $\Delta$.  Again, there is the risk of violation of isospin symmetry.  For example, when $\pi_k$ is $\pi^-$, $N'_i$ is more likely a proton than a neutron, as may be intuitively expected from the charge conservation, leading to an enhanced production of $\Delta^0$ from the collision of $p\pi^-$.  A similar consideration for $\pi^+$ leads to an enhanced production of $\Delta^+$. Since only $\Delta^0$ and $\Delta^+$ can be produced when $\pi_k$ is $\pi^0$, the $(N\pi)_{\text{nnd}}$ correlation thus enhances the production of $\Delta^0$ and $\Delta^+$ relative to that of $\Delta^-$ and $\Delta^{++}$, even in isospin-symmetric systems, which violates isospin symmetry.  It should be noted that, after the pion absorption, $\Delta'_i$ and $N'_j$ can directly interact so that the correlation between them, called $(N\Delta)_{\text{nndp}}$, may also be important when the $(N\pi)_{\text{nnd}}$ correlation is strong, even though this $N\Delta$ correlation is formally of a higher order compared to the other $(N\Delta)_{\text{nnx}}$ and $(N\Delta)_{\text{dpn}}$ correlations in Figs.~\ref{fig:nnxd} and \ref{fig:npid1}.  The effects of this $N\Delta$ correlation on the isospin violation should be similar to those of $(N\Delta)_{\text{nnx}}$ and $(N\Delta)_{\text{dpn}}$.

\subsection{Code-specific comments\label{sec:codespec}}

Details of collision treatment in participating codes have been compared in Ref.~\cite{yxzhang2018}.  For the case of only $NN$ elastic collisions, some of the deviations among the code results are associated with differences in those details.  For the sake of present work, some codes have introduced improvements above the past or chosen other options compared to what was discussed in Ref.~\cite{yxzhang2018}.  These changes are described here for completeness.

Participating codes were already employed in published studies of pion production in heavy-ion collisions.  The improvements related to pions, which were made after these publications, are also described here.  It should be noted that the physical inputs such as cross sections and decay widths used in these studies are generally different from those in the present  homework comparison.

\subsubsection{BUU-VM}

In Ref.~\cite{yxzhang2018}, the list of collision pairs is constructed by taking particles in a fixed order from the particle list that is given in the initialization, e.g., as $(p_1,p_2,\ldots,p_{640},n_{641},\ldots,n_{1280})$ in the case of the symmetric system, so that the collision pairs are chosen sequentially in the order of $(p_1,p_2),\ (p_1,p_3),\ldots,\ (p_1,n_{1280}),\ldots,\ (n_{1279},n_{1280})$ during a time step.  In the present work, however, the particle list is initialized at $t=0$ of every event in such a way that protons and neutrons are ordered evenly or randomly in the list.

When a pion is created in a $\Delta\to N\pi$ decay, it is placed at the same position as the resulting nucleon.  The pion is not absorbed by this nucleon until the nucleon is scattered by some other particle.

The production of pions and $\Delta$ in projectile fragmentation reactions was estimated with an earlier version of the code in Ref.~\cite{mallik2014}, where the isospins of particles were not treated explicitly.  The time-step size $\Delta t=0.5$ was used there.  The newest version of the code with isospins, used in Ref.~\cite{yxzhang2018}, has not been applied to study pion production.

\subsubsection{IBUU\label{sec:IBUU}}

The time dilation effect is ignored in Ref.~\cite{yxzhang2018} but it is now taken into account by the gamma factor for the Lorentz transformation from the computational frame to the two-particle center-of-mass frame, i.e., the time condition in Table IV of Ref.~\cite{yxzhang2018} has been changed to $|t_{\text{coll}}^*-t_0^*|<\frac{1}{2}\Delta t/\gamma$.  The ordering of collisions of particle pairs has also been changed so that it is randomized at every time step.

When a pion is created in a $\Delta\to N\pi$ decay, it is placed at the same position as the resulting nucleon.  Since the pion is stealth during the time step of its creation, it ends up being propagated for one time step before it is allowed to interact with other particles.

The IBUU code was used to illustrate the effects of the symmetry energy at suprasaturation densities on the $\pi^-/\pi^+$ ratio in Refs.~\cite{bali2002npa,bali2002}, and the results using the momentum-dependent nucleon potential were later compared with the FOPI data~\cite{xiao2009,mingzhang2009}. The detailed treatment for pion production in these studies can be found in Refs.~\cite{bali1993,bali1995}.  The time step of $\Delta t=0.5$ fm/$c$ in intermediate-energy heavy-ion simulations is generally used. The time dilation effect was ignored in the previous studies and in Ref.~\cite{yxzhang2018}, but it is now taken into account, as mentioned above.

\subsubsection{IQMD-BNU\label{sec:IQMD-BNU}}

The option of the collision order has changed from the fixed ordering of baryon pairs in Ref.~\cite{yxzhang2018} to an ordering that is randomized at every time step.

When two baryons collide, one of them is checked whether it has already experienced a collision with another baryon in the same time step.  The collision is allowed only if this is the first collision for it in the time step.  This extra condition is imposed in the present work, as well as in Ref.~\cite{yxzhang2018}.

An older version of the code, which is not consistent with the present work, was used in Ref.~\cite{xie2018} to study the pion--nucleon potential in Au + Au collisions at 1.5 GeV/nucleon, with a time step of $\Delta t=1.0$ fm/$c$ and a fixed ordering of baryon pairs for collisions.

\subsubsection{IQMD-IMP\label{sec:IQMD-IMP}}

No changes have been introduced in the IQMD-IMP code.  In particular, particle pairs are chosen for collisions in a fixed order, as in Ref.~\cite{yxzhang2018}.

When a pion is created in a $\Delta\to N\pi$ decay, the pion and the nucleon are placed with a distance $\bm{v}\Delta t$, where $\bm{v}$ is the relative velocity between the two particles.

The treatment of collisions and decays used in the present work and in Ref.~\cite{yxzhang2018} was used in realistic heavy-ion collisions.  In Ref.~\cite{zqfeng2010}, the high-density symmetry energy was extracted, without the isovector part of the momentum-dependent interaction.  The isospin-dependent pion--nucleon potential was proposed in Refs.~\cite{zqfeng2015,zqfeng2017}, which influences the charged pion ratio in nuclear reactions.

\subsubsection{JAM}

In this code, a causal inconsistency exists because a collision between two particles occurs when they are at two different space-time points.  In the standard setting of JAM, adopted in Ref.~\cite{yxzhang2018}, some collisions are removed to avoid causal inconsistency.  However, this setting is changed so that all collisions are included in the present work.  This change helps the code to reproduce the number of pions in an ideal Boltzmann-gas mixture.

When a pion is created in a $\Delta\to N\pi$ decay, its position is selected randomly inside the sphere with a radius of 0.5 fm centered at the resulting nucleon.  The pion is not absorbed by this nucleon until the nucleon is scattered by some other particle.

The JAM code, without the above-mentioned change related to the causal inconsistency, was used in the AMD+JAM approach in Refs.~\cite{ikeno2016,ikeno2016erratum} to study pion production in heavy-ion collisions at 300 MeV/nucleon.

\subsubsection{JQMD}

As stated in Ref.~\cite{yxzhang2018}, collision ordering is fixed but nucleons are ordered evenly in the present work to avoid isospin-dependent bias. At initialization, protons and neutrons are ordered following the pattern $(p,n,p,n,\ldots)$ for the symmetric $(\delta=0)$ system, and the pattern of the form $(n,p,n,p,n;\ n,p,n,p,n;\ \ldots)$ for the asymmetric ($\delta=0.2$) system.

When a pion is created in a $\Delta\to N\pi$ decay, it is placed at the same position as the created nucleon.  The pion is not absorbed by this nucleon until the nucleon is scattered by some other particle.

The code, before introducing the improvements described above to avoid the isospin-dependent bias, was used in Refs.~\cite{matsuda2011,iwamoto2004} with $\Delta t=1$ fm/$c$, to study pion production in heavy-ion collisions as well as the hadronic cascade induced by pions.

\subsubsection{pBUU\label{sec:pBUU}}

In pBUU, as explained in detail in Ref.~\cite{yxzhang2018}, collisions within a spatial cell volume $V_{\text{cell}}$ and a time-step interval $\Delta t$ are calculated following Monte Carlo integration of the collision integral using test particles in the cell to sample the phase-space distribution ahead of collisions.  Thus the collision can occur between any two test particles in a cell.  To prevent excessive sampling, a subsample of pairs is randomly selected for potential collisions, and they are tested for collisions at enhanced probability.  This method is designed to run at a high test-particle number.  The homework calculations were performed with $N_{\text{tp}}=1000$, as well as in Ref.~\cite{yxzhang2018}.

When a test particle has changed its identity by a collision, e.g.~from a nucleon to a $\Delta$ particle, its collisions later in the same time step are artificially turned off, i.e., it is a stealth particle until the end of the time step.  Between two time steps for collisions, all the existing particles are propagated for the time interval $\Delta t$.

In the present work, stealth particles mentioned above are changed to be `superactive' in the subsequent time step  to compensate for the reduction of the collision and decay rates they were subject to.  The collision and decay rates of such a superactive particle are increased by 50\%.  In Fig.~\ref{fig:seqsim}, a superactive particle is treated as yielding a contribution of 1.5 to the counted number of particles.  The superactive strategy is to be used in the code from now on.

In the past \cite{danielewicz1991,danielewicz1995,hong2014,tsang2017}, for energies of 300 MeV -- 2 GeV per nucleon, where pions are produced, the code pBUU was mostly used with the time step in the range $\Delta t = 0.2\text{-}0.3$ fm/$c$, usually with longer time steps at the lower energies and shorter at the higher.  The scheme for pion and $\Delta$ production for that energy range was largely unchanged since the code was put together in 1991 \cite{danielewicz1991}, except that in the late 90s the $\Delta$ and pion production cross sections and rates were switched to those from Ref.~\cite{huber1994}. At higher energies, both the time step gets shortened and the production scheme gradually changes to that from string fragmentation.

\subsubsection{RVUU}

The collision order has been changed from a fixed ordering in Ref.~\cite{yxzhang2018} to an ordering randomized at every time step.  The time dilation effect in the geometrical collision condition is now taken into account in the way of Ref.~\cite{fengli2017}, i.e.~by letting two particles collide when their distance in the computational reference frame becomes minimal during the time step and if the minimum distance $d_{\perp}^{\text{(ref)}}$ is within the range of a transformed cross section $\sigma_{\text{tot}}^{\text{(ref)}}=v'\sigma_{\text{tot}}/v$, with $v'$ defined by Eq.~\eqref{eq:vprime} and $v$ being the relative velocity in the computational reference frame.  The code is run in the parallel ensemble mode in the present work, while it was run in the full ensemble mode in Ref.~\cite{yxzhang2018}.  The total cross section is directly used in the distance criterion $\pi d_{\perp}^{\text{(ref)}2}<\sigma_{\text{tot}}^{\text{(ref)}}$ for a collision.

When a pion is created in a $\Delta\to N\pi$ decay, it is placed at the same position as the created nucleon.  The pion is not absorbed by this nucleon until the nucleon is scattered by some other particle.

In Refs.~\cite{song2015,zhenzhang2017}, the threshold effect and the effect of pion potentials in pion production were studied with the RVUU code in the parallel ensemble mode with a time step of $\Delta t=0.1$ fm/$c$.

\subsubsection{SMASH\label{sec:SMASH}}

SMASH version 1.1, which is equivalent to the current version 1.6 for the collision treatment, was used in the full-ensemble mode with $N_{\text{tp}}=100$.  In elastic collisions, only the momenta of particles are changed, with their positions unchanged.  However, in inelastic collisions, such as $NN \leftrightarrow N\Delta$ and $N\pi \rightarrow \Delta$, the outgoing particles are placed at $\frac12(\bm{r}_1 + \bm{r}_2)$, where $\bm{r}_{1,2}$ are the coordinates of incoming particles in the computational frame. The default collision algorithm applies a cross section cutoff at $\sigma_{\text{max}}/N_{\text{tp}} = 200/N_{\text{tp}}$ mb, which in this paper was increased to $\sigma_{\text{max}}/N_{\text{tp}} = 1000/N_{\text{tp}}$ mb.

Pion production in Au + Au collisions below 1$A$ GeV laboratory-frame energy was previously studied with SMASH, see Figs.~25 and 28 of Ref.~\cite{weil2016}. In this previous publication an earlier version of SMASH with fixed time steps of 0.1 fm/$c$ was used.  It was shown that the effects of mean-field potentials, Fermi motion, and Pauli blocking on $\pi^-/\pi^+$ ratio are all equally important at 0.4 GeV. These results were reproduced using the later SMASH version with the timestepless propagation \cite{oliinychenko2017}.

\subsubsection{TuQMD\label{sec:TuQMD}}

The treatment of baryon-baryon collisions is the same as in Ref.~\cite{yxzhang2018}.  When a pion is created in a $\Delta\to N\pi$ decay, its position is randomly determined inside a sphere with a radius of about 0.3 fm and centered at the resulting nucleon.  The pion is allowed to be absorbed by this nucleon, as well as by other surrounding nucleons, after they are propagated for one time step [see Fig.~\ref{fig:seqsim}(g)], if the geometrical condition is met.

Previous versions of the model were used to study pion production in heavy-ion collisions with an emphasis on the impact of pion--nucleon and Coulomb interactions \cite{maheswari1998b,fuchs1997b,zipprich1997} and of nuclear matter equation of state \cite{maheswari1998a} on pion spectra and collective flows, and impact of pion production channels on sub-threshold kaon production \cite{fuchs1997a}.  A more recent version of the model, which however omitted the factor $p^*$ in Eq.~\eqref{eq:mass-sampling} for the resonance mass sampling, was used in Refs.~\cite{cozma2016,cozma2017} to study pion production, where the time step $\Delta t=0.35$ fm/$c$ was used.

\section{$N\Delta$ system\label{sec:ndelta}}

The present transport-code comparison focuses on the tests of the collision and decay terms in a simple setup by turning off mean-field interactions and Pauli blocking.  However, compared to the case with only $NN$ elastic collisions, the collision term here is much more complicated with many input parameters.  Differences between codes may arise from various ingredients in treating the collision and decay processes and from the numerical methods and prescriptions used in different parts of individual codes.  To isolate the differences as much as possible, we limit ourselves in this section to the case with only nucleons and $\Delta$ particles by artificially turning off the decay of $\Delta$.  The spectral function of $\Delta$ still has a width.  The interpretation of results is thus much easier than in the case with pions, which is studied in Sec.~\ref{sec:ndeltapi}.  This simplification is also useful for understanding some effects due to the different ways collisions are treated in transport codes.

\subsection{Results}

\begin{figure*}
  \includegraphics[width=\textwidth]{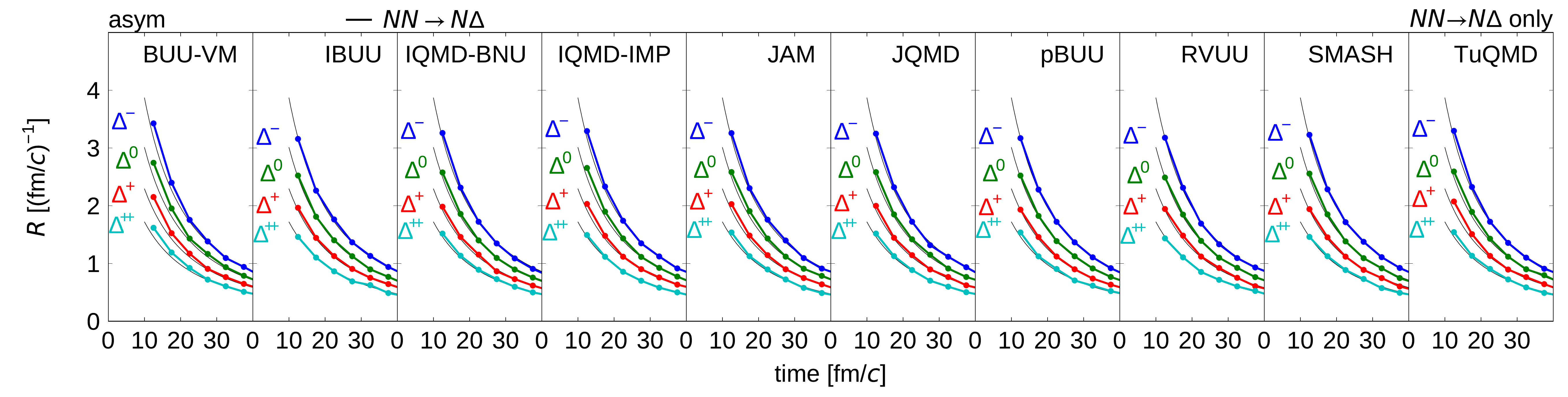}
  \caption{\label{fig:nnnd_dndta-De1P0}
    Reaction rates for $NN\to N\Delta$ processes in the asymmetric ($\delta=0.2$) system at early times $t\le 40$ fm/$c$ in the case with $NN\to N\Delta$, but without $N\Delta\to NN$ and $\Delta\to N\pi$ processes.  The points connected by lines show the rates to produce $\Delta^{-,0,+,++}$ averaged over every 5-fm/$c$ interval.  The solution of the rate equation is shown by thin lines.
  }
\end{figure*}

We first consider the code comparison for the simple case of having only the $NN\to N\Delta$ reaction, but not the backward $N\Delta\to NN$ reaction.  Starting with the initial condition of only 1280 nucleons, the number of $\Delta$ gradually increases due to the $NN\to N\Delta$ reactions that start at $t=10$ fm/$c$ as specified in the homework condition.  Figure \ref{fig:nnnd_dndta-De1P0} shows the time evolution of the $NN\to N\Delta$ reaction rates to create $\Delta^-$, $\Delta^0$, $\Delta^+$ and $\Delta^{++}$, in an asymmetric system ($\delta=0.2$).  Since the initial nucleon energy is converted to the mass of $\Delta$ resonance through the reaction $NN\to N\Delta$, the $NN$ reaction rate decreases rapidly at early times.  For comparison, we also show by thin lines the results from solving the rate equation, which assumes thermal momentum distributions at any instant of time, as described in Appendix~\ref{sec:rateeq}.  The latter assumption is a good approximation since the $NN$ elastic collision rate at $T=60$ MeV is 213 $c$/fm and is much higher than the $NN\to N\Delta$ rate.  In fact, the results of many transport codes agree with the solution of the rate equation, except for BUU-VM that has slightly higher reaction rates.  The problem of high collision rates in BUU-VM compared with other codes has already been observed in the comparison for $NN$ elastic collisions in Ref.~\cite{yxzhang2018}, with its source still unknown.

\begin{figure*}
  \includegraphics[width=\textwidth]{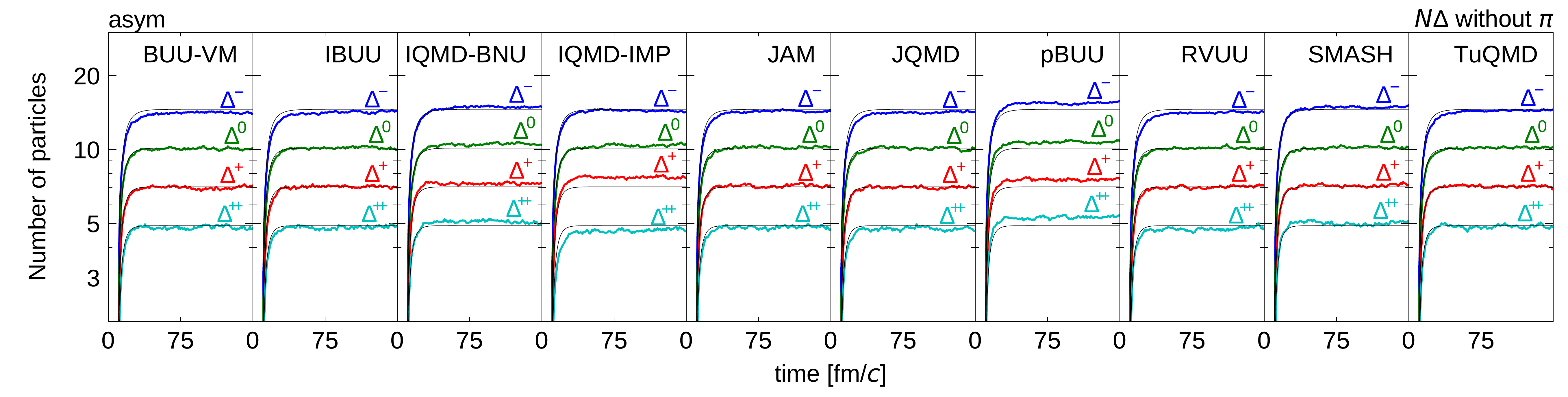}
  \caption{\label{fig:nat1a-De2P0} Time evolution of the number of $\Delta$ in the asymmetric ($\delta=0.2$) system in the case without pions but with $NN\to N\Delta$ and $N\Delta\to NN$ reaction.  Thick lines represent the results from the indicated transport codes and thin lines -- from the rate equation.}
\end{figure*}

The comparison for the case with both $NN\to N\Delta$ and $N\Delta\to NN$ reactions is represented in Fig.~\ref{fig:nat1a-De2P0}.  The colored thick lines show the time evolution of the numbers of $\Delta^-$, $\Delta^0$, $\Delta^+$ and $\Delta^{++}$ calculated by different codes for the asymmetric ($\delta=0.2$) system.  After a few tens of fm/$c$, these numbers almost reach equilibrium.  Their values are more or less equally spaced in these semi-logarithmic plots, which is expected for an ideal Boltzmann-gas mixture.  In many codes, the equilibrated numbers agree well with those from the rate-equation solution shown in the figures by thin black lines.  However, there are slight deviations for some codes.  Reasons for these deviations have not been fully understood, except for the pBUU code where the high number of $\Delta$ can be understood from its treatment of collisions.  As explained in Sec.~\ref{sec:pBUU} for this code, the $\Delta$ particle created in a time step is a stealthy particle until the end of the same time step, thus resulting in a suppression of the $N\Delta\to NN$ reaction, even though the effect is partially compensated by turning the stealth particles to be `superactive' in the next time step.  Since the rate of the $N\Delta\to NN$ reaction is 7.99 $c$/fm and the number of existing $\Delta$ is about 36.7 for the asymmetric system ($\delta=0.2$) in the ideal gas mixture in chemical equilibrium (see Appendix \ref{sec:equil}), the chance for a created $\Delta$ to be absorbed within the same time step is not negligible.%
\footnote{ The problem is expected to be smaller in usual pBUU calculations which employ a shorter time step than 0.5 fm/$c$.}

\begin{figure*}
  \includegraphics[width=\textwidth]{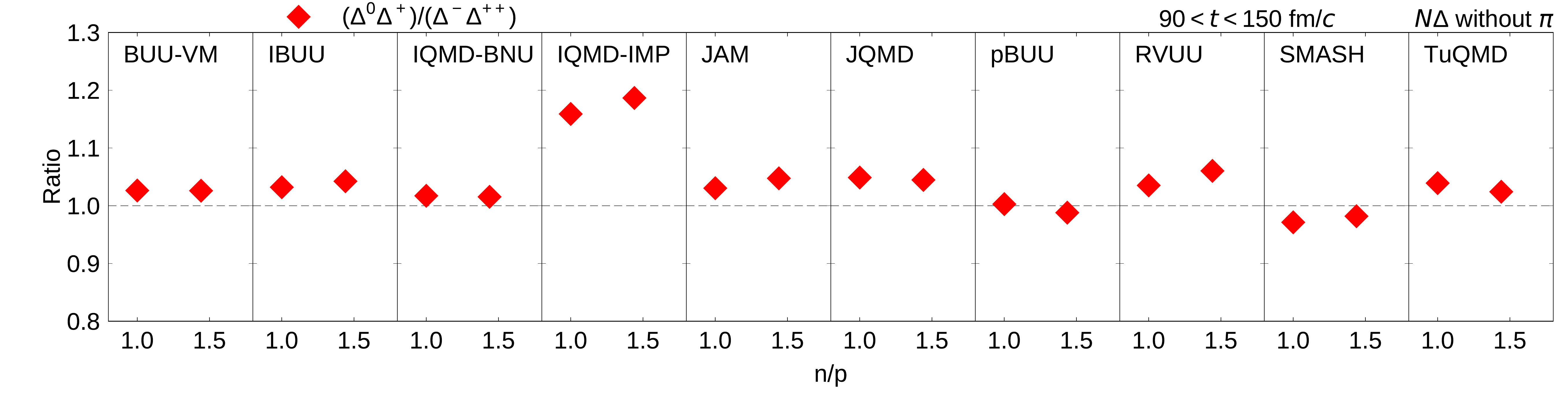}
  \caption{\label{fig:ratioa-De2P0} Ratios $(\Delta^0\Delta^+)/(\Delta^-\Delta^{++})$ averaged over late times $90<t<150$ fm/$c$ in the case without pions for the two systems ($\delta=0$ and $\delta=0.2$), versus the calculated values of $n/p$.}
\end{figure*}

A closer look at Fig.~\ref{fig:nat1a-De2P0} indicates that there are some irregularities in the spacing between the numbers of $\Delta$, and they can be more clearly seen in Fig.~\ref{fig:ratioa-De2P0} where the ratio $(\Delta^0\Delta^+)/(\Delta^-\Delta^{++})$ calculated from the numbers of  $\Delta$ in different charge states is displayed.  Although this ratio should be one for the Boltzmann gas of any asymmetry $\delta$ at chemical equilibrium, results from many codes show an excess of $\Delta^0$ and $\Delta^+$ compared to $\Delta^-$ and $\Delta^{++}$.  In most codes, this problem exists in both the asymmetric (the right point in each panel) and the symmetric system (the left point), and it can be explained by the higher-order $(N\Delta)_{\text{nnx}}$ correlation (Fig.~\ref{fig:nnxd}) induced by the way the $N\Delta\to NN$ reaction is implemented in transport codes.  Since the violation of isospin symmetry among the four species of $\Delta$, even in the symmetric ($\delta=0$) system, is unphysical, it may affect some important observables such as the $\pi^-/\pi^+$ ratio in heavy-ion collisions.  The largest violation of isospin symmetry in IQMD-IMP is most likely due to the fixed ordering of collision pairs (see Sec.~\ref{sec:IQMD-IMP}), taken from the particle list that is initialized for every event with protons first and neutrons last.  Nucleons are gradually mixed in the list by $NN\leftrightarrow N\Delta$ collisions, which however takes a long time of an order of 100 fm/$c$, and thus affects the quantities averaged over $60<t<150$ fm/$c$.

\begin{figure*}
  \includegraphics[width=\textwidth]{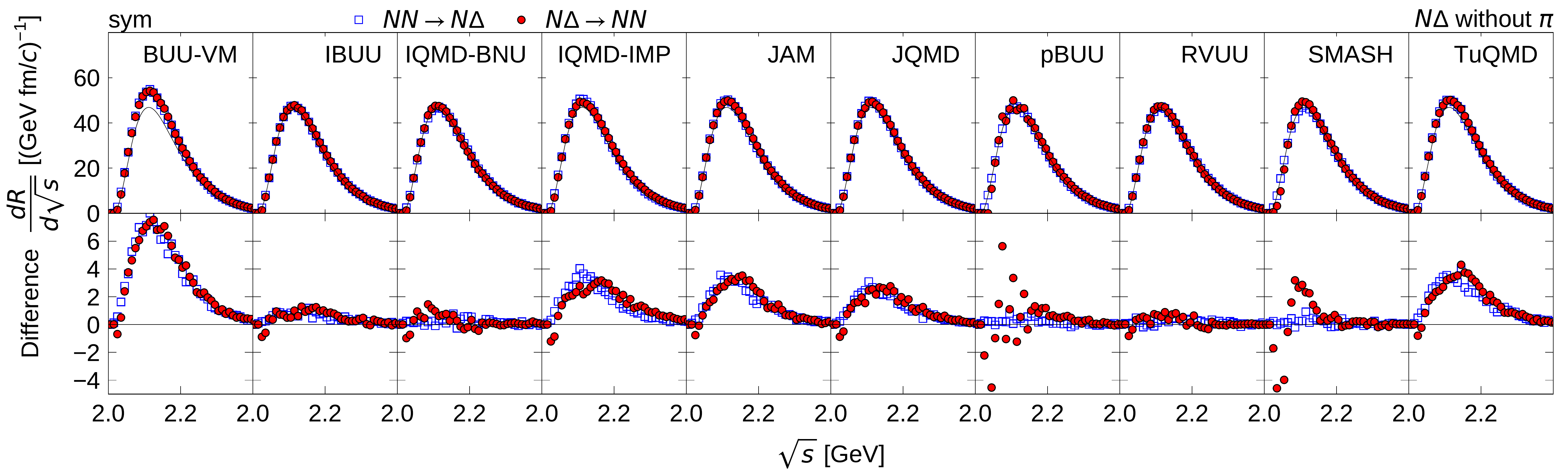}
  \caption{\label{fig:nnnd_edists-De2P0} Upper panels: Distributions of the $NN\to N\Delta$ (open squres) and $N\Delta\to NN$ (filled circles) reactions versus $\sqrt{s}$ in a symmetric ($\delta=0$) system without pions.  The reaction rates are averaged over $90<t<150$ fm/$c$. The thin line shows the $\sqrt{s}$ distribution in the ideal gas mixture.  Lower panels: Deviations of the distributions from those in the ideal gas mixture, shown in a magnified scale.}
\end{figure*}

In Fig.~\ref{fig:nnnd_edists-De2P0}, we show in the upper panels the $\sqrt{s}$ distributions of the $NN\to N\Delta$ (blue open squares) and $N\Delta\to NN$ (red filled circles) reaction rates at late times $90<t<150$ fm/$c$.  For such rates (and particle numbers in later sections) summed over the isospin channels, the quantitative results depend on the asymmetry $\delta$ of the system only weakly.  The figure is for the symmetric system, but any discussion on these results does not depend on $\delta$.  Because of the detailed balance relation, the two distributions of $NN\to N\Delta$ and $N\Delta\to NN$ should agree, and this is roughly the case in transport-code results.  The thin line in each panel is what is expected from the ideal Boltzmann-gas mixture, which can be calculated from Eq.~\eqref{eq:dlambdadsqrts} and the particle densities in the ideal gas mixture (Appendix \ref{sec:equil}).  The good agreement between the two distributions indicates that the particle momenta in transport-code calculations have reached the thermal distribution at the expected temperature.  The deviations of the transport-code results from that of the ideal gas are shown in the lower panels in a magnified scale.  It is seen that the full-ensemble BUU code (pBUU and SMASH) seems to be consistent with the expected ideal-gas rate at least for the total rates integrated over $\sqrt{s}$.  Some QMD codes (IQMD-IMP, JAM, JQMD and TuQMD) overestimate the ideal-gas value by about 10\% in a similar pattern. This could be caused by the effect of the higher-order correlations $(NN)_{\text{nnx}}$ and $(N\Delta)_{\text{nnx}}$ shown in Figs.~\ref{fig:nnx} and \ref{fig:nnxd} in Sec.~\ref{sec:hocorr}, which tend to enhance both rates.  The same effect could be expected in principle for the parallel-ensemble BUU codes, because they are supposed to work equivalently to the QMD codes for the treatment of the collision term without Pauli blocking.  Contrary to this expectation, results from IBUU, IQMD-BNU and RVUU perfectly agree with the integrated rate for the ideal gas mixture, which we have not understood well.  The higher reaction rates in BUU-VM seem to be related to those in Fig.~\ref{fig:nnnd_dndta-De1P0} for the case without the backward $N\Delta\to NN$ reaction.  In most cases, these deviations in the calculated reaction rates do not seem to affect much the equilibrated number of $\Delta$, as shown in Fig.~\ref{fig:nat1a-De2P0}.  This is possible when both the forward and backward reaction rates deviate by similar factors from the expected value.  

With a closer look at the lower panels of Fig.~\ref{fig:nnnd_edists-De2P0}, we find that the quality of agreement between the distributions of $NN\to N\Delta$ and $N\Delta\to NN$ reactions depends on the code.  Disagreements appear not only around the peak, but also in the low and high energy regions.  Four of the QMD codes (IQMD-IMP, JAM, JQMD and TuQMD) show very similar behaviors for the two rates, including the behavior of the small difference between the two rates.  The relatively large violation of the detailed balance in SMASH can be due to a cutoff imposed on cross sections (see Sec.~\ref{sec:SMASH}), which can cause a deficiency of the $N\Delta\to NN$ rate at low energies.  In all cases, the integrated rates of forward and backward reactions agree well with each other, resulting in the equilibration of the number of $\Delta$ in each code at late times displayed in Fig.~\ref{fig:nat1a-De2P0}.

\subsection{Summary of results for the $N\Delta$ system}

In the studied case of the $N\Delta$ system without pions, we have obtained reasonable agreements among codes and also with respect to the reference case of ideal gas mixture.  Some of the deviations commonly observed in many codes have been tied to the specific methods used in these codes for solving the collision term in the Boltzmann equation.  This is particularly true for the effects that result from the higher-order correlations, such as too high collision rates and the violation of isospin symmetry, although they are not very pronounced.  Whether the remaining small differences play a role in realistic heavy-ion collisions needs to be carefully studied.

\section{$N\Delta\pi$ system\label{sec:ndeltapi}}

In this section, we carry out detailed analyses of the results for the case with pions in which both $NN\leftrightarrow N\Delta$ and $\Delta\leftrightarrow N\pi$ reactions are allowed.  We focus on the equilibrium particle numbers and other quantities at late times and interpret the results based on the theoretical backgrounds described in Sec.~\ref{sec:tptheory}.

\subsection{Particle numbers\label{sec:ndeltapi-numbers}}

We have already seen from the time evolution of the numbers of $\Delta$ and $\pi$ ($N_\Delta$ and $N_\pi$) displayed in Fig.~\ref{fig:nat1a}, that the results from different codes do not agree and some of them largely deviate from the results given by the rate equation.  Comparing to Fig.~\ref{fig:nat1a-De2P0} for the case without pions, we can see that these deviations are much larger, indicating that $N_\Delta$ is strongly affected by the $\Delta\leftrightarrow N\pi$ reactions.

\begin{figure*}
  \includegraphics[width=\textwidth]{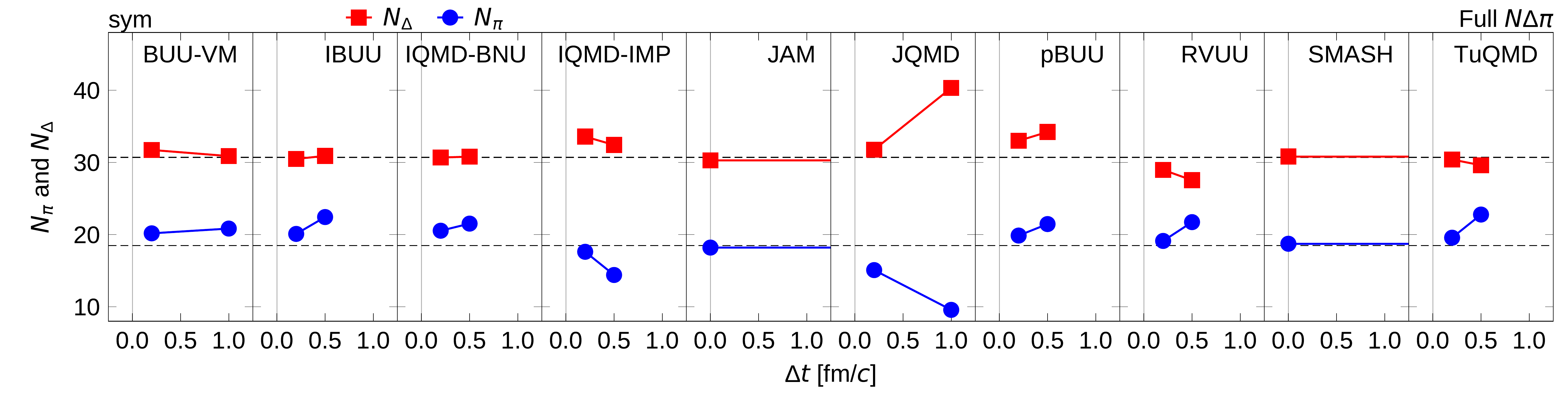}
  \caption{\label{fig:dt_nats} Dependence of the $\Delta$ (filled red squares) and $\pi$ (filled blue circles) numbers, averaged over late times $90<t<150$ fm/$c$, in the symmetric ($\delta=0$) full-$N\Delta\pi$ system.  For each code, the result with the homework time-step size $\Delta t$ chosen by the code and that with $\Delta t=0.2$ fm/$c$ are connected by a line.  The results of JAM and SMASH do not depend on the time step.  The values for the ideal Boltzmann-gas mixture are indicated with horizontal dashed lines.}
\end{figure*}

Since the $N\pi\to\Delta$ cross section is large and the lifetime of $\Delta$ is relatively short, the results in Fig.~\ref{fig:nat1a} obtained with $\Delta t$ of the choice by individual codes ($\Delta t=0.5$ or 1 fm/$c$), except for the time-step--free codes, may change if a smaller value of $\Delta t=0.2$ fm/$c$ is used. Figure~\ref{fig:dt_nats} shows the $\Delta t$ dependence of $N_\Delta$ and $N_\pi$ represented with filled red squares and blue circles, respectively, after averaging over the late times $90<t<150$ fm/$c$.  The results of time-step--free codes (JAM and SMASH) are very close to the values expected for the ideal Boltzmann-gas mixture in chemical equilibrium, represented with the dashed horizontal lines.  For the other codes, the two points at $\Delta t=0.2$ fm/$c$ and that of the choice by the code are connected by a line in Fig.~\ref{fig:dt_nats}.  Expecting the dependence on $\Delta t$ to be linear for $\Delta t$ in this range of values,\footnote{In fact, the linear dependence is confirmed in pBUU and TuQMD by calculations with various values of $\Delta t$.} we can obtain for each code the particle numbers for $\Delta t=0$ by linear extrapolation and find that they all seem to agree much better with those for the ideal gas mixture particularly for $N_\pi$.  An exception is the $N_\Delta$ in IQMD-IMP for which its value at $\Delta t=0.2$ fm/$c$ deviates from the value of the ideal gas mixture more than that at $\Delta t=0.5$ fm/$c$.  Otherwise, it seems that the problems of $N_\Delta$ and $N_\pi$ in most codes can be largely reduced by using smaller $\Delta t\to0$.

Most of the deviations at finite $\Delta t$ can be understood in detail from the treatment of the sequence of the collision process $C_k$ and the decay process $D_k$ used in individual transport codes, as summarized earlier in Fig.~\ref{fig:seqsim}.  As already discussed in Sec.~\ref{sec:seq}, the code RVUU represented in panel (a) overpredicts $N_\pi$ and underpredicts $N_\Delta$ numbers, while the code JQMD represented in panel (b) underpredicts $N_\pi$ and overpredicts $N_\Delta$ in particular at large $\Delta t$.  Since the BUU-VM code mixes events of the types given in panels (a) and (b), the predicted  $N_\Delta$ and $N_\pi$ in Fig.~\ref{fig:dt_nats} are thus close to the correct values, and exhibit a weak dependence on $\Delta t$.  For the code IBUU in panel (c) of Fig.~\ref{fig:seqsim}, where the pions created in $D_k$ are not absorbed in $C_k$ because they are treated as stealthy, a high $N_\pi$ is obtained at large $\Delta t$ in Fig.~\ref{fig:dt_nats}, as already expected in Sec.~\ref{sec:seq}.  For the code pBUU [panel (d) of Fig.~\ref{fig:seqsim}], the high $N_\Delta$ and $N_\pi$  in Fig.~\ref{fig:dt_nats} at large $\Delta t$ are likely due to some particles becoming stealthy during each time step, which has been partially remedied by the introduction of the `superactive' particles (see Sec.~\ref{sec:pBUU}).  For the codes IQMD-IMP, IQMD-BNU and TuQMD, baryon-baryon collisions $C_k^{\text{BB}}$ and meson-baryon collisions $C_k^{\text{MB}}$ are treated separately in the sequence.  For the code IQMD-IMP [panel (e) of Fig.~\ref{fig:seqsim}], the decay $D_k$ is processed in between $C_k^{\text{BB}}$ and $C_k^{\text{MB}}$.  Since the change of $N_\Delta$ by $C_k^{\text{BB}}$ processes is probably not as significant as by $D_k$ and $C_k^{\text{MB}}$, we expect the result to be similar to that from the code JQMD in panel (b).  Indeed, in Fig.~\ref{fig:dt_nats}, the behavior of $N_\pi$ in IQMD-IMP is similar to that in JQMD, but $N_\Delta$ shows a different behavior.  The treatment of collisions and decays in IQMD-BNU [panel (f)] is similar to IQMD-IMP, but the $\Delta$ particles produced in $C^{\text{BB}}_k$ are partially stealth within it and are not allowed to decay in $D_k$ for the same time step.  The pions produced in $D_k$ are also treated as stealth in $C^{\text{MB}}_k$.  Therefore the behaviors of $N_\Delta$ and $N_\pi$ may be different compared to IQMD-IMP and JQMD, as seen in Fig.~\ref{fig:dt_nats}.  The TuQMD code [panel (g)] allows for particle propagation between $C_k^{\text{BB}}$ and $C_k^{\text{MB}}$,\footnote{In the present paper, we consistently show the numbers of propagated particles for all the codes rather than the numbers at the time step boundary set by the code.} and the result in Fig.~\ref{fig:dt_nats} is similar to RVUU [panel (a)], probably because $C_k^{\text{BB}}$ does not change $N_\Delta$ much once the system reaches equilibrium.  For the time-step--free codes (JAM and SMASH), there is of course no problem in reproducing the equilibrated values of $N_\Delta$ and $N_\pi$.

It is worth mentioning that the $\Delta\leftrightarrow N\pi$ processes do not strongly impact the state of nucleons, under the conditions of the present homework.  By comparing the present full $N\Delta\pi$ system to the $N\Delta$ system studied in Sec.~\ref{sec:ndelta}, we find that the temperature of the system decreases slightly from 55.8 MeV to 54.3 MeV and the number of nucleons increases slightly from 1245.0 to 1249.3, when the pions are introduced in the symmetric ($\delta=0$) ideal gas mixture.  Correspondingly, the number $N_\Delta$ decreases from 35.0 to 30.7.  This is a relatively small change, compared to $N_\pi$ which increases from 0 to 18.4.  Therefore, one may expect that the details of $\Delta\leftrightarrow N\pi$ processes do not affect $N_\Delta$ as strongly as $N_\pi$.  For example, some codes, such as TuQMD (see Sec.~\ref{sec:TuQMD}), may include a spurious $N\pi\to\Delta$ process, particularly when $\Delta t$ is small.  This is expected to have an effect of counter-acting the $\Delta\to N\pi$ decay, but its impact on $N_\Delta$ does not seem strong in Fig.~\ref{fig:dt_nats}.

\subsection{Reaction and decay rates}

\begin{figure*}
  \includegraphics[width=\textwidth]{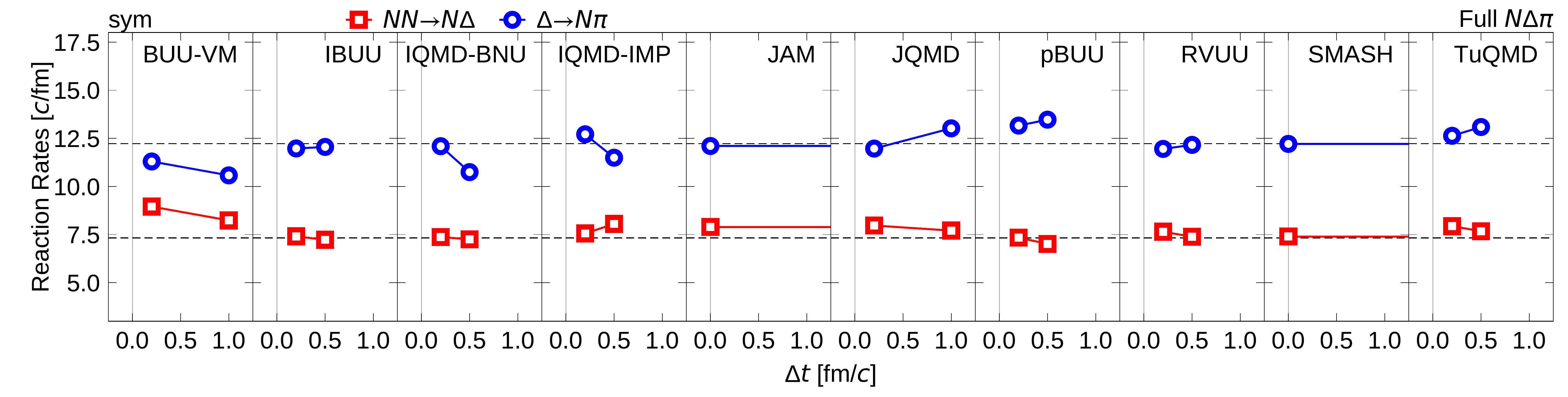}
  \caption{\label{fig:dt_ratess} Dependence on the time-step size $\Delta t$ of the $NN\to N\Delta$ reaction rate (open red squares) and the $\Delta\to N\pi$ decay rate (open blue circles), which are averaged over the late times $90<t<150$ fm/$c$, in the symmetric ($\delta=0$) full-$N\Delta\pi$ system.  For each code, the result with homework $\Delta t$ chosen by the code and that with $\Delta t=0.2$ fm/$c$ are connected by a line.  The results of JAM and SMASH do not depend on the time step.  The values for the ideal Boltzmann-gas mixture are indicated with horizontal dashed lines.}
\end{figure*}

In this subsection, we check if the reaction rates in the examined codes are correctly calculated.  The open red squares in Fig.~\ref{fig:dt_ratess} show the $\Delta t$ dependence of the $NN\to N\Delta$ rate at late times $90<t<150$ fm/$c$.  It is practically identical to the $N\Delta\to NN$ rate as far as the numbers of particles have stabilized at chemical equilibrium.  The value expected for the ideal Boltzmann-gas mixture is shown with the dashed horizontal line in Fig.~\ref{fig:dt_ratess}, calculated from Eq.~\eqref{eq:lambda_coll} using particle densities of an ideal gas mixture (Appendix \ref{sec:equil}).   We observe in general that the $\Delta t$ dependence of the $NN\to N\Delta$ rate is weak even in the codes exhibiting a strong $\Delta t$ dependence in the number $N_\Delta$.  This is understandable because $N_\Delta$ is the number of propagated $\Delta$ particles at a time-step boundary, while the number of reactions is accumulated over the computational steps (see Fig.~\ref{fig:seqsim}) and the reaction rate is thus stable even for a large value of $\Delta t$.  The small but finite deviations from the ideal-gas value are indeed not reduced by choosing a small $\Delta t$ in most codes.  In many respects, these deviations are similar to those found in the case without pions (Sec.~\ref{sec:ndelta}).  For example, many of the QMD codes predict higher rates than in the ideal gas mixture, which can be explained by the higher-order correlations of $(NN)_{\text{nnx}}$ in Fig.~\ref{fig:nnx} and $(N\Delta)_{\text{nnx}}$ in Fig.~\ref{fig:nnxd}.  In QMD and parallel-ensemble BUU codes, it is commonly observed (with an exception of IQMD-IMP) that the rate slightly decreases when $\Delta t$ is increased, which suggests weaker higher-order correlations with larger $\Delta t$, because only one collision is considered for each pair of particles in the same time step.

For the $\Delta\leftrightarrow N\pi$ processes, the $\Delta t$ dependence of the $\Delta \to N\pi$ rate is shown with open blue circles in Fig.~\ref{fig:dt_ratess}, which should be practically identical to the backward $N\pi\to\Delta$ rate.  As $\Delta t\to0$, while some codes may be converging to the right values expected for the ideal Boltzmann-gas mixture (dashed horizontal line), others do not.  For IBUU, its result agrees well with the correct value already at the large time step $\Delta t=0.5$ fm/$c$.  Results from the time-step--free codes (JAM and SMASH) also agree with those expected for the ideal gas mixture.

\begin{figure*}
  \includegraphics[width=\textwidth]{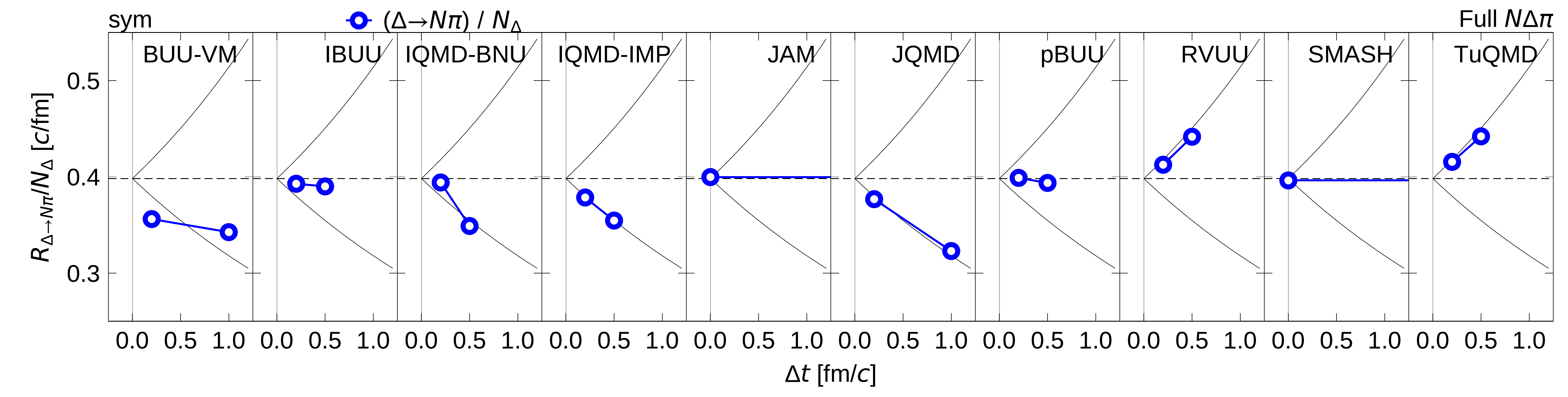}
  \caption{\label{fig:dt_dnpis} Dependence on the time-step size $\Delta t$ of the $\Delta\to N\pi$ decay rate divided by the number of the existing and propagated $\Delta$ particles, averaged over the late times $90<t<150$ fm/$c$, in the symmetric ($\delta=0$) full-$N\Delta\pi$ system.  For each code, the result with homework $\Delta t$ chosen by the code and that with $\Delta t=0.2$ fm/$c$ are connected by a line.  The results of JAM and SMASH do not depend on the time step.  The values for the ideal Boltzmann-gas mixture are indicated with horizontal dashed lines.  The lower and upper thin solid lines indicate the $\Delta t$ dependence of Eqs.~\eqref{eq:dt_dnpi1} and \eqref{eq:dt_dnpi2}, respectively, depending on the way how the collisions and decays are treated in codes.}
\end{figure*}

The relation between the calculated $\Delta\to N\pi$ rate in a code, to the way it treats particle collisions, is more clearly seen in Fig.~\ref{fig:dt_dnpis}, which shows the details of the decay rate (the open circles in Fig.~\ref{fig:dt_ratess}) divided by the $N_\Delta$ number (the filled squares in Fig.~\ref{fig:dt_nats}).  In the time-step--free codes (JAM and SMASH), it is rather trivial that this quantity $R_{\Delta\to N\pi}/N_\Delta$ agrees with that in the ideal gas (horizontal dashed line in the figure) as far as the codes reproduce the $\Delta$ mass distribution in equilibrium.  In codes that rely on time steps, the probability for the decay of a $\Delta$ particle is typically chosen to be $1-\exp(-\Gamma'\Delta t)$ in the decay procedure $D_k$ for a time-step interval $\Delta t$, where the decay rate in the computational frame $\Gamma'$ is given by Eq.~\eqref{eq:GammaInCompFrame} with Eq.~\eqref{eq:Gamma}.\footnote{
This appropriate treatment with $\Gamma'=(m_\Delta/E_\Delta)\Gamma_{\Delta\to N\pi}$ is employed in IBUU, IQMD-IMP, JQMD, TuQMD and RVUU.  In BUU-VM and IQMD-BNU, the time dilation effect is ignored by simply taking $\Gamma'=\Gamma_{\Delta\to N\pi}$.
}  In such approach, one of possible expectations for the $\Delta t$ dependence of the decay rate divided by $N_\Delta$ may be
\begin{equation}
\label{eq:dt_dnpi1}
\frac{R_{\Delta\to N\pi}}{N_\Delta}
 = \frac{\langle1-e^{-\Gamma'\Delta t}\rangle}{\Delta t}
\approx \frac{1-e^{-\langle\Gamma'\rangle\Delta t}}{\Delta t},
\end{equation}
where the average $\langle\cdot\rangle$ is over both the mass distribution and the momentum distribution of $\Delta$ particles.  This $\Delta t$ dependence calculated with the value $\langle\Gamma'\rangle$ for the ideal gas is shown by the lower thin line in each panel of Fig.~\ref{fig:dt_dnpis}.  This $\Delta t$ dependence applies to the JQMD code because, as shown in panel (b) of Fig.~\ref{fig:seqsim}, the decay procedure $D_k$ is directly considered for the $\Delta$ particles that exist at the time-step boundary and the number $N_\Delta$ is of such propagated $\Delta$ particles.  A similar $\Delta t$ dependence is expected for IQMD-IMP in panel (e) of Fig.~\ref{fig:seqsim}, because the $NN\leftrightarrow N\Delta$ processes in $C_k^{\text{BB}}$ do not change the number of $\Delta$ on average, as long as the sum $N_\Delta+N_\pi$, which is not affected either by $C_k^{\text{MB}}$ or $D_k$, has reached equilibrium at these late times.  In IQMD-BNU, the decay rate could be lower than IQMD-IMP because some $\Delta$ particles are stealth in $D_k$, but the decay rate at small $\Delta t$ can also be larger because the time dilation effect is ignored in IQMD-BNU.  On the other hand, for the code RVUU in panel (a), the number of $\Delta$ particles considered in $D_k$ is not the same as the number $N_\Delta$ of propagated $\Delta$ particles, since it has been increased by the $N\pi\to\Delta$ process in $C_k$, on the average by a factor $e^{\Gamma'\Delta t}$, provided it balances with the decay factor $e^{-{\Gamma'\Delta t}}$ at these late times.  For RVUU, we therefore expect
\begin{equation}
\label{eq:dt_dnpi2}
\frac{R_{\Delta\to N\pi}}{N_\Delta}
=\frac{\langle e^{\Gamma'\Delta t}(1-e^{-\Gamma'\Delta t})\rangle}{\Delta t}
\approx\frac{e^{\langle\Gamma'\rangle\Delta t}-1}{\Delta t},
\end{equation}
which is shown by the upper thin line in each panel of Fig.~\ref{fig:dt_dnpis}.  We note that the decay rate in RVUU actually follows this expected line.  Since the baryon-baryon collisions $C_k^{\text{BB}}$ do not affect the number of $\Delta$ on average in panel (g) of Fig.~\ref{fig:seqsim}, the TuQMD result is expected to follow the same line as confirmed in Fig.~\ref{fig:dt_dnpis}.  The very low decay rate in BUU-VM is difficult to understand.

\begin{figure*}
  \includegraphics[width=\textwidth]{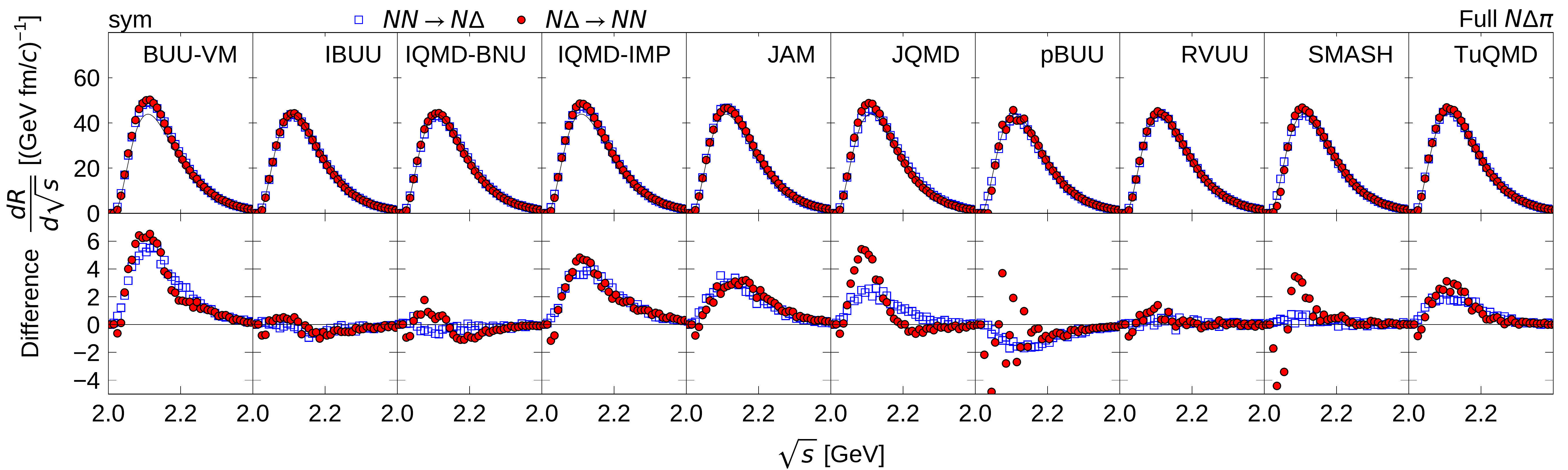}
  \caption{\label{fig:nnnd_edists}
    Upper panels: Distributions of the $NN\to N\Delta$ (open squares) and $N\Delta\to NN$ (filled circles) reactions in $\sqrt{s}$, in the symmetric ($\delta=0$) full-$N\Delta\pi$ system.  The reaction rates are averaged over $90<t<150$ fm/$c$. The thin curve shows the $\sqrt{s}$ distribution for the ideal gas mixture in chemical equilibrium.  Lower panels:  The deviations of the distributions from those in the ideal gas mixture, shown in a magnified scale.
  }
\end{figure*}

\begin{figure*}
  \includegraphics[width=\textwidth]{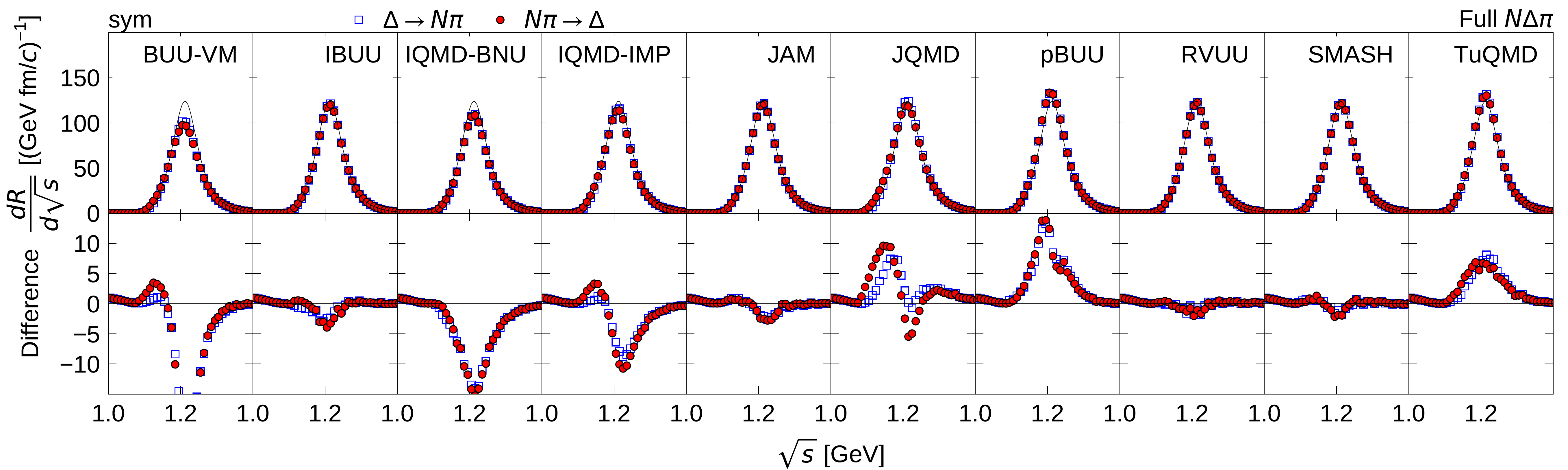}
  \caption{\label{fig:dnpi_edists}
    Upper panels: The $\sqrt{s}$ distributions of the $\Delta\to N\pi$ decay (open squares) and the $N\pi\to N$ reaction (filled circles) in $\sqrt{s}$, in the symmetric ($\delta=0$) full-$N\Delta\pi$ system.  The rates are averaged over $90<t<150$ fm/$c$. The thin curve shows the $\sqrt{s}$ distribution for the ideal gas mixture in chemical equilibrium.  Lower panels: The deviations of the distributions from those in the ideal gas mixture, shown in a magnified scale.
  }
\end{figure*}

For more insight into the codes, we show in Fig.~\ref{fig:nnnd_edists} the $\sqrt{s}$ distributions of the $NN\leftrightarrow N\Delta$ rates, as in Fig.~\ref{fig:nnnd_edists-De2P0} for the $N\Delta$ system.  The agreements and deviations of these distributions for different codes are similar to those in the case without pions, and similar discussions in Sec.~\ref{sec:ndelta} can be applied here.  An exception is that the quality of the agreement between the $\sqrt{s}$ distributions of the forward $NN\to N\Delta$ and backward $N\Delta\to NN$ rates becomes worse in JQMD when pions are introduced.  For the $\Delta\leftrightarrow N\pi$ processes, the $\sqrt{s}$ distributions are similarly shown in Fig.~\ref{fig:dnpi_edists}.  The overall deviations of the integrated rates from the ideal gas case have already been discussed for Fig.~\ref{fig:dt_ratess}.  The agreement of the forward and backward distributions is generally very good, except in JQMD.  With a closer look, however, we notice that the $N\pi\to\Delta$ rate in the low $\sqrt{s}$ region tends to be overestimated compared to the ideal-gas value or to the $\Delta\to N\pi$ rate, in most of the QMD and parallel-ensemble BUU codes.  This could be due to, e.g., the higher-order $(N\pi)_{\text{dx}}$ correlation (see Fig.~\ref{fig:npidx}), whose behavior and significance may depend on the details in the code.  In some codes (IQMD-IMP and JQMD), this deviation in the $N\pi\to\Delta$ rate seems to be visibly affecting the mass distribution of existing $\Delta$ particles which is overestimated by these codes in the low mass region.  These extra $\Delta$ particles correspond to the excess of the $N\Delta\to NN$ rate in the low $\sqrt{s}$ region in Fig.~\ref{fig:nnnd_edists} compared to the $NN\to N\Delta$ rate, in contrast to the case without pions (Fig.~\ref{fig:nnnd_edists-De2P0}).  In JQMD, when the time step $\Delta t$ is reduced from 1 fm/$c$ to 0.2 fm/$c$, we find that the differences between $\Delta\leftrightarrow N\pi$ rates and between $NN\leftrightarrow N\Delta$ rates decrease to a level similar to other codes.

\subsection{Isotopic ratios}

In Sec.~\ref{sec:digest}, we have already shown the results of the three isotopic ratios defined in Eq.~\eqref{eq:piratios}.  As shown by blue circles in Fig.~\ref{fig:dt_piratioa_late}, the time-step dependence of the $\pi$ ratio is strong, so that the results with a usual time step, such as $\Delta t=0.5$ fm/$c$, have not converged sufficiently well.  The results of different codes deviate from those of the ideal gas mixture differently.  On the other hand, the agreement among codes seems very good for the $\pi$-like ratio, which is most directly related to the $\pi^-/\pi^+$ observable in heavy-ion collisions.  The results seem to have converged already at large values of $\Delta t$.  We can now explain how the $\pi$-like ratio can be reliably predicted in spite of the unsatisfactory description of the $\pi$ ratio and the $N_\pi$ and $N_\Delta$ numbers, at least when the system has reached equilibrium.

We recall the $\Delta t$ dependence of the $\pi$ ratio in Fig.~\ref{fig:dt_piratioa_late} (blue circles), which is strongly correlated with the total pion number $N_\pi$ in Fig.~\ref{fig:dt_nats} (blue circles).  We have already seen in Sec.~\ref{sec:ndeltapi-numbers} that the latter is affected by how the sequence of collisions $C_k$ and decays $D_k$ is ordered (see Fig.~\ref{fig:seqsim}).  For example, for codes that allow the decays $D_k$ just before the propagation at the time-step boundary, the number $N_\pi$ tends to be high, especially when $\Delta t$ is large.  We also need to realize that the $\pi$ ratio is increased by $D_k$ (and decreased by $C_k$) because the $\pi^-/\pi^+$ ratio of the newly produced pions by $D_k$ is exactly the $\Delta$($\pi$-like) ratio, which is much higher than the $\pi^-/\pi^+$ ratio of existing pions (i.e.\ the $\pi$ ratio) as seen in Fig.~\ref{fig:dt_piratioa_late}.  Therefore, codes that allow $D_k$ just before the time-step boundary predict a relatively high value of the $\pi$ ratio, which explains the reason for its correlation with $N_\pi$.

On the other hand, the $\pi$-like ratio remains constant under $D_k$.  In the case of equilibrium, the change of particle numbers by $C_k$ is canceled by the change caused by $D_k$, which means that $C_k$ also does not change the $\pi$-like ratio.  Therefore, during the computational steps of collisions and decays (Fig.~\ref{fig:seqsim}), the $\pi$-like ratio stays constant on average, without any dependence on the choice in the order of $C_k$ and $D_k$.

This consideration suggests that transport codes can predict reliably the $\pi$-like ratio (green diamonds in Fig.~\ref{fig:dt_piratioa_late}) even with a large time step $\Delta t$ in spite of large deviations in the $\pi$ ratio and in the numbers $N_\Delta$ and $N_\pi$.  However, as mentioned before, the deviations in the numbers of propagated particles may result in their different time evolutions in heavy-ion collisions, which can in principle indirectly affect the isotopic ratios.

\subsection{Violation of isospin symmetry}

\begin{figure*}
  \includegraphics[width=\textwidth]{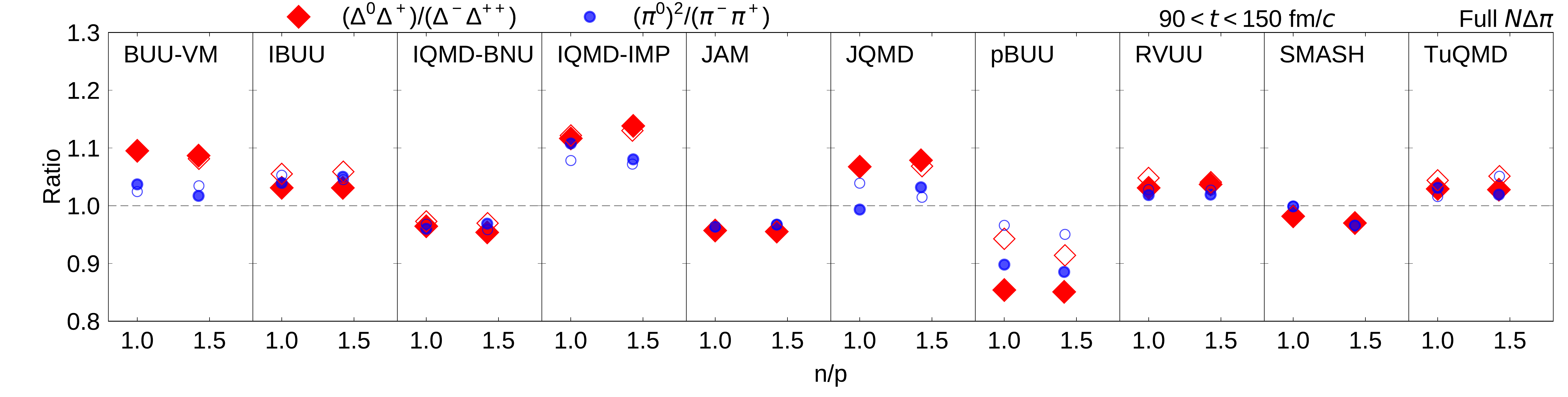}
  \caption{\label{fig:ratioa} Ratios $(\Delta^0\Delta^+)/(\Delta^-\Delta^{++})$ (diamonds) and $(\pi^0)^2/(\pi^-\pi^+)$ (circles), averaged over late times $90<t<150$ fm/$c$, in the full-$N\Delta\pi$ system ($\delta=0$ and $\delta=0.2$), versus the calculated values of $n/p$.  Filled symbols are the results calculated with the homework time-step parameter $\Delta t$ chosen by the code, while open symbols are with $\Delta t=0.2$ fm/$c$.}
\end{figure*}

For a more accurate description of the $\pi$-like ratio, we need to better understand the remaining uncertainties in the isotopic ratios.  In principle, uncertainties might have been caused by some code-specific problems which have appeared in the above analyses as unexplained behaviors of individual codes.  In this subsection, we focus on a commonly observed issue of the violation of isospin symmetry.

In Fig.~\ref{fig:ratioa}, the ratio $(\Delta^0\Delta^+)/(\Delta^-\Delta^{++})$, which measures the excess of $\Delta^{0,+}$ relative to $\Delta^{-,++}$, is shown by the red filled diamond for the symmetric ($\delta=0$) and asymmetric ($\delta=0.2$) systems, versus the calculated $n/p$ value.  Compared with the ratio of one in ideal Boltzmann-gas mixtures of any isospin asymmetry $\delta$, the calculated values from most codes, at the late times $90<t<150$ fm/$c$ in symmetric and asymmetric systems, are significantly different from the expected value.  Since the $\pi^-/\pi^+$ ratio for pions from the decays of $\Delta^0$ and $\Delta^+$ (or from $\Delta^-$ and $\Delta^{++}$) is the same as the ratio $\Delta^0/\Delta^+$ (or $\Delta^-/\Delta^{++}$) and $\Delta^0/\Delta^+ < \Delta^-/\Delta^{++}$ in the asymmetric system, an excess of $\Delta^{0,+}$ results in a smaller value of $\Delta$($\pi$-like) ratio, which qualitatively explains the tendency for most codes to underestimate the $\Delta$($\pi$-like) ratio as shown in Fig.~\ref{fig:dt_piratioa_late}.  In Fig.~\ref{fig:ratioa}, the excess of existing $\pi^0$ relative to $\pi^{-,+}$ is shown in the form of $(\pi^0)^2/(\pi^-\pi^+)$ by blue filled circles, which is well correlated to the excess of $\Delta^{0,+}$ as expected.  The corresponding open symbols are these ratios for a smaller time-step size $\Delta t=0.2$ fm/$c$.  Reducing $\Delta t$ is seen to only improve the results from pBUU, but not those from other codes.  The ratios in SMASH are very close to 1, which is most likely related to the weaker correlations in this full-ensemble BUU code.

The excess of $\Delta^0$ and $\Delta^+$ relative to $\Delta^-$ and $\Delta^{++}$ indicates a violation of isospin symmetry, and this is most likely due to the higher-order correlations induced from the geometrical method used in treating particle collisions in these codes.  In the case without pions shown in Fig.~\ref{fig:ratioa-De2P0}, the violation of the isospin symmetry has been attributed to the $(N\Delta)_{\text{nnx}}$ correlation (Fig.~\ref{fig:nnxd}) as discussed in Sec.~\ref{sec:nnxd}.  In the present case with pions, this correlation is expected to become weaker because the $\Delta$ particle ($\Delta_j'$ in Fig.~\ref{fig:nnxd}) is allowed to decay before it can collide again with the nucleon.  Adding the effects related to pions, the amount of isospin symmetry violation is expected to depend on individual codes as speculated in Sec.~\ref{sec:npid}.  In the case of JAM, only the correlations in Figs.~\ref{fig:nnxd}, \ref{fig:npidx} and \ref{fig:npid1} have effects, since the correlations in Fig.~\ref{fig:npid2} are absent due to its adopted prescription for treating particle collisions and decays.  Although the effects of isospin symmetry violation from the $(N\Delta)_{\text{dpn}}$ correlation in Fig.~\ref{fig:npid1} is rather complicated, the JAM result seems to imply that the net effect of these correlations is to slightly reduce the $(\Delta^0\Delta^+)/(\Delta^-\Delta^{++})$ ratio.  In other codes that use the geometrical method for treating particle collisions, the $(N\pi)_{\text{nnd}}$ correlation (Fig.~\ref{fig:npid2}) can also induce the violation of isospin symmetry, which is expected to increase the $(\Delta^0\Delta^+)/(\Delta^-\Delta^{++})$ ratio (see Sec.~\ref{sec:npid}).  Compared to the case without pions shown in Fig.~\ref{fig:ratioa-De2P0}, the results shown in Fig.~\ref{fig:ratioa} clearly indicate that the effect of the $(N\pi)_{\text{nnd}}$ correlation is strong so that the excess of $\Delta^0$ and $\Delta^+$ is often serious.

\begin{figure*}
  \includegraphics[width=\textwidth]{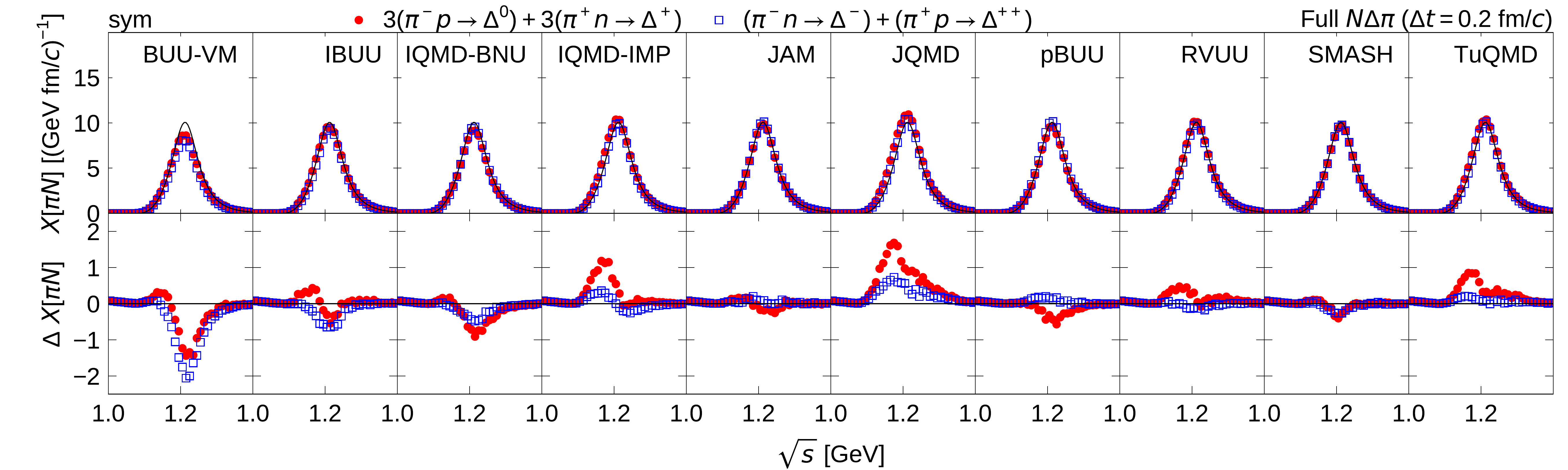}
  \caption{\label{fig:npid_dndeics}
    Upper panels: Distributions of the $\pi N\to\Delta$ rates per active pion in $\sqrt{s}$, namely $X[\pi N]=(dR[\pi N\to\Delta]/d\sqrt{s})/N_{\pi}^{\text{active}}$, defined for different isospin channels, in the symmetric ($\delta=0$) full-$N\Delta\pi$ system with the time-step size $\Delta t=0.2$ fm/$c$.  Both the rates and pion numbers are averaged over $90<t<150$ fm/$c$.  The combination $3X[\pi^-p]+3X[\pi^+n]$ is shown with red filled circles and $X[\pi^-n]+X[\pi^+p]$ with blue open squares.  The distribution for the ideal gas mixture is shown with the thin curve.  Lower panels: The deviations from the ideal gas mixture, shown in a magnified scale.
  }
\end{figure*}

The above interpretation of the isospin symmetry violation caused by the $(N\pi)_{\text{nnd}}$ correlation (Fig.~\ref{fig:npid2}) is further supported by considering some combinations of the $N\pi\to\Delta$ rates per existing pion, for different isospin channels shown in Fig.~\ref{fig:npid_dndeics},
\begin{equation}
  \label{eq:pin-per-pi}
  X[\pi N] = \frac{1}{N_{\pi}^{\text{active}}}
  \frac{dR[N\pi\to\Delta]}{d\sqrt{s}},
\end{equation}
for $\pi\in\{\pi^-, \pi^0, \pi^+\}$ and $N\in\{n,p\}$.  As illustrated in Fig.~\ref{fig:seqsim}, the number of existing pions in many codes gradually decreases during the progress of computational steps in the collision procedure $C_k$, and therefore the number of such existing pions does not necessarily agree with the number $N_\pi$ at the time-step boundaries.  Furthermore, some pions are stealth in IBUU and pBUU.  Therefore, the average number of pions that are actually participating in collisions may be better given by
\begin{equation}
  N_\pi^{\text{active}}=N_\pi\pm
  \frac12 \sum_{N'\in\{n,p\}}R[N'\pi\to\Delta]\Delta t,
\end{equation}
where the positive sign is for IQMD-IMP and JQMD, and negative for other codes, depending on the treatment of particle collisions and decays in individual codes.  Note that $\Delta t=0$ for JAM and SMASH.  For BUU-VM and pBUU, $N_\pi^{\text{active}}=N_\pi$ is assumed, because either the results from two methods are mixed (BUU-VM) or because `superactive' particles are introduced to compensate for the stealth particles (pBUU).  This number of active pions, $N_\pi^{\text{active}}$, is used in Eq.~\eqref{eq:pin-per-pi} to calculate the rates per existing pion.  The rates $X[\pi N]$ of different reaction channels are related by the isospin Clebsh-Gordan factors, e.g., we would expect $X[\pi^-p]=\frac13X[\pi^-n]=X[\pi^+n]=\frac13X[\pi^+p]$ in the symmetric ($\delta=0$) system.  In Fig.~\ref{fig:npid_dndeics}, $3X[\pi^-p]+3X[\pi^+n]$ is shown by red filled circles and $X[\pi^-n]+X[\pi^+p]$ is shown by blue open squares.  The agreement with the distribution in the ideal Boltzmann-gas mixture (thin curve) justifies our understanding on the collision prescriptions used in these codes.  With a closer look in the lower panels, which show the deviations from the ideal-gas distribution in a magnified scale, we notice that red filled circles and blue open squares agree well only in SMASH.  In many codes (BUU-VM, IBUU, IQMD-IMP, JQMD, RVUU, TuQMD), $X[\pi^-p]$ and $X[\pi^+n]$ are relatively too large compared to $X[\pi^-n]$ and $X[\pi^+p]$, which is consistent with the discussions of the $(N\pi)_{\text{nnd}}$ correlation in Sec.~\ref{sec:npid}, in that there is an excess of $\Delta^{0,+}$ as exactly seen in Fig.~\ref{fig:ratioa}.  In JAM and pBUU, the effect is opposite so that $X[\pi^-p]$ and $X[\pi^+n]$ are relatively too small.  In the case of JAM, this can be qualitatively explained by the higher-order $(N\pi)_{\text{dx}}$ correlation arising as in Fig.~\ref{fig:npidx} which is expected to enhance the $\pi^-n$ and $\pi^+p$ collisions relative to the other pion absorption channels.  However, its effect on the JAM result does not seem very strong, as also discussed in Sec.~\ref{sec:npid}.  In IQMD-BNU, the relatively good agreement of the red filled circles and blue open squares indicates a weak $(N\pi)_{\text{nnd}}$ correlation, which could be because in this code the $\Delta$ particle and the pion are treated as stealth for a time step after their creations (see Fig.~\ref{fig:seqsim}(f) and Fig.~\ref{fig:npid2}).

In codes with time steps, particles are usually propagated during the series of reactions shown in Fig.~\ref{fig:npid2}.  In the methods for panels (a), (c), (f) and (g) of Fig.~\ref{fig:seqsim}, the pion is always propagated after $\Delta_j\to N_j'\pi_k$, which may weaken the correlation significantly, in particular for large $\Delta t$, since the pion tends to move fast.  In another method represented in panel (b) of Fig.~\ref{fig:seqsim}, the $\Delta$ particle is always propagated after $N_iN_j\to N_i'\Delta_j$, but it does not move as much as a pion would, and thus a strong correlation may remain during any subsequent reactions of $\Delta_j\to N_j'\pi_k$ and $N_i'\pi_k\to\Delta_i'$, both of which can take place in the same time step.  In the method of panel (e), the three processes of $N_iN_j\to N_i'\Delta_j$, $\Delta_j\to N_j'\pi_k$ and $N_i'\pi_k\to\Delta_i'$ can take place in the same time step without propagation of any particle.  This is likely the reason for the larger difference between red filled circles and blue open squares in Fig.~\ref{fig:npid_dndeics} in IQMD-IMP and JQMD than in other codes.  For IBUU, RVUU and TuQMD, the difference is smaller for a larger $\Delta t=0.5$ fm/$c$ (not shown) probably because of the weakening of correlations.  This example shows that problems related to correlations and the collision--decay sequence can affect the results in a complicated way.

\subsection{Summary of results from the equilibrated $N\Delta\pi$ system\label{sec:ndeltapi-summary}}

\begin{figure*}
\includegraphics[width=\textwidth]{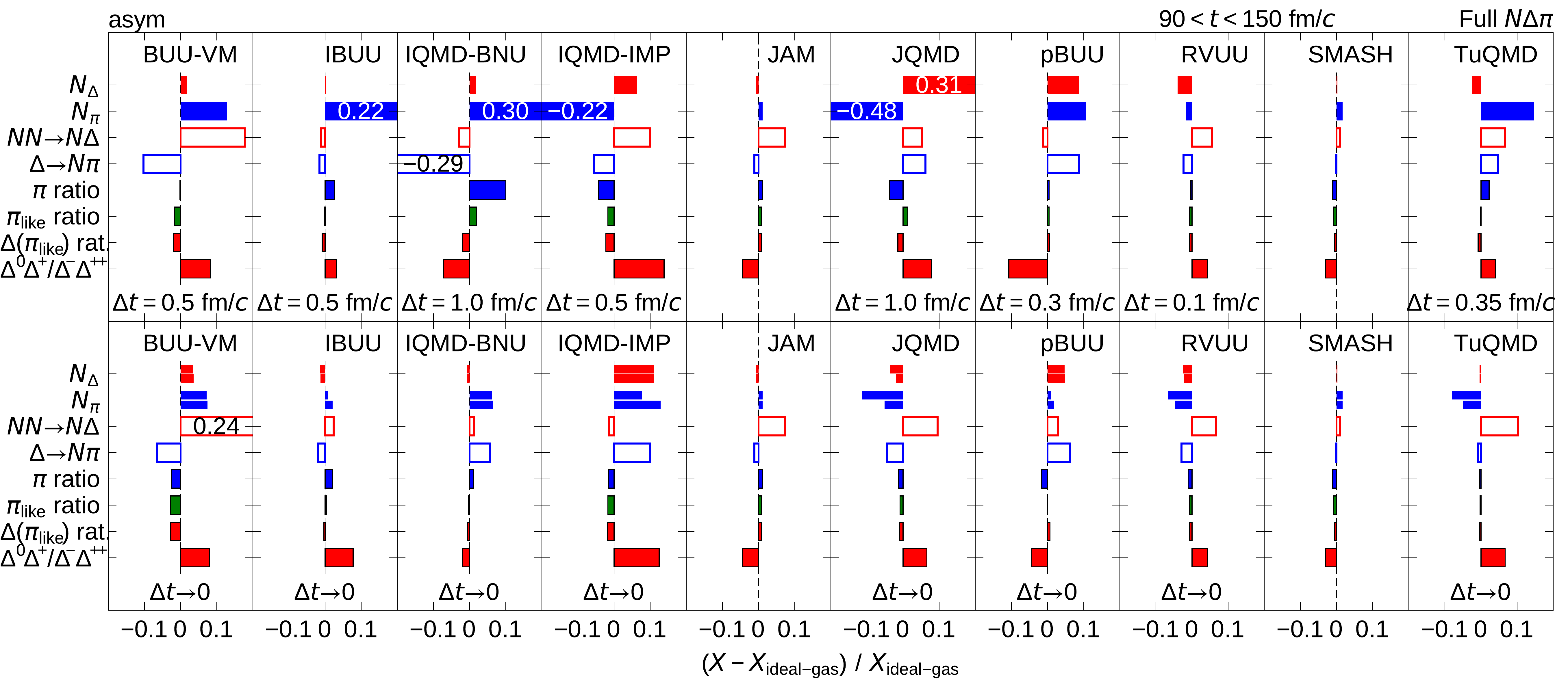}
\caption{\label{fig:dtsummary} Relative deviation from the value in the ideal Boltzmann-gas mixture for various quantities such as the numbers of particles, the collision and decay rates and the isotopic ratios, averaged over $90<t<150$ fm/$c$ in the asymmetric system ($\delta=0.2$).  The deviations of the three pion ratios have been magnified by a factor 5.  The upper panel for each code shows the case of a time-step size $\Delta t$ which was used in calculations published in the past (see Sec.~\ref{sec:codespec}).  The lower panel shows the limiting case of $\Delta t\to 0$.  These results have been obtained by linear interpolation or extrapolation of the homework results for $\Delta t=0.2$ fm/$c$ and $\Delta t$ chosen by the code, except for JAM and SMASH which do not rely on time step.  For each of the particle numbers $X=N_\Delta$ and $N_\pi$ in the lower panel, two bars indicate the extrapolated values obtained assuming a linear $\Delta t$ dependence of $X$ (upper bar), and a linear dependence of $1/X$ (lower bar).
}
\end{figure*}

The performance of codes in the box comparison for the equilibrated system is summarized in Fig.~\ref{fig:dtsummary}, which shows the relative deviations of various quantities from the values in the ideal Boltzmann-gas mixture.  The upper panel displays the case of the time-step size $\Delta t$ which was used in calculations published in the past (see Sec.~\ref{sec:codespec}), while the lower panel shows the limiting case of $\Delta t\to 0$.  The homework results for $\Delta t=0.2$ fm/$c$ and those for $\Delta t$ chosen by the code were linearly interpolated or extrapolated to obtain these cases.

\subsubsection{General summary}

Accurate calculation of the numbers of $\Delta$ and $\pi$ in a nuclear system is difficult for codes relying on time steps, if a usual time-step size, such as $\Delta t=0.5$ fm/$c$, is chosen.  This issue is due to the method used in handling the collision--decay sequence within each time step.  The problem can be improved either by taking the limit of $\Delta t\to0$ (see the lower panels of Fig.~\ref{fig:dtsummary} with a caution about the linear extrapolation\footnote{The linear extrapolation to $\Delta t\to 0$ in Fig.~\ref{fig:dtsummary} is somewhat ambiguous particularly for the number of pions ($N_\pi$), as indicated in the lower panel by the two blue bars for $N_\pi$, which show two cases of extrapolations assuming that $N_\pi$ is linear in $\Delta t$ (upper bar) and that $1/N_\pi$ is linear in $\Delta t$ (lower bar).}), by introducing a better method to handle the sequence, or by employing a time-step--free method.  It is understandable for many codes, even at $\Delta t\to 0$, that collision rates such as the $NN\to N\Delta$ rate are affected by correlations which are inevitably induced by the geometrical treatment of particle collisions.  Also, the correlations can affect the isotopic ratios, particularly the unphysical excess of $\Delta^0$ and $\Delta^+$ relative to $\Delta^-$ and $\Delta^{++}$ seen in many codes (see $\Delta^0\Delta^+/\Delta^-\Delta^{++}$ in Fig.~\ref{fig:dtsummary}), which can cause a systematic underestimation of the charged pion ratio.  The problem cannot be removed by choosing a small $\Delta t$.  It also does not seem easy to predict the number of pions with an accuracy better than 5\%, because it is probably affected by the details of the correlations in the pion absorption process.  However, these problems may be improved by carefully choosing a prescription to avoid repeated spurious collisions in the geometrical method of treating particle collisions.  On the other hand, the full-ensemble BUU codes are practically free from these effects due to correlations.
 
Even with the above-mentioned problems, we may be in a fortunate situation that transport codes can rather precisely predict the $\pi$-like ratio, which is defined in Eq.~(\ref{eq:piratios}c) and is most directly related to the $\pi^-/\pi^+$ observable in heavy-ion collisions.  There is a good reason why the issue of the collision--decay sequence does not affect much the $\pi$-like ratio in equilibrated systems.  Because of the complexity of correlations, particularly those caused by pions, the effect may appear differently in cases other than the equilibrated one studied here.  It is therefore of interest to study this effect in the early non-equilibrium stage of the present box comparison (in Sec.~\ref{sec:noneq}) and in real heavy-ion collisions.  It should be kept in mind that the problem in the numbers of propagated particles due to the issue of the collision--decay sequence affects the propagation under the mean field, which can then affect indirectly the $\pi$-like ratio in heavy-ion collisions.

\subsubsection{Diagnostic results for individual codes}

For the BUU-VM code, the problem of a too high $NN$ elastic collision rate has already been seen in the previous comparison of Ref.~\cite{yxzhang2018}, and a similar problem is seen here again as too high $NN\leftrightarrow N\Delta$ rates.  Also the decay rate per existing $\Delta$ is too low.  These basic problems in this code are still under investigation.

The two time-step--free codes (JAM and SMASH) generally reproduce well the expected values for the ideal gas mixture.  In particular, SMASH is almost free from effects of correlations because it uses the full ensemble method.  The only problem found here is the violation of the detailed balance due to a cutoff imposed on cross sections, which is probably not a problem in practical heavy-ion collisions.  On the other hand, JAM suffers from the effects of correlations due to the geometrical method for particle collisions, when the code is run with one test particle per physical particle.  However, the code seems to have been constructed carefully so as to go around serious influences of correlations, compared to other QMD codes participating in the present comparison.

Two of the QMD codes (JQMD and TuQMD), which predicted largely different values with finite time-step sizes $\Delta t$, have converged to give consistent results in the limit of $\Delta t\to 0$, taking into account the ambiguity in extrapolation.  The deviations from the ideal-gas values in the pion number, in the $NN\leftrightarrow N\Delta$ rate and in the $(\Delta^0\Delta^+)/(\Delta^-\Delta^{++})$ ratio are most likely explained as originating from the correlations induced by the geometrical method.  The RVUU code, which was run in the parallel ensemble mode, gives results at $\Delta t\to 0$ similar to those given by JQMD and TuQMD, though the effects of correlations seem a little weaker in the $NN\leftrightarrow N\Delta$ rate and the $(\Delta^0\Delta^+)/(\Delta^-\Delta^{++})$ ratio.  In these three codes, the suppression of $N_\pi$ (compared to the case of JAM) may also be understood as an effect of the $(N\pi)_{\text{nnd}}$ correlation, which is absent in JAM.  The slightly suppressed $N_\Delta$ in JQMD and RVUU, compared to the case of JAM and TuQMD, may also be explained e.g.\ from the different strengths of the $(N\Delta)_{\text{dpn}}$ correlation of Fig.~\ref{fig:npid1}, which can depend on how the position of the pion is determined after $\Delta\to N\pi$ (see Sec.~\ref{sec:codespec} for details).  The $\Delta\to N\pi$ rate is naturally correlated with $N_\Delta$.  Thus, for these four codes, after eliminating the effects of finite $\Delta t$, we have reached a rather complete understanding of the results affected by correlations.  Of course, the effects of correlations in these and other codes have to be carefully monitored in other cases of applications.

Results from the IBUU code at $\Delta t\to 0$ could be similar, in principle, to those from JQMD, RVUU and TuQMD, because this code uses the parallel ensemble method.  However, the $NN\leftrightarrow N\Delta$ rate is similar to the ideal-gas value, and therefore is smaller than expected for the codes suffering from correlations.  An $NN$ collision rate smaller than expected was also seen for this code in the previous comparison of Ref.~\cite{yxzhang2018}.  An advantage of this code is that the dependence on the time step $\Delta t$ is weak for many quantities, except for the number of pions.

The results from the IQMD-BNU code, at $\Delta t\to 0$, are surprising in that they are quite similar to the ideal-gas values, even though one would expect effects of correlations for this QMD code.  It would be very interesting if we could know how this code goes around the effects of correlations.  The deviations in the $\Delta$ decay rate and the pion number are fully explained as originating from the omission of the time dilation effect in the decay of $\Delta$.  It should be straightforward to improve the code to consider time dilation.

IQMD-IMP is the only QMD code, participating in this comparison, that tests collisions for the particle pairs taken in a fixed order from a particle list which is initialized asymmetrically, e.g.~by listing protons first and neutrons later.  This can cause unphysical asymmetries, such as in the present comparison, so this choice of the code should be changed.  However, this effect of asymmetry should disappear at $\Delta t\to 0$.  When $\Delta t$ is reduced, there appears another problem that the numbers of both $\Delta$ particles and pions increase beyond the values in the ideal gas mixture.  We also see that the $N\Delta\to NN$ rate decreases (in spite of the increased number of $\Delta$) and the violation of the detailed balance in $NN\leftrightarrow N\Delta$ becomes more serious, when $\Delta t$ is reduced.  These behaviors, seen only in this code, have to be understood.

The pBUU code uses volume cells to process collisions, in contrast to the other codes participating in the present comparison.  An advantage of this method is that it is free from the issue of correlations.  Numerical errors originating from the finite number of test particles as well as from the finite volume-cell and time-step sizes have to be controlled well.  The pBUU code processes particles according to a list ordered according to particle type, similarly to IQMD-IMP, but does so with the order randomly changed from lowest towards highest isospin or reverse.  Still leaving intermediate isospins always in the middle seems to impact ratios of species with intermediate to extreme isospin as evidenced in Fig.~\ref{fig:npid_dndeics} for finite $\Delta t$.  Still, the present results indicate that numerical errors are controlled within an acceptable range when a typical time-step size of $\Delta t=0.2$--$0.3$ is chosen.

\section{Comparisons at early times\label{sec:noneq}}

In this section, we focus our study on the non-equilibrium effects at early times of a few tens of fm/$c$.  In our homework, the system at $t=0$ is composed of only nucleons and is evolved for the first 10 fm/$c$ only with $NN$ elastic collisions, after which the $\Delta$ and pion production sets in.  As Fig.~\ref{fig:nat1a} indicates, it takes a few tens of fm/$c$ before the system reaches equilibrium.  In the case of heavy-ion collisions at energies for which pions get produced, the violent phase of the reaction ends within a few tens of fm/$c$. To evaluate the accuracy of transport codes, it is therefore important to also carry out the box comparison at early times.

As we have briefly seen in Sec.~\ref{sec:digest}, the uncertainties in transport-code results may be larger in non-equilibrium than in equilibrated systems.  Although a complete identification of the sources of problems is beyond the scope of this paper, we try here to figure out as much as possible where any uncertainty comes from.  The information here is also useful for improving individual codes.

In the following comparisons, we use the solution of the rate equation as a reference.  The rate equation assumes thermal momentum distributions at any instant with a common temperature for all the particle species (Appendix \ref{sec:rateeq}).  Therefore, the solution is not necessarily completely consistent with the exact solution of the Boltzmann equation \eqref{eq:boltz}, because of non-thermal or non-equilibrium effects.  With this in mind, we still use this reference to compare the results of different transport codes and to clarify the non-equilibrium effects in these results.

\subsection{$N\Delta$ system\label{sec:noneq-ndelta}}

\begin{figure*}
  \includegraphics[width=\textwidth]{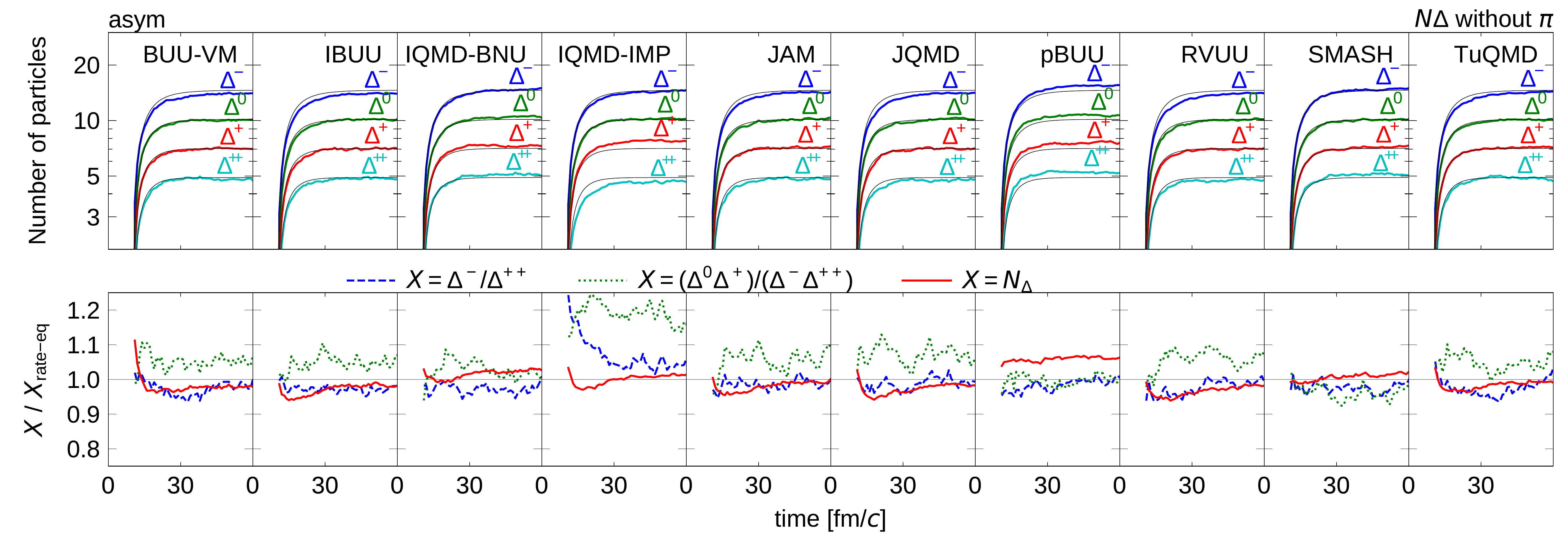}
  \caption{\label{fig:nata-De2P0} Upper panels: Time evolution of the number of $\Delta$ in the asymmetric ($\delta=0.2$) system, at $t<60$ fm/$c$, in the case without pions but with $NN\to N\Delta$ and $N\Delta\to NN$ processes included.  Lower panels: For the quantities $X=\Delta^-/\Delta^{++}$ (blue dashed line), $X=(\Delta^0\Delta^+)/(\Delta^-\Delta^{++})$ (green dotted line) and $X=N_\Delta$ (red solid line), their ratios to the rate-equation solutions, $X/X_{\text{rate-eq}}$, are shown as functions of time.}
\end{figure*}

In this subsection, we make comparisons for the case without pions, i.e., with the decay of $\Delta$ particles turned off even though the spectral function of $\Delta$ has a width.  In the upper panels of Fig.~\ref{fig:nata-De2P0}, the time evolutions of the numbers of the four species of $\Delta$ are represented with the colored thick lines at early times $t<60$ fm/$c$, for the asymmetric ($\delta=0.2$) system.  These transport-code results may be compared with the solutions of the rate equation which are shown by thin black lines.  At $t=60$ fm/$c$, the numbers of $\Delta$ have already reached equilibrium almost completely.  The equilibrated values have been discussed in detail in Sec.~\ref{sec:ndelta}.  We pay attention here to the behaviors at early times, before equilibrium is reached.

For a closer comparison of the time evolution of the number of $\Delta$ ($N_\Delta$), the red solid line in the lower panels of Fig.~\ref{fig:nata-De2P0} shows the ratio of $N_\Delta$ calculated by transport codes to the same quantity in the rate-equation solution, namely $X/X_{\text{rate-eq}}$ with $X=N_\Delta$.  Without any exception, the results for this ratio are slightly increasing in this time span except at the very beginning.  This means that the increase of $N_\Delta$ toward equilibrium in transport codes is slightly slower than in the rate equation.  In QMD and parallel-ensemble BUU codes, this is likely due to the extra $N\Delta\to NN$ collisions by the higher-order $(N\Delta)_{\text{nnx}}$ correlation (Fig.~\ref{fig:nnxd} in Sec.~\ref{sec:hocorr}), which has the effect of counter-acting the $NN\to N\Delta$ reaction.  The effect is weak in full-ensemble BUU codes (pBUU and SMASH), i.e.\ the slope is small.  For the results of many codes, the agreement with the rate equation is good at the beginning right after inelastic collisions set in, which is reasonable because the higher-order correlations have not been established yet because they need additional scatterings by other particles.

\begin{figure*}
  \includegraphics[width=\textwidth]{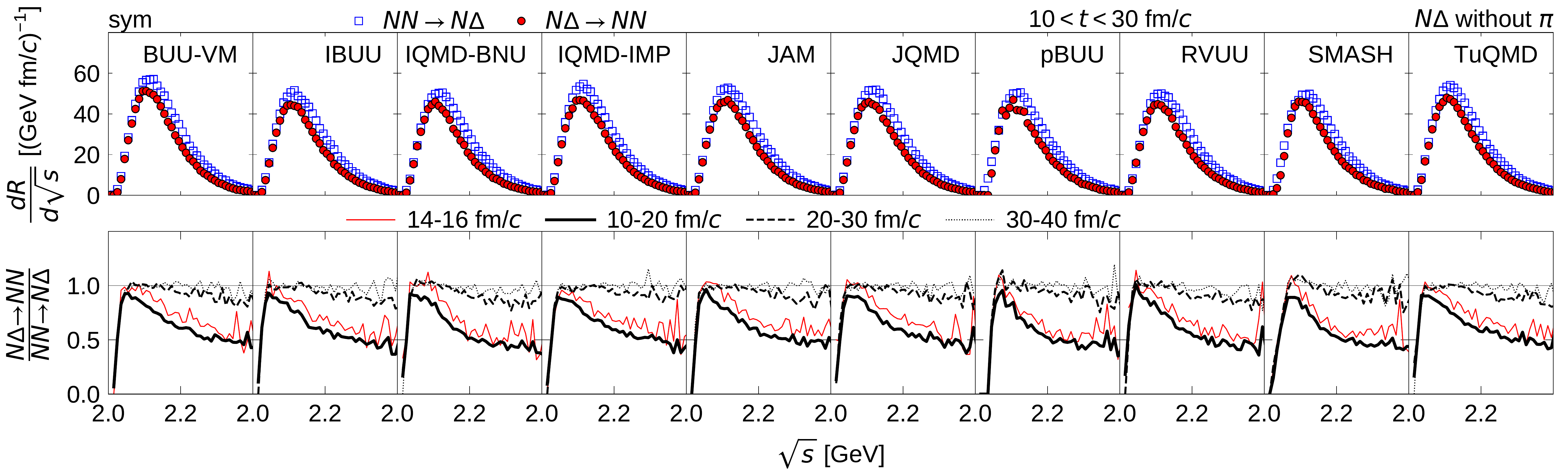}
  \caption{\label{fig:nnnd_early_edists-De2P0} Upper panels: Distributions of the $NN\to N\Delta$ (open squres) and $N\Delta\to NN$ (filled circles) reaction rates in $\sqrt{s}$, in the symmetric ($\delta=0$) system without pions.  The reaction rates are averaged over $10<t<30$ fm/$c$.  Lower panels: The $\sqrt{s}$ dependence of the backward-to-forward ratio $(N\Delta\to NN)/(NN\to N\Delta)$ for different indicated time spans.}
\end{figure*}

Distributions of the $NN\leftrightarrow N\Delta$ rates in $\sqrt{s}$, averaged over $10<t<30$ fm/$c$, are shown in the upper panels of Fig.~\ref{fig:nnnd_early_edists-De2P0}.  As expected, the overall integrated rates of the forward and backward reactions do not agree with each other in this early time span because the numbers of particles have not yet equilibrated and more $\Delta$ particles need to be produced than absorbed.  Furthermore, we notice that the shapes of these distributions are different, which suggests that nucleons and $\Delta$ particles do not follow the thermal momentum distributions at a common temperature.  This point is clearer in the lower panels, which show the backward-to-forward ratio of the reaction rates, for different time spans represented with different line styles.  For an early time span of $10<t<20$ fm/$c$ (thick solid line), the backward rate is much lower than the forward rate at high energies, while they are almost balanced at low values of $\sqrt{s}$.\footnote{
During the finite time span from $t=10$ to 20 fm/$c$, the temperature drops from $T=60$ to 56.4 MeV in the rate equation for this system.  Since the forward rate decreases and the backward rate rapidly increases as the time progresses, the backward-to-forward ratio for this time span can depend on $\sqrt{s}$ in principle even for the rate-equation solution.  However, the red thin line in the lower panels of Fig.~\ref{fig:nnnd_early_edists-De2P0} shows the ratio for a narrower time span from $t=14$ to 16 fm/$c$, which clearly indicates that the $\sqrt{s}$ dependence at the earliest times is not an artifact of the finite time span.
}  As the time progresses, the forward and backward rates balance gradually as shown by the dashed line ($20<t<30$ fm/$c$) and the dotted line ($30<t<40$ fm/$c$).  As we have already seen in Fig.~\ref{fig:nnnd_edists-De2P0}, the detailed balance is achieved eventually at late times.  The low backward rate in the lowest energy bin may be due to the truncation of the $N\Delta\to NN$ cross section on account of the finite size of the box in the present homework.  The characteristic $\sqrt{s}$ dependence in the lower panels of Fig.~\ref{fig:nnnd_early_edists-De2P0} is observed very consistently in almost all codes.  Therefore we believe that it indicates a real non-equilibrium effect, which is described by transport codes, but not by the rate equation.

Some isotopic ratios of $\Delta$ are shown in the lower panels of Fig.~\ref{fig:nata-De2P0}.  The excess of $\Delta^0$ and $\Delta^+$ relative to $\Delta^-$ and $\Delta^{++}$ is represented by the green dotted line as the $X/X_{\text{rate-eq}}$ ratio with $X=(\Delta^0\Delta^+)/(\Delta^-\Delta^{++})$.  For this quantity, $X_{\text{rate-eq}}$ is very close to 1, and therefore the line practically shows the current value of $X$.  As already discussed for late times in Sec.~\ref{sec:ndelta}, the excess of $\Delta^0$ and $\Delta^+$ can be interpreted as due to the higher-order $(N\Delta)_{\text{nnx}}$ correlation (Fig.~\ref{fig:nnxd}) in QMD and parallel-ensemble BUU codes.  Compared to the equilibrium case in Fig.~\ref{fig:ratioa-De2P0}, the amount of the excess seems large and the effect is clearly seen at early times, even with statistical fluctuations, in almost all the QMD and parallel-ensemble BUU codes.  The strong effect at early times may be related to the non-equilibrium effect discussed in the context of Fig.~\ref{fig:nnnd_early_edists-De2P0}.  The deficiency of high-energy $N\Delta\to NN$ collisions increases the relative importance of low-energy $N\Delta$ collisions in which the higher-order correlation is supposed to be strong such as shown by the condition of Eq.~\eqref{eq:hocorr-cond}.

The blue dashed line in the lower panels of Fig.~\ref{fig:nata-De2P0} shows the $X/X_{\text{rate-eq}}$ ratio of $X=\Delta^-/\Delta^{++}$.  In all codes except for IQMD-IMP, this ratio is lower than 1, which means that the $\Delta^-/\Delta^{++}$ ratio in transport codes is lower than that in the rate-equation solution.  This is possible because of the asymmetric strength of the effect of the $(N\Delta)_{\text{nnx}}$ correlation in asymmetric environments, as discussed in Sec.~\ref{sec:nnxd}.  The effect is within a few percent in most cases, but it can directly affect the isotopic ratios such as the $\pi^-/\pi^+$ ratio, when $\Delta$ resonances are allowed to decay.  It does not contradict the isospin symmetry, though it is absent in the naive rate equation.

\subsection{$N\Delta\pi$ system\label{sec:noneq-ndeltapi}}

\begin{figure*}
  \includegraphics[width=\textwidth]{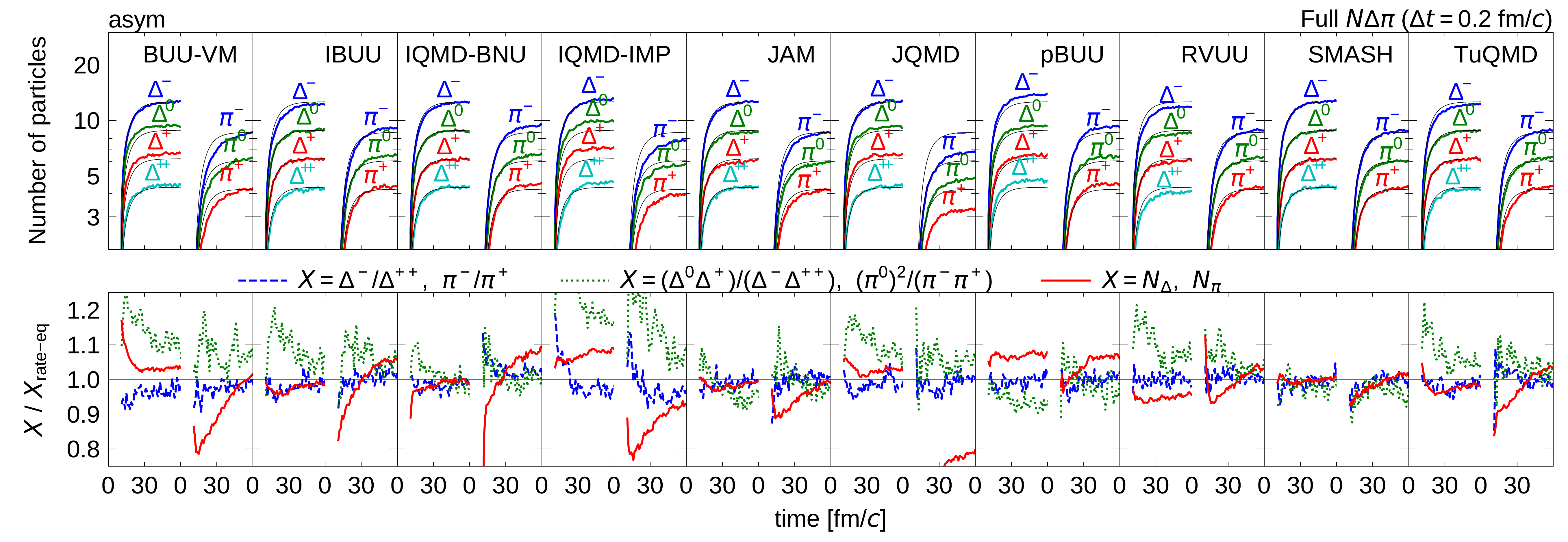}
  \caption{\label{fig:nata} Upper panels: Time evolution of the numbers of $\Delta$ and $\pi$ in the asymmetric ($\delta=0.2$) full-$N\Delta\pi$ system at $t<60$ fm/$c$.  The solution of the rate equation is represented with thin curves.  Lower panels: For the quantities $X=\Delta^-/\Delta^{++}$ or $\pi^-/\pi^+$ (blue dashed line), $X=(\Delta^0\Delta^+)/(\Delta^-\Delta^{++})$ or $(\pi^0)^2/(\pi^-\pi^+)$ (green dotted line) and $X=N_\Delta$ or $N_\pi$ (red solid line), their ratios to the rate-equation solutions, $X/X_{\text{rate-eq}}$, are shown as functions of time.  Left and right sides of each panel represent $\Delta$ and $\pi$, respectively.}
\end{figure*}

When the pions are introduced with the $\Delta\leftrightarrow N\pi$ processes, we have already seen that some problems appear prominently in the comparisons presented in Sec.~\ref{sec:ndeltapi}, for the equilibrated numbers of particles and their isotopic ratios.  We here continue the comparisons by focusing on early times before the system equilibrates.  As we have learned in Sec.~\ref{sec:ndeltapi}, one of the sources of the problems is the issue of the collision--decay sequence within one time step.  Correspondingly we mainly show here the results with a small time step size $\Delta t=0.2$ fm/$c$ in order to minimize the impact of this issue.  We should still expect that some effects of a finite size of $\Delta t=0.2$ fm/$c$ may remain, and that the problems due to higher-order correlations cannot be reduced by choosing a small $\Delta t$.

The upper panels of Fig.~\ref{fig:nata} display the time evolution of the numbers of the isospin species of $\Delta$ and $\pi$ in terms of colored thick lines.  The numbers of $\Delta$ particles ($N_\Delta$) and those of pions ($N_\pi$) are shown side by side, in the same way as in Fig.~\ref{fig:nat1a}, but now for the early times $0<t<60$ fm/$c$ and for the case with $\Delta t=0.2$ fm/$c$.  At $t=60$ fm/$c$, the numbers $N_\Delta$ and $N_\pi$ have almost equilibrated in most codes.  As we already know, the deviations of these equilibrium values from the solution of the rate equation, shown by thin lines in the figure, are mainly due to the issue of the collision--decay sequence.  At early times in the upper panels of Fig.~\ref{fig:nata}, the increase of $N_\pi$ toward equilibrium is slower than in the rate-equation solution in many codes.  This is more clearly seen in the lower panels, where, for some quantities $X$, the line shows the ratio of the value in the transport-code result to that in the rate equation.  The red solid lines represent the $X/X_{\text{rate-eq}}$ ratios for $X=N_\Delta$ (left) and $X=N_\pi$ (right).  The strongly increasing $X/X_{\text{rate-eq}}$ for $X=N_\pi$, as a function of time, indicates a suppressed increase of $N_\pi$ toward equilibrium, which is observed in all the QMD and parallel-ensemble BUU codes.  This could be expected at least partly due to the existence of $N\pi$ correlations which enhance the $N\pi\to\Delta$ process such as in Figs.~\ref{fig:npidx} and \ref{fig:npid2}.  However, there may be other reasons because the suppression of the rise in $N_\pi$ is also observed in full-ensemble BUU codes (pBUU and SMASH) though it is weaker than in other codes.  The behaviors of the $X/X_{\text{rate-eq}}$ ratio for $X=N_\Delta$ are similar to those in the case without pions (Fig.~\ref{fig:nata-De2P0}).

\begin{figure*}
  \includegraphics[width=\textwidth]{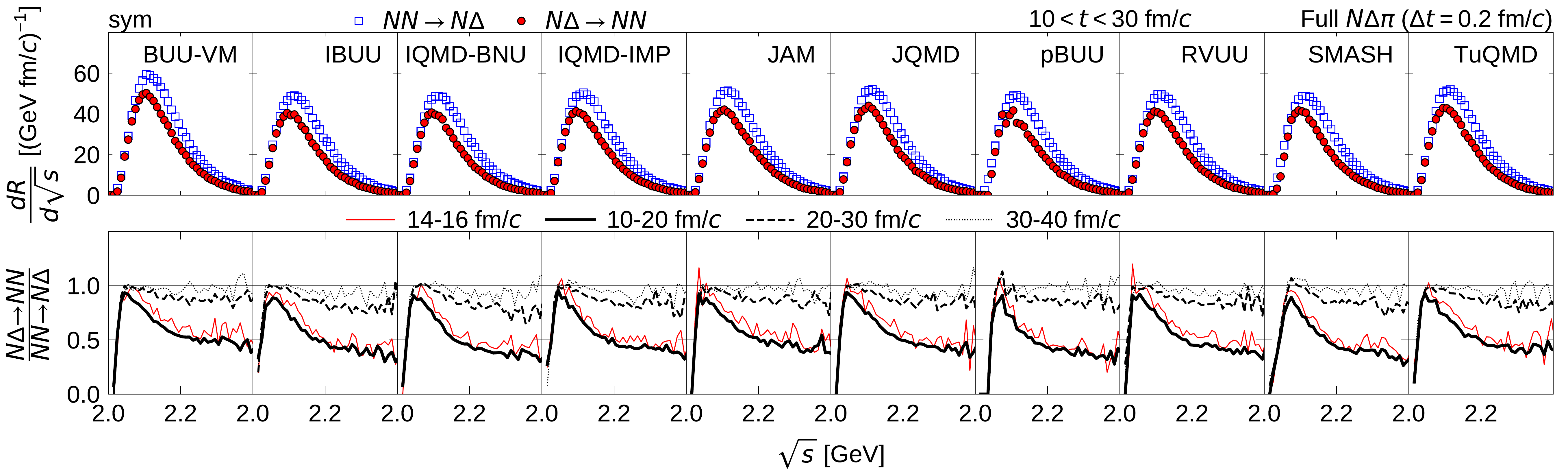}
  \caption{\label{fig:nnnd_early_edists} Upper panels: Distributions of the $NN\to N\Delta$ (open squres) and $N\Delta\to NN$ (filled circles) reaction rates in $\sqrt{s}$, in the symmetric ($\delta=0$) full-$N\Delta\pi$ system.  The reaction rates are averaged over $10<t<30$ fm/$c$.  Lower panels: The $\sqrt{s}$ dependence of the backward-to-forward ratio $(N\Delta\to NN)/(NN\to N\Delta)$ calculated over different indicated time spans.}
\end{figure*}

Figure \ref{fig:nnnd_early_edists} shows the $\sqrt{s}$ distributions of the $NN\to N\Delta$ and $N\Delta\to NN$ processes in the upper panels, and the backward-to-forward ratio in the lower panels.  In the same way as in Fig.~\ref{fig:nnnd_early_edists-De2P0} for the case without pions, these results from all the codes consistently indicate that the momentum distribution of $\Delta$ has not been thermalized yet at these early times.  The results in Fig.~\ref{fig:nnnd_early_edists} are quite similar to those in Fig.~\ref{fig:nnnd_early_edists-De2P0}, though we may see in the lower panels that establishing the detailed balance might be slightly slower when pions are introduced, judging from comparison of the two figures from early ($10<t<20$ fm/$c$; thick solid lines), through intermediate ($20 < t < 30$ fm/$c$; dashed lines), to later times ($30<t<40$ fm/$c$; dotted lines).

\begin{figure*}
  \includegraphics[width=\textwidth]{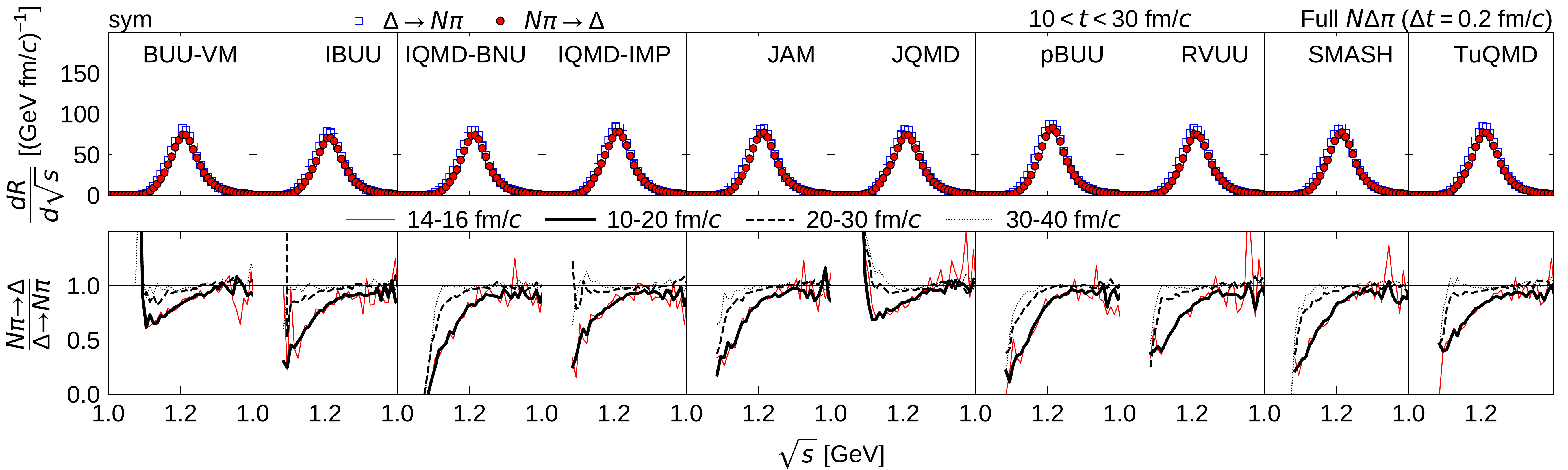}
  \caption{\label{fig:dnpi_early_edists} Distributions of the $\Delta\to N\pi$ (open squres) and $N\pi\to\Delta$ (filled circles) reaction rates in $\sqrt{s}$, in the symmetric ($\delta=0$) full-$N\Delta\pi$ system.  The reaction rates are averaged over the time span of $10<t<30$ fm/$c$.  Lower panels: The $\sqrt{s}$ dependence of the backward-to-forward ratio $(\Delta\to N\pi)/(N\pi\to\Delta)$ calculated for different indicated time spans.}
\end{figure*}

The same analysis is done in Fig.~\ref{fig:dnpi_early_edists} for the $\Delta\to N\pi$ and $N\pi\to\Delta$ processes.  The evidence of a non-equilibrium effect is clearly seen again, in particular in the lower panels.  In contrast to the case for $NN\leftrightarrow N\Delta$, the rates quickly balance within the high-energy part in all the codes.  The low-energy part takes a relatively long time of the order of a few tens of fm/$c$, before the detailed balance is established.  All the codes show the same qualitative feature and therefore we believe that this non-equilibrium effect is also a physical one described by transport codes, but not by the rate equation.  However, the code dependence of the effect is stronger for $\Delta\leftrightarrow N\pi$ than for $NN\leftrightarrow N\Delta$.  In particular, the behaviors of BUU-VM and JQMD are significantly different from the other codes.  The observed effect implies that the high-momentum part of the pion momentum distribution is enhanced compared to the low-momentum part at these early times.  Since high-momentum pions can be strongly absorbed because of the high relative velocities with nucleons, the effect can enhance the pion absorption rate, which is consistent with the slow increase of the number $N_\pi$ observed in Fig.~\ref{fig:nata}, even in parallel-ensemble BUU codes.  The time scale of the non-equilibrium effect is also similar to that of the suppression of the increase in $N_\pi$.  This non-equilibrium effect is closely related to the mass dependence of the decay width $\Gamma(m)$.  In fact, in similar calculations with a constant $\Gamma(m)=115$ MeV, no significant $\sqrt{s}$ dependence is observed in the backward-to-forward ratio for $\Delta\leftrightarrow N\pi$.

\begin{figure*}
  \includegraphics[width=\textwidth]{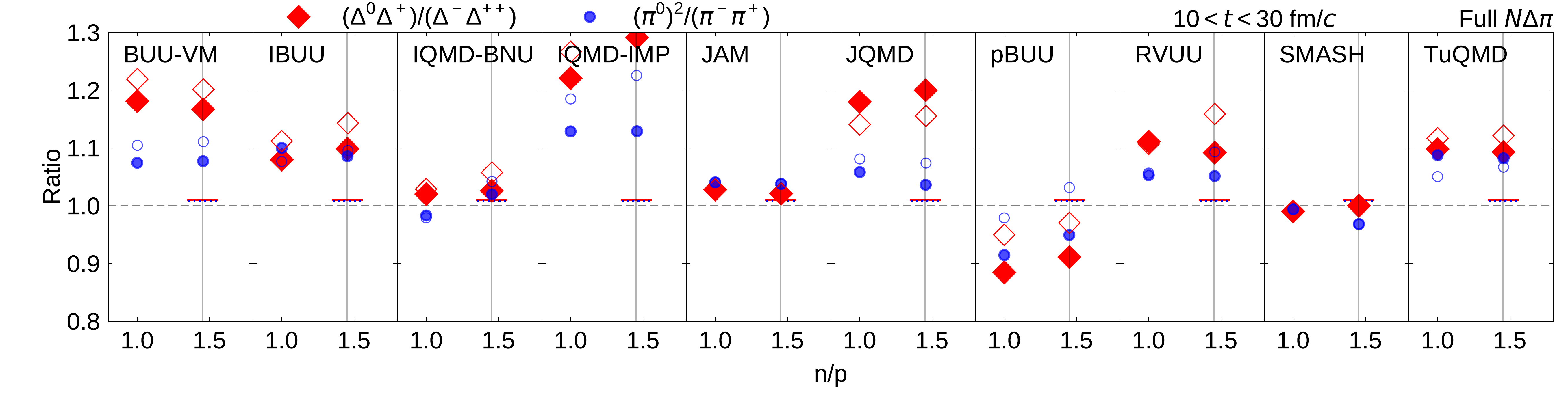}
  \caption{\label{fig:ratioa_early} Ratios $(\Delta^0\Delta^+)/(\Delta^-\Delta^{++})$ (diamonds) and $(\pi^0)^2/(\pi^-\pi^+)$ (circles) averaged over early times $10<t<30$ fm/$c$, in the symmetric ($\delta=0$) and asymmetric ($\delta=0.2$) full-$N\Delta\pi$ systems, shown with the calculated values of $n/p$ for the horizontal axis.  Filled symbols represent the results calculated with the homework time-step parameter $\Delta t$ chosen by the code, while open symbols are with $\Delta t=0.2$ fm/$c$.  The vertical line indicates the value of $n/p$ in the rate-equation solution for the $\delta=0.2$ system.  The corresponding ratios of the rate-equation solution are shown with short horizontal lines for the $\delta=0.2$ system.}
\end{figure*}

As mentioned before in other similar cases, the excess of $\Delta^0$ and $\Delta^+$ relative to $\Delta^-$ and $\Delta^{++}$ indicates a violation of isospin symmetry for some reason in the transport codes, such as uncontrolled effects of higher-order correlations.  In the lower panels of Fig.~\ref{fig:nata}, the $X/X_{\text{rate-eq}}$ ratio is shown for $X=(\Delta^0\Delta^+)/(\Delta^-\Delta^{++})$ (left) and $X=(\pi^0)^2/(\pi^-\pi^+)$ (right) with green dotted lines.  In most of QMD and parallel-ensemble BUU codes, the excess of $\Delta^0$ and $\Delta^+$ is very strong during the first several tens of fm/$c$, as compared to the case without pions (Fig.~\ref{fig:nata-De2P0}).  The same quantity $(\Delta^0\Delta^+)/(\Delta^-\Delta^{++})$ is shown in Fig.~\ref{fig:ratioa_early} with red diamonds for the averaged numbers of $\Delta$ in the early times $10<t<30$ fm/$c$.  The results of both symmetric ($\delta=0$) and asymmetric ($\delta=0.2$) systems are shown against the calculated value of $n/p$.  As we have seen in other cases, deviations from 1 appear independently of the isospin asymmetry of the system, which evidences an unphysical violation of isospin symmetry.  Furthermore, the deviations here are about twice as large as those in the situation of equilibrium at late times (Fig.~\ref{fig:ratioa}) in many codes.  The filled red diamonds represent the results with the homework time step $\Delta t$ originally chosen by individual codes, while the open red diamonds represent those with $\Delta t=0.2$ fm/$c$.  In the majority of codes, the deviations become slightly more serious when $\Delta t$ is reduced.  In the figure, it is observed that the excess of $\pi^0$, shown by blue circles, is correlated with the excess of $\Delta^0$ and $\Delta^+$, as may be expected.

The codes having a strong excess of $\Delta^0$ and $\Delta^+$ in Fig.~\ref{fig:ratioa_early} correspond to those having the $(N\pi)_{\text{nnd}}$ correlation (Fig.~\ref{fig:npid2}), which is qualitatively expected to induce such a violation of isospin symmetry.  Because of the observed coincidence of time scales, we may guess that the enhanced violation at early times may be due to a combined effect of the correlations and the non-equilibrium effect.  In contrast to QMD and parallel-ensemble BUU codes, almost no excess or suppression of $\Delta^0$ and $\Delta^+$ is seen in two full-ensemble BUU codes (pBUU with $\Delta t=0.2$ fm/$c$ and SMASH), which is consistent with the weakness of correlations in these codes.  The behavior of the JAM result (and maybe the IQMD-BNU result) is different from the other QMD codes, which is understandable because the $(N\pi)_{\text{nnd}}$ correlation (Fig.~\ref{fig:npid2}) does not exist in JAM, as a result of its prescribed method of treating collisions.  However, other correlations should exist in the JAM results, and the good agreement may be due to cancellations of many complicated effects.

The three isotopic ratios of Eq.~\eqref{eq:piratios} have been compared in Fig.~\ref{fig:dt_piratioa_early} for the numbers of particles averaged over the early times $10<t<30$ fm/$c$.  As already mentioned in Sec.~\ref{sec:digest}, the agreement of the $\pi$-like ratio among codes is not as good then as at late times.  While significant $\Delta t$ dependence of the $\pi$ ratio (blue circles) is already expected as at late times, it is now important to understand the $\Delta t$ dependence of the $\Delta$($\pi$-like) ratio (red squares).  Its dependence on $\Delta t$ may be due to some effect of the collision--decay sequence, since the isotopic ratios do not necessarily remain constant under $C_k$ at early times, because various collision and decay channels have not balanced out yet.  However, it does not seem easy to explain the $\Delta t$ dependence observed in different codes based on the adopted prescriptions for the collision--decay sequence.  

Another possible idea to interpret the $\Delta t$ dependence of the isotopic ratios is that the $\Delta t$ dependence of correlations can be important at early times, because they are now much stronger than at late times.  For example, as has been mentioned in the context of Fig.~\ref{fig:npid_dndeics}, the $(N\pi)_{\text{nnd}}$ correlation strengthens when $\Delta t$ is reduced to 0.2 fm/$c$ in IBUU, RVUU and TuQMD, while this correlation is already strong even with a large $\Delta t$ in IQMD-IMP and JQMD.  These different situations have been explained based on the adopted prescriptions for the collision--decay sequence.  In the present case at early times, the excess of $\Delta^0$ and $\Delta^+$ in the lower panels of Fig.~\ref{fig:nata} (green dotted line) has increased in many codes (except for JQMD) when $\Delta t$ is reduced to 0.2 fm/$c$.  Similar outcome is observed in Fig.~\ref{fig:ratioa_early} when comparing the red filled diamonds and the red open diamonds.  However, in order to explain the $\Delta$($\pi$-like) ratio, we should also pay attention to the $\Delta^-/\Delta^{++}$ ratio shown by the blue dashed line in the lower panels of Fig.~\ref{fig:nata}.  As discussed before, this ratio is likely affected by the asymmetric strength of the effect of some correlations in asymmetric environments.  For the codes of BUU-VM and IBUU, in which the $\Delta$($\pi$-like) and other ratios decrease when $\Delta t$ is reduced in Fig.~\ref{fig:dt_piratioa_early}, it is consistently found that $(\Delta^0\Delta^+)/(\Delta^-\Delta^{++})$ increases and $\Delta^-/\Delta^{++}$ decreases at the earliest times when $\Delta t$ is reduced, which is consistent with the increasing effect of correlations.  These codes then predict relatively low values of the $\Delta$($\pi$-like) ratio and of the $\pi$-like ratio when the limit of $\Delta t\to 0$ is taken in Fig.~\ref{fig:dt_piratioa_early}.  In contrast, in IQMD-BNU, IQMD-IMP, JQMD, pBUU, RVUU and TuQMD, the $\Delta^-/\Delta^{++}$ ratio increases when $\Delta t$ is reduced, which is the main origin of the $\Delta t$ dependence of the $\Delta$($\pi$-like) ratio in these codes.  These codes tend to predict relatively high values of the $\Delta$($\pi$-like) ratio and of the $\pi$-like ratio for $\Delta t\to 0$.  We have not understood yet what causes these different behaviors of the $\Delta^-/\Delta^{++}$ ratio when $\Delta t$ is changed.

\subsection{Summary of results from early times}

The $\sqrt{s}$ distributions of $NN\leftrightarrow N\Delta$ and $\Delta\leftrightarrow N\pi$ reactions suggest the existence of the non-equilibrium or non-thermal effect in the momentum distributions of both $\Delta$ particles and pions, at early times of a few tens of fm/$c$.  It is likely one of the reasons why the increase in the number of pions is slower in transport codes, than in the rate equation that assumes thermal momentum distributions.  At these early times, many codes show a strong excess of $\Delta^0$ and $\Delta^+$ compared to $\Delta^-$ and $\Delta^{++}$, which is a signature of violation of isospin symmetry caused by correlations.  The correlations then affect the isotopic ratios, such as the $\pi$-like ratio, at early times.  The dependence of the correlations on the time-step size $\Delta t$ seems significant at these early times.  We should, however, better understand the $\Delta t$ dependence of the $\Delta^-/\Delta^{++}$ ratio which may also be affected by correlations.  The reliability of the code predictions is not as good as in the equilibrated situation at late times, in particular when the results are extrapolated to the smallest time step $\Delta t\to 0$.  The agreement may be improved, e.g., if all the codes are modified to remove the $(N\pi)_{\text{nnd}}$ correlation (Fig.~\ref{fig:npid2}).  The effects of other sources of correlations should also be studied carefully, because their significance may depend on the situation where they occur.

\section{Conclusions\label{sec:concl}}

We have compared results from various transport codes, for the production of pions and $\Delta$ resonances, under well-defined conditions of a system in a box with periodic boundary conditions.  One important result is that the $\pi$-like ratio [Eq.~(\ref{eq:piratios}c)] for the ideal gas mixture is reproduced well by all participating transport codes, when the system reaches equilibrium in the box.  This makes it promising to use transport models to extract the high-density symmetry energy from heavy-ion collision data.  However, we have also encountered disagreements in some other quantities, such as the numbers of $N_\Delta$ and $N_\pi$, and the $\pi$ ratio [Eq.~(\ref{eq:piratios}a)].  A rather consistent understanding of these results has been achieved after the detailed analyses in Secs.~\ref{sec:ndelta}, \ref{sec:ndeltapi} and \ref{sec:noneq}, which have already been summarized at the end of each section. The problems have been understood as originating from the processes relevant for $\Delta$ resonances and pions, in which baryons can change their identities and mesons can be created and annihilated.  Figure \ref{fig:dtsummary} gave a good impression of the rather satisfactory agreement achieved between most of the codes in the limit of a small time step.

Two important concepts were found to be relevant for explaining the behaviors of the transport-code results and to understand the sources of remaining uncertainties.  One concerns the ordering in the collision--decay sequence within the same time step, and the other concerns the correlations induced by using the geometrical method for treating collisions.  The former is found to affect the dependence of some quantities, such as $N_\pi$ and the $\pi$ ratio, on the time-step size $\Delta t$, although this problem should disappear in the limit of $\Delta t\to 0$.  The latter affects the results in at least two ways.  One is to increase the collision rates, which, however, does not seem to cause further problems when the system reaches equilibrium in our study.  The other effect due to the correlations, on the other hand, can lead to a violation of the isospin symmetry in the transport codes and thus affect some isotopic ratios.  This problem cannot be reduced by choosing a small $\Delta t$, and can sometimes become stronger when $\Delta t$ is reduced.  As a general remark, the question of correlations is an issue between the assumptions of the Boltzmann or rate equation, which is a one-body theory with a Markovian collision term, and of the simulations for heavy-ion collisions, which may want to include some kind of correlations to describe physical phenomena and observables. These physical issues have to be better understood.

The present comparison has been carried out for the simplest case of a system in a box without mean-field interactions and the Pauli blocking.  If our goal were limited to this case, and to a comparison with the ideal-gas Boltzmann distributions, the best solution would be to use a time-step--free code (JAM or SMASH) to avoid the problem due to the collision--decay sequence, and to solve the Boltzmann equation using a full-ensemble BUU code (pBUU or SMASH) to avoid the problem of correlations.  However, for applications to heavy-ion collisions, over-optimizing the codes for this simplest case may not be desirable.  One may still need models with physical fluctuations and correlations, like the Boltzmann-Langevin approach or the QMD approach, to investigate the effects of, for example, cluster formation in realistic heavy-ion collisions.  We should also consider that the Pauli blocking, which was deactivated in this work, plays important roles in realistic systems, and that the accuracy of evaluation of the blocking factor is affected by fluctuations, depending on the strategies in codes, as was seen in Ref.~\cite{yxzhang2018}.

Fortunately, the uncertainty of $\pm2\%$ in the $\pi$-like ratio in the studied case when the system has equilibrated is small.  On the other hand, the remaining uncertainty due to the above-mentioned problems seems to depend on the various conditions.  At early times before reaching thermal and chemical equilibrium, we have found that the predictions on the $\pi$-like ratio are not as stable as in the equilibrated case, though the uncertainty is still within an acceptable range.  From the basis of the present study under tightly controlled conditions in a simple case, it will be helpful to understand, at least partly, the influence of the various effects in full heavy-ion collisions.  Such a study is currently ongoing.

\begin{acknowledgments}
  A.~Ono acknowledges support from Japan Society for the Promotion of Science KAKENHI Grants No.~JP24105008 and No.~JP17K05432.
  J.~Xu was supported by the Major State Basic Research Development Program (973 Program) of China under Contract No.~2015CB856904, and the National Natural Science Foundation of China under Grant No.~11475243 and No.~11421505.
  M.~Colonna acknowledges the support from the European Unions Horizon 2020 research and innovation programme under Grant Agreement No.~654002.
  P.~Danielewicz was supported by the U.S.\ Department of Energy Office of Science under Grant No.~DE-SC0019209.
  C.M.~Ko was supported by the US Department of Energy under Contract No.~DE-SC0015266 and the Welch Foundation under Grant No.~A-1358. 
  Y.J.~Wang was supported by the National Natural Science Foundation of China under Grants No.~11847315, No.~11875125, and No.~11505057.
  H.H.~Wolter was supported by the Universe Cluster of the German Research Foundation (DFG).
  L.W.~Chen was supported by the National Natural Science Foundation of China under Grant No.~11625521.
  B.A.~Li is supported by the U.S.\ Department of Energy under Award No.~DE-SC0013702.
  T.~Ogawa acknowledges support from Japan Society for the Promotion of Science KAKENHI Grant No.~JP26790072.
  D.~Oliinychenko was supported by the U.S.\ Department of Energy, Office of Science, Office of Nuclear Physics, under contract number DE-AC02-05CH11231 and received support within the framework of the Beam Energy Scan Theory (BEST) Topical Collaboration.
  P.~Danielewicz, M.B.~Tsang, C.M.~Ko and B.A~Li are also supported by the CUSTIPEN (China-U.S.\ Theory Institute for Physics with Exotic Nuclei) under the US Department of Energy Grant No.~DE-SC0009971.
  The computation by the JAM code was carried out at the HOKUSAI supercomputer system of RIKEN.

  \par
  The authors acknowledge the support by the organizers of the Transport Workshop held in Busan, South Korea, September 14--15, 2018, which was organized for the discussions on the present work and related subjects.
  The writing committee (consisting of the first nine authors) would like to acknowledge the generous financial support from the director of the Facility for Rare Isotope Beams (FRIB) in hosting writing sessions at the 2017 International Collaboration in Nuclear Theory (ICNT).
  Writing sessions by J.X., A.O., M.B.T.\ and Y.X.Z.\ were also hosted at the Chinese Institute of Atomic Energy (CIAE), with support from CIAE funding and State Administration of Foreign Experts Affairs, No.~T170517002.
  Other writing sessions by A.O., J.X., C.M.K., M.B.T.\ and Y.X.Z.\ were hosted at Sun Yat-sen University.
  The stay of A.O.\ during the ICNT meeting was supported by National Science Foundation under Grant No.~PHY-1430152 (JINA Center for the Evolution of the Elements).
\end{acknowledgments}

\appendix
\section{Equilibrated ideal Boltzmann-gas mixture\label{sec:equil}}
Since the Boltzmann equation does not take into account the fermionic and bosonic characters of baryons and mesons in the present homework, the corresponding equilibrium distribution is that of a relativistic Boltzmann gas.  At temperature $T$ and chemical potential $\mu_\alpha$, the phase-space distribution of the particle species $\alpha\in N,\pi,\Delta$ is
\begin{equation}
  \label{eq:f-thermal}
  g_\alpha f_\alpha(p)=\tilde{g}_\alpha e^{-(\sqrt{m_\alpha^2+p^2}-\mu_\alpha)/T},
\end{equation}
where the spectral function for $\Delta$ has been absorbed in the degeneracy factor,
\begin{equation}
  \tilde{g}_\alpha=\begin{cases}
    g_\alpha & \alpha\in N, \pi\\
    g_\alpha A(m_\alpha) & \alpha\in \Delta.
  \end{cases}
\end{equation}
The number density of this particle species is
\begin{equation}
  \rho_\alpha=g_\alpha \int\frac{d^3p}{(2\pi)^3} f_\alpha(p)
  =\tilde{g}_\alpha 
     \frac{e^{\mu_\alpha/T}}{2\pi^2}m_\alpha^2 T K_2(m_\alpha/T),
\end{equation}
where $K_n$ is the $n$-th order modified Bessel function of the second kind.  The average energy per particle is
\begin{equation}
\begin{split}
  e(T,m_\alpha) &=\frac{g_\alpha}{\rho_\alpha} \int\frac{d^3 p}{(2\pi)^3}
                      \sqrt{m_\alpha^2+p^2}\, f_\alpha(p)\\
  \label{eq:e-per-particle}
  &=m_\alpha K_1(m_\alpha/T)/K_2(m_\alpha/T)+3T.
\end{split}
\end{equation}
At chemical equilibrium, the chemical potentials for $\pi$ and $\Delta$ are related to those of neutron ($\mu_n$) and proton ($\mu_p$) by
\begin{equation}
\label{eq:chemeq-pi}
\begin{split}
\mu_{\pi^-} &= \mu_n-\mu_p,\\
  \mu_{\pi^0} &= 0,\\
  \mu_{\pi^+} &= \mu_p-\mu_n
\end{split}
\end{equation}
and
\begin{equation}
\label{eq:chemeq-delta}
\begin{split}
  \mu_{\Delta^-}&= 2\mu_n-\mu_p,\\
  \mu_{\Delta^0}&= \mu_n,\\
  \mu_{\Delta^+}&= \mu_p,\\
  \mu_{\Delta^{++}}&= 2\mu_p-\mu_n.
  \end{split}
\end{equation}
Using above equations, we can determine the values of $T$, $\mu_n$, and $\mu_p$ from the total energy, baryon number and charge in the system.

Since the $\Delta$ resonances with different masses are  
considered as distinguished particles labeled by the index $\alpha$, their number in a particular charge state, such as the $\Delta^0$, is given by
\begin{equation}
 \begin{split}
   N_{\Delta^0}
   &=\sum_{\alpha\in\Delta^0}\rho_\alpha V\\
   &=4\frac{e^{\mu_n/T}}{2\pi^2}VT
     \int_{m_N+m_\pi}^\infty\frac{dm}{2\pi} A(m)
     m^2 K_2(m/T)
 \end{split}
\end{equation}
with $V$ being the volume of the system.

\section{Rate equation\label{sec:rateeq}}

Characteristics of an equilibrated system are discussed in Appendix \ref{sec:equil}.  Here we discuss rate equations where we assume a system that is equilibrated kinematically but not chemically.  With this we consider here the rate equations for the number densities $\{\rho_\alpha(t);\ \alpha\in N,\pi,\Delta\}$ as functions of time.  Without chemical equilibrium, Eqs.~\eqref{eq:chemeq-pi} and \eqref{eq:chemeq-delta} are not satisfied.  Moreover, since $\Delta$ particles with different masses are distinguished by the index $\alpha$, there are an uncountable number of independent variables in the absence of chemical equilibrium.  However, with the momentum distributions of all  particles $\alpha$ described by the thermal distributions [Eq.~\eqref{eq:f-thermal}], the temperature $T$ can be determined from the densities $\{\rho_\alpha\}$ and total energy $E_{\text{total}}$ by the energy conservation relation,
\begin{equation}
  \sum_{\alpha}e(T,m_\alpha)\rho_\alpha=E_{\text{total}}/V,
\end{equation}
with $e(T,m_\alpha)$ given by Eq.~\eqref{eq:e-per-particle}.

With the thermal distribution of Eq.~\eqref{eq:f-thermal} or
\begin{equation}
  f_\alpha(p)=\rho_\alpha \frac{2\pi^2}{m_\alpha^2TK_2(m_\alpha/T)}
  e^{-(1/T)\sqrt{m_\alpha^2+p^2}},
\end{equation}
integrating the Boltzmann equation \eqref{eq:boltz} over the momentum yields the rate equation for $\rho_\alpha(t)$,
\begin{equation}
  \frac{d\rho_\alpha}{dt}
  = \sum_\beta \sum_{\gamma\le\delta}R_{\alpha\beta\leftrightarrow\gamma\delta}
  + \sum_\beta\sum_\gamma R_{\alpha\beta\leftrightarrow\gamma}
  + \sum_{\beta\le\gamma} R_{\alpha\leftrightarrow\beta\gamma}
\end{equation}
with
\begin{align}
  R_{\alpha\beta\leftrightarrow\gamma\delta}
  &=\lambda_{\gamma\delta\to\alpha\beta}(1+\delta_{\alpha\beta})
  \frac{\rho_\gamma\rho_\delta}{1+\delta_{\gamma\delta}}
  -\lambda_{\alpha\beta\to\gamma\delta}\rho_\alpha\rho_\beta,
  \\
  R_{\alpha\beta\leftrightarrow\gamma}
  &=\lambda_{\gamma\to\alpha\beta}\rho_\gamma
  -\lambda_{\alpha\beta\to\gamma}\rho_\alpha\rho_\beta,
  \\
  R_{\alpha\leftrightarrow\beta\gamma}
  &=\lambda_{\beta\gamma\to\alpha}\rho_\beta\rho_\gamma
  -\lambda_{\alpha\to\beta\gamma}\rho_\alpha.
\end{align}
The temperature-dependent coefficients $\lambda$ are the average values of $v'\sigma$ and $\Gamma'$ in Eqs.~\eqref{eq:icoll-22}, \eqref{eq:icoll-21} and \eqref{eq:icoll-12} for thermal distributions.  They can be expressed as
\begin{align}
  \lambda_{\alpha\beta\to\gamma\delta}
  &=\int_{M_{\text{th}}}^\infty
    \frac{d\lambda_{\alpha\beta\to\gamma\delta}}{d\sqrt{s}} d\sqrt{s},
  \label{eq:lambda_coll}\\
  \frac{d\lambda_{\alpha\beta\to\gamma\delta}}{d\sqrt{s}} 
  &=\frac{[p^*(s,m_\alpha,m_\beta)]^2 s K_1(\sqrt{s}/T)
    \sigma_{\alpha\beta\to\gamma\delta}}
  {m_\alpha^2 m_\beta^2 T K_2(m_\alpha/T) K_2(m_\beta/T)},\label{eq:dlambdadsqrts}
  \\    
  \lambda_{\alpha\beta\to\gamma}
  &=\frac{[p^*(m_\gamma^2,m_\alpha,m_\beta)]^2
    m_\gamma^2 K_1(m_\gamma/T)\sigma_{\alpha\beta\to\gamma}}
  {m_\alpha^2 m_\beta^2 T K_2(m_\alpha/T) K_2(m_\beta/T)},
  \\
  \lambda_{\alpha\to\beta\gamma}
  &=\frac{K_1(m_\alpha/T)}{K_2(m_\alpha/T)}\Gamma_{\alpha\to\beta\gamma},
\end{align}
with the function $p^*$ defined in Eq.~\eqref{eq:pincm}.  The lower bound of the $\sqrt{s}$ integration is $M_{\text{th}}=\max(m_\alpha+m_\beta,\ m_\gamma+m_\delta)$.

\bibliography{ono_nucl}

\end{document}